\documentclass[aps,preprint,groupedaddress]{revtex4}
\usepackage{amsfonts}
\usepackage{mathrsfs}
\usepackage{graphicx}
\usepackage{subfigure}
\usepackage{amsmath}
\usepackage{amssymb}
\usepackage{amsbsy}
\usepackage{bm}
\usepackage{xcolor}

\usepackage{algorithm}
\usepackage{algpseudocode}

\newcommand{\nin}{N_{\rm in}}

\begin{document}

\title{Ensemble perspective for understanding temporal credit assignment}
\author{Wenxuan Zou}
\thanks{W.Z. and C.L. contributed equally to this work.}
\author{Chan Li}
\thanks{W.Z. and C.L. contributed equally to this work.}
\author{Haiping Huang}
\email{huanghp7@mail.sysu.edu.cn}
\affiliation{PMI Lab, School of Physics,
Sun Yat-sen University, Guangzhou 510275, People's Republic of China}
\date{\today}

\begin{abstract}
Recurrent neural networks are widely used for modeling spatio-temporal sequences in both 
nature language processing and neural population dynamics. However, understanding the temporal credit 
assignment is hard. Here, we propose that each individual connection in the recurrent computation is modeled by
a spike and slab distribution, rather than a precise weight value. We then derive the mean-field algorithm to train
the network at the ensemble level. The method is then applied to classify handwritten digits when pixels are read in
sequence, and to the multisensory integration task that is a fundamental cognitive function of animals. 
Our model reveals important connections that determine the overall performance of the network.
The model also shows how spatio-temporal information is processed through the hyper-parameters of the distribution, and moreover reveals
distinct types of emergent neural selectivity. To provide a mechanistic analysis of the ensemble learning, we first derive an analytic solution of
the learning at the infinitely-large-network limit. We then carry out a low-dimensional projection of
both neural and synaptic dynamics, analyze symmetry breaking in the parameter space, and finally demonstrate the role of stochastic plasticity in the recurrent computation.
Therefore, our study sheds light on mechanisms of how weight uncertainty impacts the temporal credit assignment in recurrent neural networks
from the ensemble perspective.
\end{abstract}

 \maketitle

\section{Introduction}
Recurrence is ubiquitous in the brain. Neural networks with reciprocally connected recurrent units are called recurrent neural networks (RNN). Because of feedback supports provided by
these recurrent units, this type of neural networks is able to maintain information about sensory inputs across temporal domains, and thus 
plays an important role in processing time-dependent sequences, thereby widely used as a basic computational block in nature language processing~\cite{LSTM-1997,Graves-2013,Seq-2014,Bah-2015,Cho-2014,Chung-2014} and 
even modeling brain dynamics in various kinds of neural circuits~\cite{Sussillo-2020,Murray-2019,Buonomano-2009,Miconi-2017}.

Training RNNs is in general very challenging, because of the intrinsic difficulty in capturing long-term dependence of the sequences.
Advanced architectures commonly introduce gating mechanisms, e.g., the long short-term memory network (LSTM) with multiplicative gates controlling
the information flow across time steps~\cite{LSTM-1997}, or a simplified variant---gated recurrent unit network (GRU)~\cite{Chung-2014}. All these recurrent 
neural networks are commonly trained by backpropagation through time (BPTT)~\cite{Elman-1990,Werbos-1990}, which sums up all gradient (of the loss function) contributions over
all time steps of a trial, to update the recurrent network parameters. The training is terminated until a specific network yields a satisfied generalization accuracy on
unseen trials (time-dependent sequences). This specific network is clearly a point-estimate of the candidate architecture realizing the desired computational task.
A recent study of learning (spatial) credit assignment suggests that an ensemble of candidate networks, instead of the traditional point-estimate, can be 
successfully learned at a cheap computational cost, and particularly yields statistical properties of the trained network~\cite{Li-2020}.
The ensemble training is achieved by only defining a spike and slab (SaS) probability distribution for each network connection, which offers an opportunity to
look at the relevance of each connection to the macroscopic behavioral output of the network. Therefore, we expect that a similar perspective applies to
the RNNs, and an ensemble of candidate networks can also emerge during training of the SaS probability distributions for RNNs.

Weight uncertainty is a key concept in studying neural circuits~\cite{Prob-2013}. The stochastic nature of computation appears not only in the earlier sensory layers but also in deep layers of internal dynamics.
Revealing the underlying mechanism about how the uncertainty is combined with the recurrent dynamics becomes essential to interpret the behavior of RNNs.
In particular, addressing how a RNN learns a probability distribution over weights could provide potential insights towards understanding of learning in the brain.
There may exist a long-term dependence in the recurrent dynamics, and both directions of one connection may carry distinct information in a spatio-temporal domain. How a training at the ensemble level combines
these intrinsic properties of RNNs thus becomes intriguing. In this work, we derive a mean-field method to train RNNs, considering a weight distribution for each direction of connection.
We then test our method on both MNIST dataset in a sequence-learning setting~\cite{Le-2015} and multi-sensory integration tasks~\cite{Shams-2008,MSI-2009,MSIrv-2017}. The multi-sensory integration tasks are
 relevant to computational modeling of cognitive 
experiments of rodents and primates.

By analyzing the distribution of the model parameters, and its relationship with the computational task, and moreover
the selectivity of each recurrent unit, we are able to provide mechanistic
understanding about the recurrent dynamics in both engineering applications and
computational neuroscience modeling. Our method learns the statistics of the RNN ensemble, producing a dynamic architecture
that adapts to temporally varying inputs, which thereby goes beyond traditional training protocols focusing on a stationary network topology.
The hyper-parameters governing the weight distribution reveal which credit assignments are critical to the network behavior, 
which further explains distinct functions of each computational layer and the emergent neural selectivity. Therefore, our ensemble theory can be used
as a promising tool to explore internal dynamics of widely used RNNs.

To reveal the underlying mechanisms about how the ensemble learning works, we carry out a low-dimensional projection of both neural and synaptic dynamics, which displays non-trivial 
hidden structures related to the success of the learning. Moreover, we observe how symmetry breaking emerges during learning in the hyper-parameter space, and design a toy model of learning to
show the role of the stochastic plasticity capturing the weight uncertainty in the accuracy of the recurrent computation. In addition, we derive an analytic solution of the learning when the network is sufficiently large,
which explains
the lazy regime of the ensemble training. These mechanistic analyses provide deep insights towards understanding the temporal credit assignment.

\section{Model and Ensemble Training}
In this study, we consider a recurrent neural network
processing a time-dependent input $\bm{x}(t)$ of time length $T$. The input signal is sent to
the recurrent reservoir via the input weight matrix $\bm{W^{\rm in}}$. The number of input 
units $\nin$ are determined
by design details of the task, and $w_{ij}^{\rm in}$ denotes 
the weight value for the connection from input unit $j$ to recurrent node $i$. The neural responses of recurrent units are represented by an 
activity vector $\bm{r}(t)$ at time step $t$. We define $w_{ij}$ 
as the connection from node $j$ to node $i$ in the reservoir.
In general, $w_{ij} \neq w_{ji}$, indicating different weights for different
directions of information flow. In addition, we do not preclude the self-interaction $w_{ii}$, 
which can be
used to maintain the representation encoded in the history of the internal dynamics~\cite{Hu-2018}. 
The specific
statistics of this self-interaction can be determined by the learning shown below.
The internal activity $\bm{r}(t)$ is read out through the output matrix $\bm{W^{\rm out}}$ 
in the form of the time-dependent output $\bm{y}(t)$ whose cardinality is determined by the specific task. 

We first define $h_i(t)$ as the time-dependent synaptic current of neuron $i$.
The dynamics of the recurrent network can thus be summarized as follows, 
\begin{subequations}\label{rnndyn}
\begin{align}
 h_{i}(t+1)&=(1-\alpha)h_{i}(t)+\alpha u_{i}(t+1)+\sqrt{2\alpha\sigma^{2}} n_i ,\\
 u_{i}(t+1)&=\sum_{j=1}^{N}{w_{ij}r_{j}(t)}+\sum_{j=1}^{\nin}{w_{ij}^{\rm in}x_{j}(t+1)},\\
 r_i(t)&=\phi(h_i(t)),\\
 z_{k}(t)&=\sum_{i=1}^{N}{w_{ki}^{\rm out}r_{i}(t)}, \\
 y_{k}(t)&=f(z_{k}(t)),
\end{align}
\end{subequations}
where $u(\cdot)$ is the pre-activation function, $\phi( \cdot )$ denotes the nonlinear transfer
function, for which
we use the rectified linear unit (ReLU) function
for all tasks. $\alpha=\frac{\Delta t}{\tau}$, where $\Delta t$ is a small
time interval (e.g., emerging from discretization of a continuous dynamics~\cite{OU-1996}), 
and $\tau$ denotes the time constant of the dynamics. $n_i\sim\mathcal N (0, 1)$ 
indicates a normally distributed random number with zero mean and unit variance,
sampled independently at every time step, and $\sigma$ controls the strength
of the recurrent noise intrinsic to the network. This intrinsic noise is present
in modeling cognitive tasks,
but absent for engineering
applications. Considering a 
Gaussian white noise and a baseline ($\bm{x}_0$), we write the input vector $\bm{x}(t)$ as 
\begin{equation}\label{rnnin}
\bm{x}(t) =\bm{x}_{0} + \bm{x}_{t}^{\rm task}+\sqrt{2\sigma_{\rm in}^2/\alpha}\boldsymbol{\xi},
\end{equation}
where the input has been written in a discrete time form, $\bm{x}_t^{\rm task}$ denotes the sequence of the task,
and $\xi_i\sim\mathcal{N}(0,1)$, which is independently sampled at each time step. 
The noise term is 
also called the external sensory noise of strength $\sigma_{\rm in}$,
commonly observed in brain circuits~\cite{Song-2016,MSI-2012}, but is absent in
engineering applications.

For a MNIST classification task, $f(\cdot)$ is chosen to be the softmax function, 
because the softmax function can specify the probability over all the classes at the
last time step $T$, i.e., $y_{k}(T) = \frac{e^{z_{k}(T)}}{\sum_{j}{e^{z_{j}(T)}}}$.
We define $\hat{y}_{k}$ as the target label (one-hot form assuming a single peak of probability at the precise location of that digit), and use the cross
entropy $\mathcal{L} = -\sum_{k}{\hat{y}_{k}\ln{y_{k}(T)}}$ as the loss function to be minimized.
For a multisensory integration task, $f(\cdot)$ is an identity function, and the mean squared error (MSE) is chosen to 
be the objective function, as is commonly used in computational neuroscience studies~\cite{Miconi-2017,Song-2016}.

\begin{figure}
\centering
     \includegraphics[bb=3 3 707 282,width=0.8\textwidth]{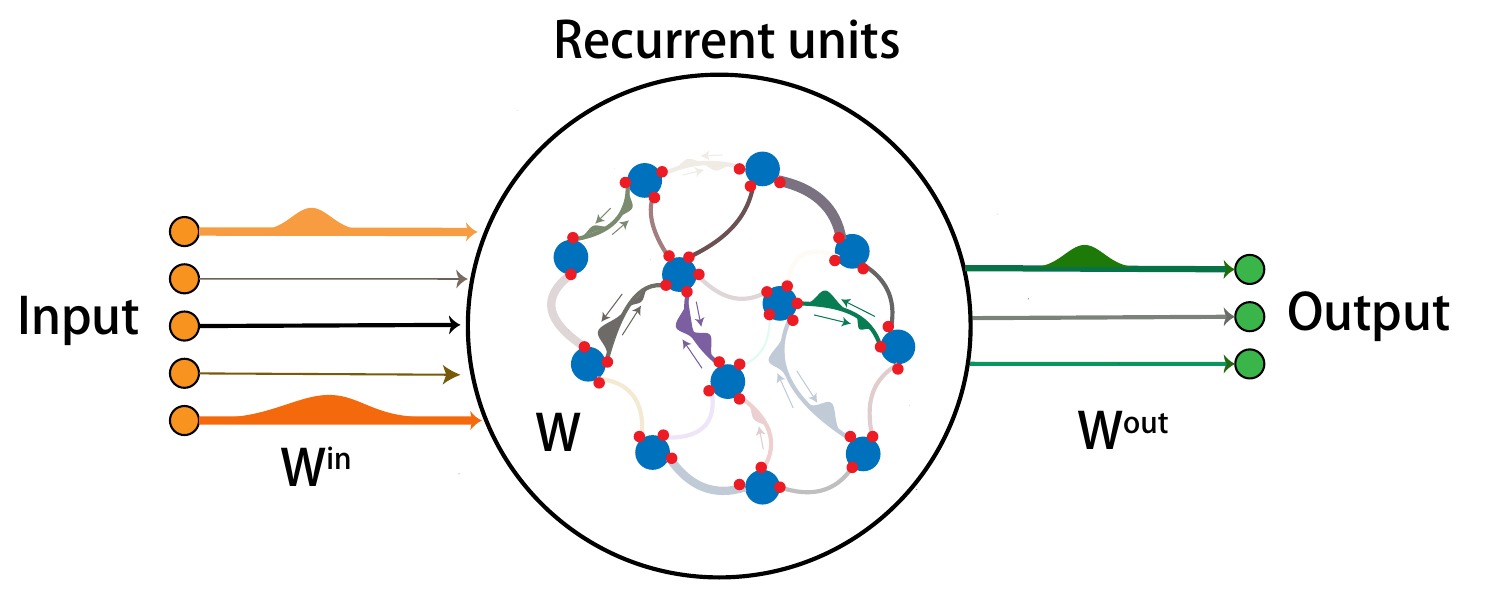}
  \caption{
  Illustration of a RNN learning the temporal credit assignment.
 Each connection is described by the spike and slab distribution;
the spike indicates the corresponding synaptic weight absent for a task, while
the slab represents a Gaussian distribution responsible for the 
weight uncertainty. The Gaussian slab is displayed in the plot, and the arrow in the reservoir
shows different directions of information flow. Different sizes of connection show 
how probable these connections should 
be present during recurrent computation.
  }\label{model}
\end{figure}

To search for the optimal random network ensemble for time-dependent tasks 
in recurrent neural networks, we model the statistics of the weight matrices by 
the SaS probability distribution~\cite{Li-2020} as follows,
\begin{subequations}
\begin{align}
P(w_{ij}^{\rm in})&=\pi_{ij}^{\rm in}\delta(w_{ij}^{\rm in})+(1-\pi_{ij}^{\rm in})\mathcal{N}(w_{ij}^{\rm in}|m_{ij}^{\rm in},\Xi_{ij}^{\rm in}),\\
P(w_{ij})&=\pi_{ij}\delta(w_{ij})+(1-\pi_{ij})\mathcal{N}(w_{ij}|m_{ij},\Xi_{ij}),\\
P(w_{ki}^{\rm out})&=\pi_{ki}^{\rm out}\delta(w_{ki}^{\rm out})+(1-\pi_{ki}^{\rm out})\mathcal{N}(w_{ki}^{\rm out}|m_{ki}^{\rm out},\Xi_{ki}^{\rm out}).
\end{align}
\end{subequations}
The spike mass at $\delta(\cdot)$ is related to 
the network compression,
indicating the necessary weight resources required for a specific task. In other words, this term drives the sparsity of the working network.
The continuous slab, $\mathcal {N} (w_{ij}|m_{ij}, \Xi_{ij})$, denotes
the Gaussian distribution with mean $m_{ij}$ and variance  $\Xi_{ij}$, 
characterizing the weight uncertainty when the corresponding connection 
can not be absent (see Fig.~\ref{model}). The SaS distribution was first introduced in studying Bayesian variable selection in regression problems~\cite{sas-1988}.
Here, we adapt the distribution to learn a recurrent neural network with both sparse architectures and weight uncertainty supporting stochastic synaptic plasticity.

Next, we derive the learning equations about how the SaS parameters are updated, based on mean-field approximation. More precisely, 
we consider the average over the statistics
of the network ensemble during training. Notice that, 
the first and second moments of the weight ${w}_{ij}$ for three sets of weights can 
be written in a common form as $\mu_{ij}=(1-\pi_{ij})m_{ij}$
and $\varrho_{ij}=(1-\pi_{ij})((m_{ij})^{2}+\Xi_{ij})$.
Given a large fan-in, the pre-activation $u_{i}(t)$ and the output $z_{i}(t)$ can 
be re-parametrized by using standard Gaussian random variables
, i.e., it is reasonable to assume that they are subject to 
$\mathcal{N}(u_{i}(t)| G^{\rm in}_{i}(t)+G_{i}^{\rm rec}(t-1), \sqrt{(\Delta_{i}^{\rm in}(t))^{2}+(\Delta_{i}^{\rm rec}(t-1))^{2}})$ 
and $\mathcal{N}(z_{i}(t)| G_{i}^{\rm out}(t), \Delta_{i}^{\rm out}(t))$, respectively, 
according to the central-limit-theorem. Note that when the number of fan-in 
to a recurrent unit is small,
the central-limit-theorem may break. In this situation, we take the deterministic limit. 
Therefore, the mean-field dynamics of the model becomes
\begin{subequations}\label{MFP}
\begin{align}
h_{i}(t+1)&=(1-\alpha)h_{i}(t)+\alpha u_{i}(t+1)+\sqrt{2\alpha\sigma^{2}}n_i ,\\
u_{i}(t+1)&=G_{i}^{\rm rec}(t)+G_{i}^{\rm in}(t+1)+\epsilon_{i}^{\rm u}(t+1)\sqrt{(\Delta_{i}^{\rm in}(t+1))^{2}+(\Delta_{i}^{\rm rec}(t))^{2}},\\
r_i(t)&=\phi(h_i(t)),\\
z_{k}(t)&=G_{k}^{\rm out}(t)+\epsilon_{k}^{\rm out}(t)\Delta_{k}^{\rm out}(t),\\
y_{k}(t)&=f(z_{k}(t)), 
\end{align}
\end{subequations}
where 
\begin{subequations}
\begin{align}
G_{i}^{\rm in}(t+1)&=\sum_{j}{\mu_{ij}^{\rm in}x_{j}(t+1)},\\
G_{i}^{\rm rec}(t+1)&=\sum_{j}{\mu_{ij}r_{j}(t+1)},\\
G_{k}^{\rm out}(t+1)&=\sum_{i}{\mu_{ki}^{\rm out}r_{i}(t+1)},\\
(\Delta_{i}^{\rm in}(t+1))^{2}&=\sum_{j}{(\varrho_{ij}^{\rm in}-(\mu_{ij}^{\rm in})^{2})(x_{j}(t+1))^{2}},\\
(\Delta_{i}^{\rm rec}(t+1))^{2}&=\sum_{j}{(\varrho_{ij}-(\mu_{ij})^{2})(r_{j}(t+1))^{2}},\\
(\Delta_{k}^{\rm out}(t+1))^{2}&=\sum_{i}{(\varrho_{ki}^{\rm out}-(\mu_{ki}^{\rm out})^{2})(r_{i}(t+1))^{2}}.
\end{align}
\end{subequations}
Note that $\{\bm{\epsilon}^{\rm u}(t)\}$ and $\{\bm{\epsilon}^{\rm out}(t)\}$ are both independent
random variables 
sampled from the standard Gaussian distribution with zero mean and unit variance, which are quenched 
for every single training mini-epoch and also time-step dependent, maintaining the same sequence of values 
in both feedforward and backward computations.

Updating the network parameters $(\pmb{\theta}_{ik}^{\rm in}, \pmb{\theta}_{ik}, \pmb{\theta}_{ki}^{\rm out})$ 
can be achieved by the gradient descent on the objective function $\mathcal{L}$.
First of all, we update $\pmb{\theta}_{ki}^{\rm out}\equiv(m_{ki}^{\rm out},\pi_{ki}^{\rm out},\Xi_{ki}^{\rm out})$.
\begin{subequations}
\begin{align}
\frac{\partial \mathcal{L}}{\partial m_{ki}^{\rm out}}&=\sum_{t=0}^{T}\frac{\partial \mathcal{L}}{\partial z_{k}(t)} \frac{\partial z_{k}(t)}{\partial m_{ki}^{\rm out}},\\
\frac{\partial \mathcal{L}}{\partial \pi_{ki}^{\rm out}}&=\sum_{t=0}^{T}\frac{\partial \mathcal{L}}{\partial z_{k}(t)} \frac{\partial z_{k}(t)}{\partial \pi_{ki}^{\rm out}},\\
\frac{\partial \mathcal{L}}{\partial \Xi_{ki}^{\rm out}}&=\sum_{t=0}^{T}\frac{\partial \mathcal{L}}{\partial z_{k}(t)} \frac{\partial z_{k}(t)}{\partial \Xi_{ki}^{\rm out}},
\end{align}
\end{subequations}
where $\frac{\partial \mathcal{L}}{\partial z_{k}(t)}$ is related to the form of the loss function.
For categorization tasks, $f(\cdot)$ is chosen to be the 
softmax function, and we use the cross entropy as our objective function, for which 
 $\frac{\partial \mathcal{L}}{\partial z_{k}(t)}=(y_{k}(T)-\hat{y}_{k})\delta_{t,T}$. It is worth noticing that
 for multi-sensory integration tasks, this derivative does not vanish at intermediate time steps and become thus time-dependent.
The other term $\frac{\partial z_{k}(t)}{\partial \theta_{ki}^{\rm out}}$ can be 
directly computed, showing how sensitive the network activity is read out under the change of
the SaS parameters in the output layer:
\begin{subequations}
\begin{align}
\frac{\partial z_{k}(t)}{\partial m_{ki}^{\rm out}}&=(1-\pi_{ki}^{\rm out})r_{i}(t)+\frac{\epsilon_{k}^{\rm out}(t) (\mu_{ki}^{\rm out}\pi_{ki}^{\rm out})(r_{i}(t))^{2}}{\Delta_{k}^{\rm out}},\\
\frac{\partial z_{k}(t)}{\partial \pi_{ki}^{\rm out}}&=-m_{ki}^{\rm out}r_{i}(t)+\frac{\epsilon_{k}^{\rm out}(t)((m_{ki}^{\rm out})^{2}(1-2\pi_{ki}^{\rm out})-\Xi_{ki}^{\rm out})(r_{i}(t))^{2}}{2\Delta_{k}^{\rm out}},\\
\frac{\partial z_{k}(t)}{\partial \Xi_{ki}^{\rm out}}&=\frac{\epsilon_{k}^{\rm out}(t)(1-\pi_{ki}^{\rm out})(r_{i}(t))^{2}}{2\Delta_{k}^{\rm out}}.
\end{align}
\end{subequations}

Next, we derive the learning equation for the hyper-parameters in the reservoir and input 
layer, i.e., $\pmb{\theta}_{ij}\equiv(m_{ij},\pi_{ij},\Xi_{ij})$ and 
$\pmb{\theta}_{ij}^{\rm in}\equiv(m_{ij}^{\rm in},\pi_{ij}^{\rm in},\Xi_{ij}^{\rm in})$.
To get a general form, we set $\tilde{m}_{ij}\equiv(m_{ij}, m_{ij}^{\rm in}), \tilde{\pi}_{ij}\equiv(\pi_{ij},  \pi_{ij}^{\rm in})$, 
and $ \tilde{\Xi}_{ij}\equiv(\Xi_{ij}, \Xi_{ij}^{\rm in})$. Based on the chain rule, we then arrive at the following equations,
\begin{subequations}
\begin{align}
\frac{\partial \mathcal{L}}{\partial \tilde{m}_{ij}}&=\sum_{t=0}^{T}\frac{\partial \mathcal{L}}{\partial h_{i}(t)} \frac{\partial h_{i}(t)}{\partial u_{i}(t)} \frac{\partial u_{i}(t)}{\partial \tilde{m}_{ij}} = \sum_{t=0}^{T}\alpha \delta_{i}(t)\frac{\partial u_{i}(t)}{\partial \tilde{m}_{ij}},\\
\frac{\partial \mathcal{L}}{\partial \tilde{\pi}_{ij}}&=\sum_{t=0}^{T}\frac{\partial \mathcal{L}}{\partial h_{i}(t)} \frac{\partial h_{i}(t)}{\partial u_{i}(t)} \frac{\partial u_{i}(t)}{\partial \tilde{\pi}_{ij}}=\sum_{t=0}^{T}\alpha \delta_{i}(t) \frac{\partial u_{i}(t)}{\partial \tilde{\pi}_{ij}} ,\\
\frac{\partial \mathcal{L}}{\partial \tilde{\Xi}_{ij}}&=\sum_{t=0}^{T}\frac{\partial \mathcal{L}}{\partial h_{i}(t)} \frac{\partial h_{i}(t)}{\partial u_{i}(t)} \frac{\partial u_{i}(t)}{\partial \tilde{\Xi}_{ij}} =\sum_{t=0}^{T}\alpha  \delta_{i}(t)  \frac{\partial u_{i}(t)}{\partial \tilde{\Xi}_{ij}},
\end{align}
\label{eq1}
\end{subequations}
where we have defined $\delta_{i}(t)\equiv\frac{\partial \mathcal{L}}{\partial h_{i}(t)}$. The auxiliary variable $\delta_{i}(t)$ can be
computed by the chain rule once again, resulting in a BPTT equation of the error signal starting from
the last time-step $T$:
\begin{equation}
\delta_{i}(t)=\sum_{j}{\frac{\partial\mathcal{L}}{\partial h_{j}(t+1)} \frac{\partial h_{j}(t+1)}{\partial h_{i}(t)}}+\sum_{k}{\frac{\partial\mathcal{L}}{\partial z_{k}(t)} \frac{\partial z_{k}(t)}{\partial r_{i}(t)}}\phi'(h_i(t)),
\end{equation}
where $t = 0, 1 ,2, ... , T-1$, and $\phi'(\cdot)$ 
denotes the derivative of the transfer function. The first summation in $\delta_{i}(t)$ can be directly expanded as
\begin{subequations}
\begin{align}
&\sum_{j}{\frac{\partial\mathcal{L}}{\partial h_{j}(t+1)} \frac{\partial h_{j}(t+1)}{\partial h_{i}(t)}}
=(1-\alpha)\delta_{i}(t+1)+\sum_{j}{\alpha \delta_{j}(t+1)\frac{\partial u_{j}(t+1)}{\partial h_{i}(t)}}, \\
&\frac{\partial u_{j}(t+1)}{\partial h_{i}(t)}=\frac{\partial u_j(t+1)}{\partial r_i(t)}\frac{\partial r_i(t)}{\partial h_i(t)}=(1-\pi_{ji})m_{ji}\phi'(h_i(t))+\epsilon_{j}^{\rm u}(t+1) \frac{(\varrho_{ji}-(\mu_{ji})^{2})r_{i}(t)\phi'(h_i(t))}{\sqrt{((\Delta_{j}^{\rm in}(t+1))^{2}+(\Delta_{j}^{\rm rec}(t))^{2})}}.
\end{align}
\end{subequations}
The second summation in $\delta_{i}(t)$ is given by
\begin{equation}
\sum_{k}{\frac{\partial\mathcal{L}}{\partial z_{k}(t)} \frac{\partial z_{k}(t)}{\partial h_{i}(t)}}=\sum_{k}\frac{\partial\mathcal{L}}{\partial z_{k}(t)}\times \left[\mu_{ki}^{\rm out}+\epsilon_{k}^{\rm out}(t) \frac{(\varrho_{ki}^{\rm out}-(\mu_{ki}^{\rm out})^{2})r_{i}(t)}{\Delta_{k}^{\rm out}(t)}\right]\phi'(h_i(t)).
\end{equation}
It is worth noting that for the last time-step $T$, the error signal $\delta_{i}(T)$ 
is written as
\begin{equation}
\delta_{i}(T)=\sum_{k}{\frac{\partial\mathcal{L}}{\partial z_{k}(T)}\times \left[\mu_{ki}^{\rm out}+\epsilon_{k}^{\rm out}(T) \frac{(\varrho_{ki}^{\rm out}-(\mu_{ki}^{\rm out})^{2})r_{i}(T)}{\Delta_{k}^{\rm out}(T)}\right]\phi'(h_i(T))}.
\end{equation}
To compute Eq.~(\ref{eq1}), we need to work out the following derivatives, which characterize the 
sensitivity of the pre-activation under the change of 
hyper-parameters $\pmb{\theta}_{ij}$ and $\pmb{\theta}^{\rm in}_{ij}$. We summarize the results as follows, 
\begin{subequations}
\begin{align}
\frac{\partial u_{i}(t)}{\partial m_{ij}^{\rm in}}&=(1-\pi_{ij}^{\rm in})x_{j}(t)+\epsilon_{i}^{\rm u}(t)\frac{\mu_{ij}^{\rm in}\pi_{ij}^{\rm in}(x_{j}(t))^{2}}{\sqrt{(\Delta_{i}^{\rm in}(t))^{2}+(\Delta_{i}^{\rm rec}(t-1))^{2}}},\\
\frac{\partial u_{i}(t)}{\partial \pi_{ij}^{\rm in}}&=-m_{ij}^{\rm in}x_{j}(t)+\epsilon_{i}^{\rm u}(t)\frac{((m_{ij}^{\rm in})^{2}(1-2\pi_{ij}^{\rm in})-\Xi_{ij}^{\rm in})(x_{j}(t))^{2}}{2\sqrt{(\Delta_{i}^{\rm in}(t))^{2}+(\Delta_{i}^{\rm rec}(t-1))^{2}}},\\
\frac{\partial u_{i}(t)}{\partial \Xi_{ij}^{\rm in}}&=\epsilon_{i}^{\rm u}(t)\frac{(1-\pi_{ij}^{\rm in})(x_{j}(t))^{2}}{2\sqrt{(\Delta_{i}^{\rm in}(t))^{2}+(\Delta_{i}^{\rm rec}(t-1))^{2}}},\\
\frac{\partial u_{i}(t)}{\partial m_{ij}}&=(1-\pi_{ij})r_{j}(t-1)+\epsilon_{i}^{\rm u}(t)\frac{\mu_{ij}\pi_{ij}(r_{j}(t-1))^{2}}{\sqrt{(\Delta_{i}^{\rm in}(t))^{2}+(\Delta_{i}^{\rm rec}(t-1))^{2}}},\\
\frac{\partial u_{i}(t)}{\partial \pi_{ij}}&=-m_{ij}r_{j}(t-1)+\epsilon_{i}^{\rm u}(t)\frac{((m_{ij})^{2}(1-2\pi_{ij})-\Xi_{ij})(r_{j}(t-1))^{2}}{2\sqrt{(\Delta_{i}^{\rm in}(t))^{2}+(\Delta_{i}^{\rm rec}(t-1))^{2}}},\\
\frac{\partial u_{i}(t)}{\partial \Xi_{ij}}&=\epsilon_{i}^{\rm u}(t)\frac{(1-\pi_{ij})(r_{j}(t-1))^{2}}{2\sqrt{(\Delta_{i}^{\rm in}(t))^{2}+(\Delta_{i}^{\rm rec}(t-1))^{2}}}.
\end{align}
\end{subequations}

\begin{figure}
     \includegraphics[bb=10 26 270 269,width=0.4\textwidth]{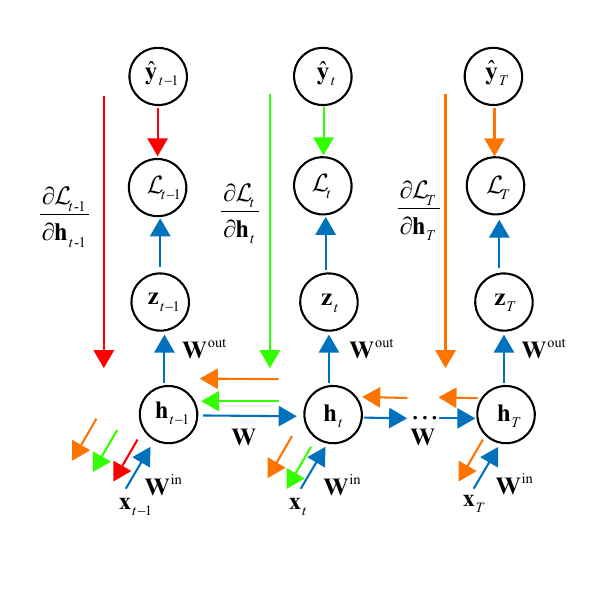}
  \caption{Illustration of the generalized back-propagation through time. 
$\bm{x}_{t}$ is a time-dependent input into the network, 
resulting in a sequence of hidden representations $\bm{r}(t)$ [i.e., $\phi(\bm{\mathrm{h}}_{t})$] and a time-dependent loss $\mathcal{L}_{t}$. 
The error signal $\frac{\partial \mathcal{L}_{t}}{\partial \bm{\mathrm{h}}_{t}}$ propagates back
from the last time-step $T$ to the first time step, yielding the temporally accumulated 
gradients for hyper-parameters to be updated. The contribution of
$\frac{\partial \mathcal{L}_{t}}{\partial \bm{\mathrm{z}}_{t}}$ is not shown in this plot. $\bm{x}_t$, $\mathbf{z}_t$ and $\mathbf{h}_{t}$ 
indicate $\bm{x}(t)$, $\mathbf{z}(t)$ and $\mathbf{h}(t)$ in the main text, 
respectively.
    }\label{gBP}
\end{figure}

In this learning process, our model learns a RNN ensemble to realize the time-dependent computation, in contrast to the standard BPTT algorithm which gives only a
point-estimate of RNN weights. In particular, if we set $\bm{\pi}=0$ and $\bm{\Xi}=0$, our learning equation reduces to the standard BPTT. Hence,
our model can be thought of as a generalized version of BPTT (i.e., gBPTT), as shown in Fig.~\ref{gBP}, where each weight parameter should be understood as the SaS hyper-parameters,
corresponding to our ensemble setting.

The size of the candidate network space can be captured 
by the network entropy $S=-\int_{\mathbb{R}^{\mathcal{D}}}P(\bm{w})\ln{P(\bm{w})}d\bm{w}$,
where $\mathcal{D}$ is the number of weight parameters in the network. 
If we assume the joint distribution of weights to be factorized across individual
connections, the overall energy $S$ can be obtained by summing up the entropy of individual 
weights as $S=\sum_{\ell}{S_{\ell}}$. The entropy of 
each directed connection ${\ell}$ is derived as follows~\cite{Li-2020}:
\begin{equation}\label{enEq}
S_{\ell}=-\pi_{\ell}\ln{[\pi_{\ell}\delta(0)+(1-\pi_{\ell})\mathcal{N}(0|m_{\ell}, \Xi_{\ell})]}-\frac{1-\pi_{\ell}}{\mathcal{B}}\sum_{s}{\Gamma(\epsilon_{s})},
\end{equation}
 where $\Gamma(\epsilon_{s})=\ln{[\pi_{\ell} \delta(m_{\ell}+\sqrt{\Xi_{\ell}}\epsilon_{s})+\frac{1-\pi_{\ell}}{\sqrt{\Xi_{\ell}}}\mathcal{N}(\epsilon_{s}|0, 1)]}$
 and $\mathcal{B}$ denotes the number of standard Gaussian variables $\epsilon_{s}$. 
 This model entropy is just an approximate estimate of the true value whose exact computation is impossible.
 
 If $\pi_{\ell}=0$, the probability distribution of $w_{\ell}$ 
 can be written as $P(w_{\ell})=\mathcal{N}(w_{\ell}|m_{\ell}, \Xi_{{\ell}})$, and
 the entropy $S_{\ell}$ can be analytically computed as $\frac{1}{2}\ln{(2\pi e \Xi_{\ell})}$. 
 If $\Xi_{\ell}=0$, the Gaussian distribution reduces to a Dirac delta function, 
 and the entropy becomes an entropy of discrete random variables, 
 $S_{\ell}=-\pi_{\ell}\ln{\pi_{\ell}}-(1-\pi_{\ell})\ln{(1-\pi_{\ell})}$. However, there may exist a mixture of discrete and continuous contributions to the entropy.
 Therefore, the entropy value can take negative values. In practice, we use $\delta(x)=\lim_{a\to0^+}\frac{1}{\sqrt{2\pi a}}e^{-\frac{x^2}{2a}}$
 to approximate the delta-peak with a small value of $a$. The stochasticity of the SaS distribution can be also decomposed into 
 two levels: one at the choice of connection that is absent and the other at the Gaussian slab itself. The former one is a discrete entropy characterized by
 the spike probability, and the latter is a continuous entropy just characterized by the variance of the Gaussian slab.
\begin{figure}
     \includegraphics[bb=2 3 777 273,width=0.9\textwidth]{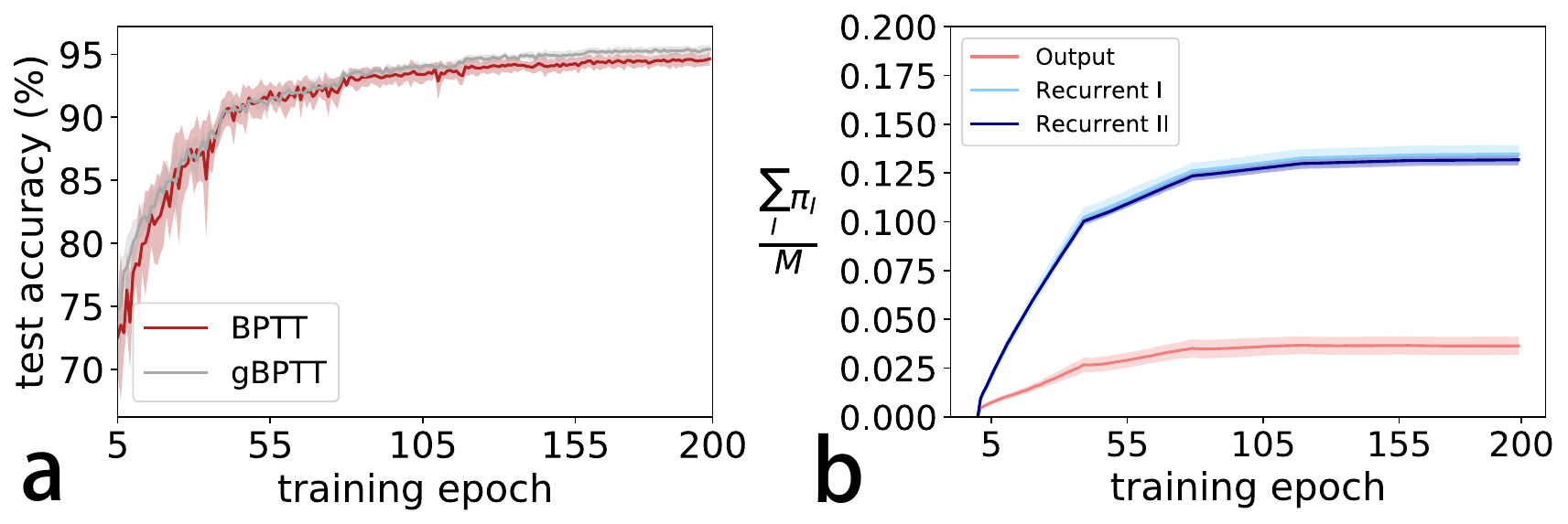}
  \caption{
  Properties of trained RNN models for the pixel-by-pixel MNIST classification.
  (a) Test accuracy versus training epoch. The network of $N=100$ neurons 
  is trained on the full MNIST dataset ($60\,000$ images), and tested on 
  the unseen data of $10\,000$ images. Five independent runs of the algorithm
  are considered. There are no baseline inputs. Other parameters for training are as follows:
  $\alpha=0.1$, $\ell_2$ regularization strength is $10^{-4}$, and the initial learning rate ${\rm lr}_0=0.001$. (b) Evolution of the network sparsity per connection. 
  Recurrent ${\rm I}$ denotes the case of $w_{ij}$ ($i<j$) along which the
  recurrent feedback passes from neuron $j$ to neuron $i$; while recurrent
  ${\rm II}$ denotes the case of $w_{ji}$ ($j>i$) along which the recurrent feedback
  passes from neuron $i$ to neuron $j$. The fluctuation is estimated from 
  five independent runs [the same training conditions as in (a)]. 
  }\label{pixel}
\end{figure}

\begin{figure}
     \includegraphics[bb=5 6 1262 813,width=0.9\textwidth]{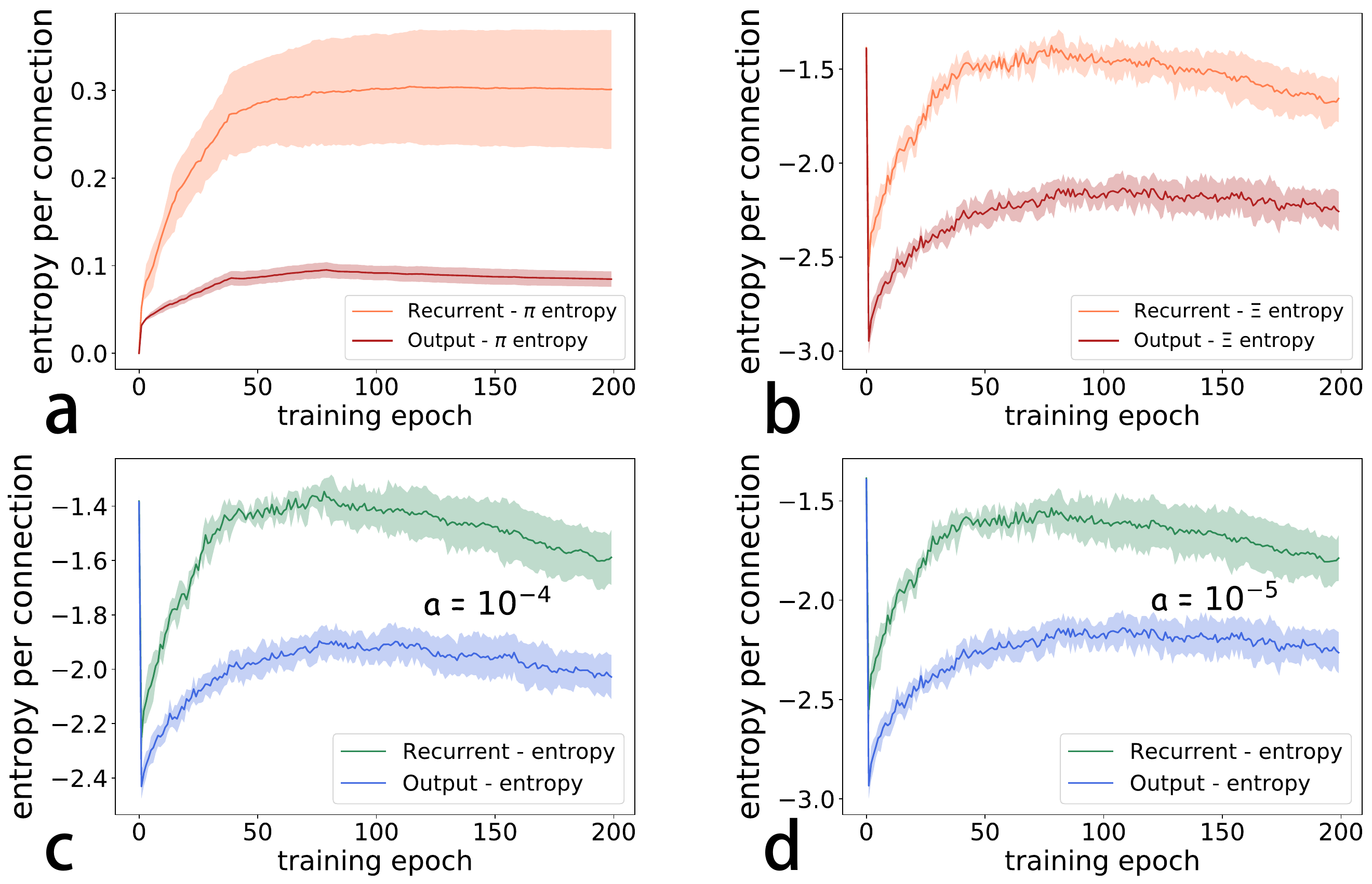}
  \caption{
  Evolution of entropy per connection for the pixel-by-pixel MNIST classification. The training conditions
  are the same as in Fig.~\ref{pixel}. Five independent runs are considered.
  The input weight is deterministic without any stochasticity.
  In addition, to avoid divergence, we relax all values of $\Xi$ smaller than $a$ to be $a$ in the computation of the entropy.
  Our results are not sensitive to the choice of $a$, once the value of $a$ is small.
  (a,b) We decompose the variability of the SaS distribution into two levels: the 
  variability of selecting zero
  synapses ($\pi$-entropy) and the variability inside the Gaussian slab ($\Xi$-entropy). 
  (c,d) Evolution of entropy per
connection ($\mathcal{B}= 100$). We use
the Gaussian distribution of small variance (indicated by $a$) to approximate the delta-peak. The value of $a$ does not affect the qualitative behavior of the entropy profile.
  }\label{pixel-entro}
\end{figure}
\section{Results}
In this section, we show applications of our ensemble theory of temporal
credit assignment to both engineering and computational cognitive tasks, i.e., pixel-by-pixel MNIST digit 
classification, and the multisensory integration demonstrating 
the benefit of multiple sources of information for decision making. 
We will explore in detail rich properties of trained RNN model accomplishing 
the above computational tasks of different nature.
\subsection{Pixel-by-Pixel MNIST digit classification}
Training RNNs is hard because of long-term dependency in the sequence 
of inputs. A challenging task of long-term dependency is to train a RNN
to classify the MNIST images when the $784$ pixels are fed into the network
one by one (note that $\bm{x}_0=0$), where the network is required to predict the category of the image
after seeing all the $784$ pixels. This task displays a long range of dependency, 
because the network reads one pixel at a single time-step in a scan-line 
order from the top left pixel to the bottom right pixel, 
and the information of as long as $784$ time steps must be maintained 
before the final decision. We apply the vanilla RNN with $N=100$ 
recurrent units to achieve this challenging goal. Because the input 
size is one ($\nin=1$) in this task, and thus the set of hyper-parameters $(\pi^{\rm in},\Xi^{\rm in})$ 
are set to zero (i.e., the deterministic limit). The entire MNIST dataset is divided into mini-batches for stochastic
gradient descent (SGD), and we apply cross entropy as the objective function to be minimized.
In this task, the error signal appears only after the network reads all the $784$ pixels.
In the current setting, we ignore the noise terms in the dynamics equations [Eq.~(\ref{rnndyn})
and Eq.~(\ref{rnnin})]. Surprisingly, although working at the ensemble level, our model
can achieve a comparable or even better performance than the traditional BPTT method, 
as shown in Fig.~\ref{pixel} (a). 

Our simulation reveals that the sparsity per connection, $\frac{1}{M}\sum_{\ell}{\pi_{\ell}}$
(a total of $M$ directed connections in the network), achieves a larger value in the recurrent layer 
compared with the output layer; this behavior does not depend on the specific 
directions of the coupling [Fig.~\ref{pixel} (b)]. During training, the sparsity
level grows until saturation, suggesting that the training may remove or minimize
the impacts of irrelevant information across time steps. Interestingly, the entropy
per connection in the recurrent layer also increases at the early stage of the training,
but decreases at the late stage (Fig.~\ref{pixel-entro}), showing that the training is able
to reorganize the information landscape shaped by the recurrent feedbacks. The network at
the end of training becomes more deterministic (e.g., $\Xi$ gets close to zero, or the
spike probability goes away from one half). The discrete $\pi$-entropy always increases until saturation 
at a certain level. The entropy profile in the output layer shows a similar behavior yet with a lower
entropy value. This is consistent with distinct roles of 
recurrent and readout layers. The recurrent layer (or computing reservoir)
searches for an optimal way of processing temporally complex sensory inputs,
while the output layer implements the decoding of the signals hidden in the reservoir dynamics. 
Note that two types of deterministic weights yield vanishing individual 
entropy values, i.e., (i) $\pi$= 1 (unimportant (UIP) weight); 
(ii) $\pi$ = 0, $\Xi = 0$ and $m\neq 0$ (very important (VIP) weight).

\begin{figure}
     \includegraphics[bb=1 33 1127 1113,width=0.9\textwidth]{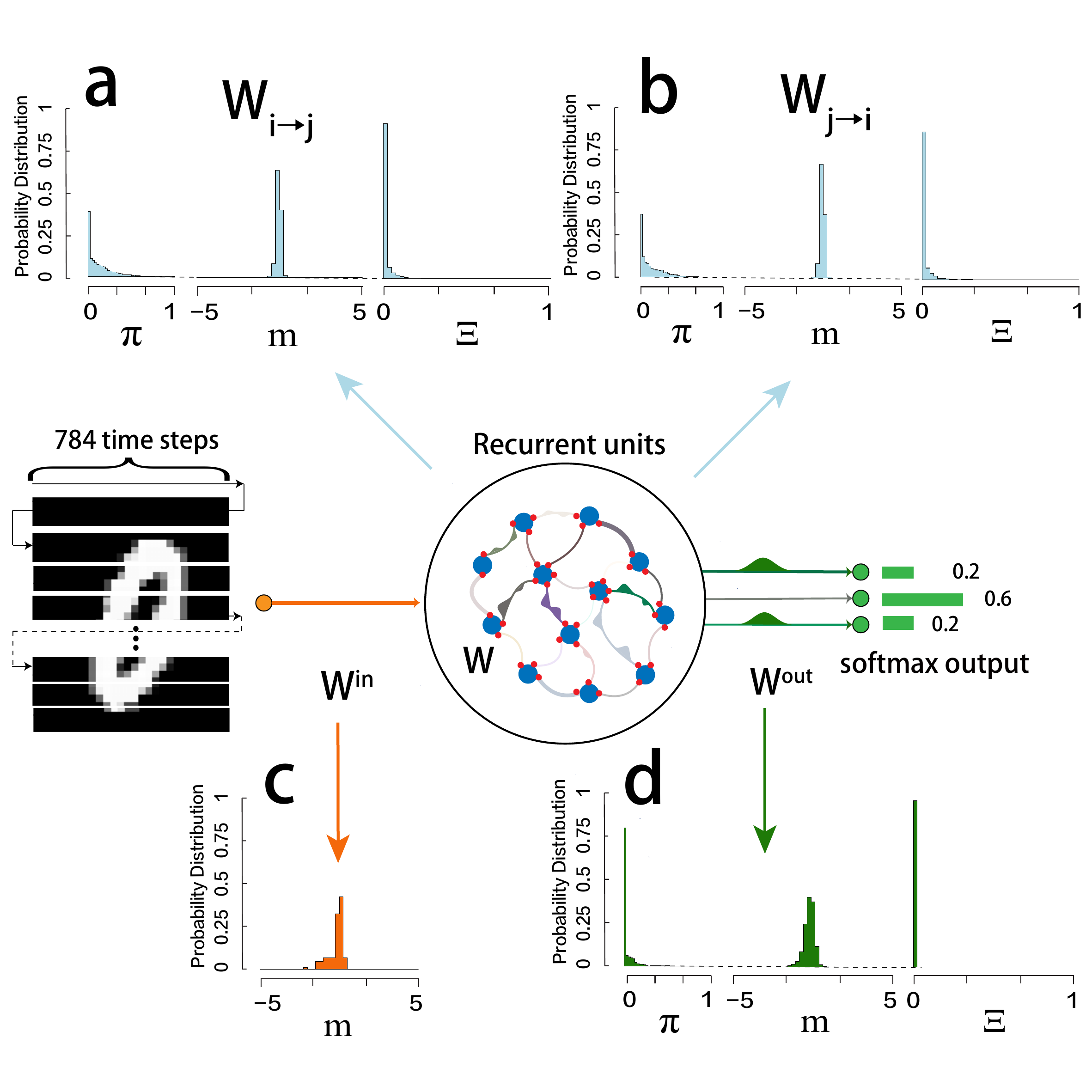}
  \caption{
 Distributions of hyper-parameters $(\pi, m, \Xi)$ in the 
 trained network (pixel-by-pixel MNIST classification).  
 The training conditions are the same as in Fig.~\ref{pixel}. In (a,b), $i<j$ is assumed.
 }\label{distribution}
\end{figure}

\begin{figure}
     \includegraphics[bb=4 3 1495 427,width=0.9\textwidth]{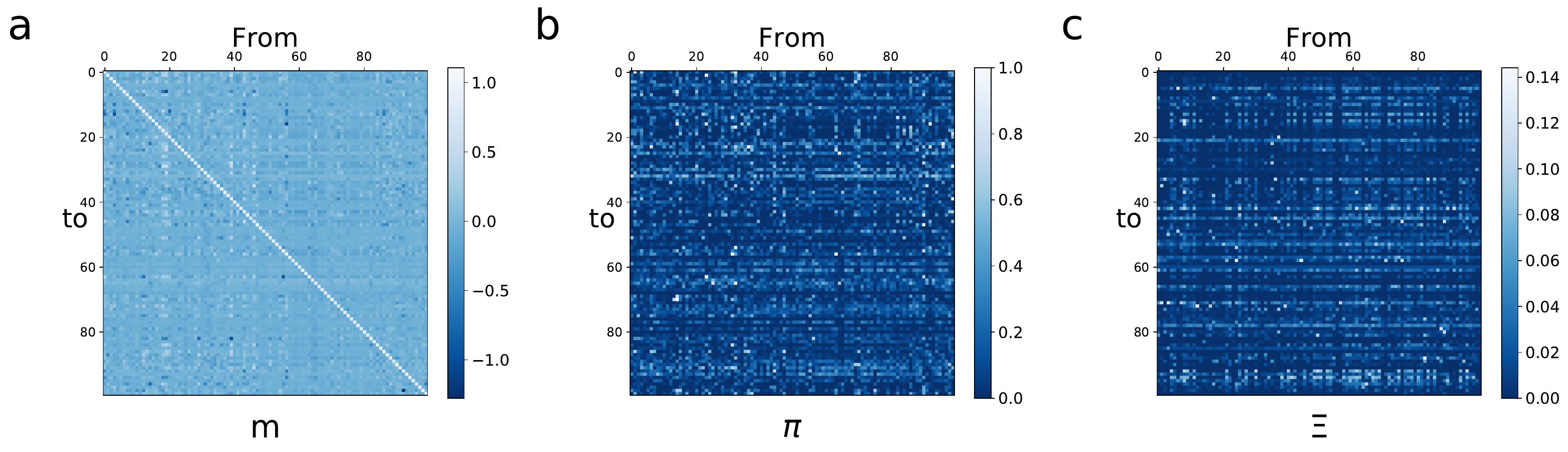}
  \caption{
 Heterogeneous structures of trained RNN networks. Training conditions are the same as 
 in Fig.~\ref{pixel}. Hyper-parameters $(\pi, m, \Xi)$ are plotted in the matrix form
 with the dimension $N\times N$, where $N$ indicates the number of recurrent neurons. 
 The element of these matrices, say $m_{ij}$, denotes the hyper-parameter value 
 for the directed link from neuron $j$ to neuron $i$.
  }\label{matrix}
\end{figure}

To study the network behavior from the perspective of hyper-parameter
distributions, we plot the distribution of three sets of parameters $(\bm{m}, \bm{\Xi}, \bm{\pi})$ 
for the output layer, two different directions of connections in the recurrent layer, 
and $\bm{m}$ of the input layer  ($(\pi^{\rm in},\Xi^{\rm in})$ are set to zero as explained before),
as shown in Fig.~\ref{distribution}. The distribution of spike probability $\bm{\pi}$ has
the shape of $L$ for all layers. The extreme at $\bm{\pi}=0$ indicates that the corresponding 
synaptic weight carries important feature-selective information, and are thus critical to 
the network performance. The shape of $\bm{\pi}$-distribution profile reflects the task 
difficulty for the considered network (given the same initialization condition), since 
we observe that in a simpler task for the network to read $28$ pixels at each time step 
(rather than one pixel by one pixel), the profile of $\bm{\pi}$-distribution develops
a U shape, with the other peak at $\pi=1$, implying the emergence of a significant 
fraction of UIP weights.
In addition, $\bm{\Xi}$ has an L-shaped distribution which peaks at zero, suggesting
that the corresponding weight distribution takes a deterministic value of $m$, which
becomes the weight value of that connection.

As for the readout layer, the mean of the continuous slab becomes
much more dispersed, compared with that of the recurrent layer, 
while the statistics profile for $\bm{\pi}$ and $\bm{\Xi}$ becomes
more converged, which is in an excellent agreement with the entropy
profile shown in Fig.~\ref{pixel-entro}. We thus conclude that the 
recurrent layer has a greater variability than the output layer. 
This great variability allows the recurrent layer to transform
the complex sensory inputs with hierarchical spatio-temporal structures
in a flexible way. Nevertheless, the minimal variability makes the 
readout behavior more robust. It is therefore interesting in future
works to address the precise mechanism underlying how the statistics 
of the connections in a RNN facilitates the formation of latent dynamics for computation.

We next look at the specific profile of individual hyper-parameters 
in a matrix form (Fig.~\ref{matrix}). We
find that the diagonal (self-interaction) of $\bm{m}$ emerges from gBPTT
(note that the diagonals of $\bm{\pi}$ and $\bm{\Xi}$ nearly vanish),
demonstrating the significant role of self-interactions in maintaining
long-term memory and thereby the learning performance, in accord with 
the heuristic strategy used in a recent study~\cite{Hu-2018}. There 
also appears heterogeneity in the hyper-parameter matrices, i.e., some
directed connections play a more important role than the others. 
In particular, some connections can be eliminated for saving computation resources. 

\begin{figure}
     \includegraphics[bb=2 2 762 260,width=0.9\textwidth]{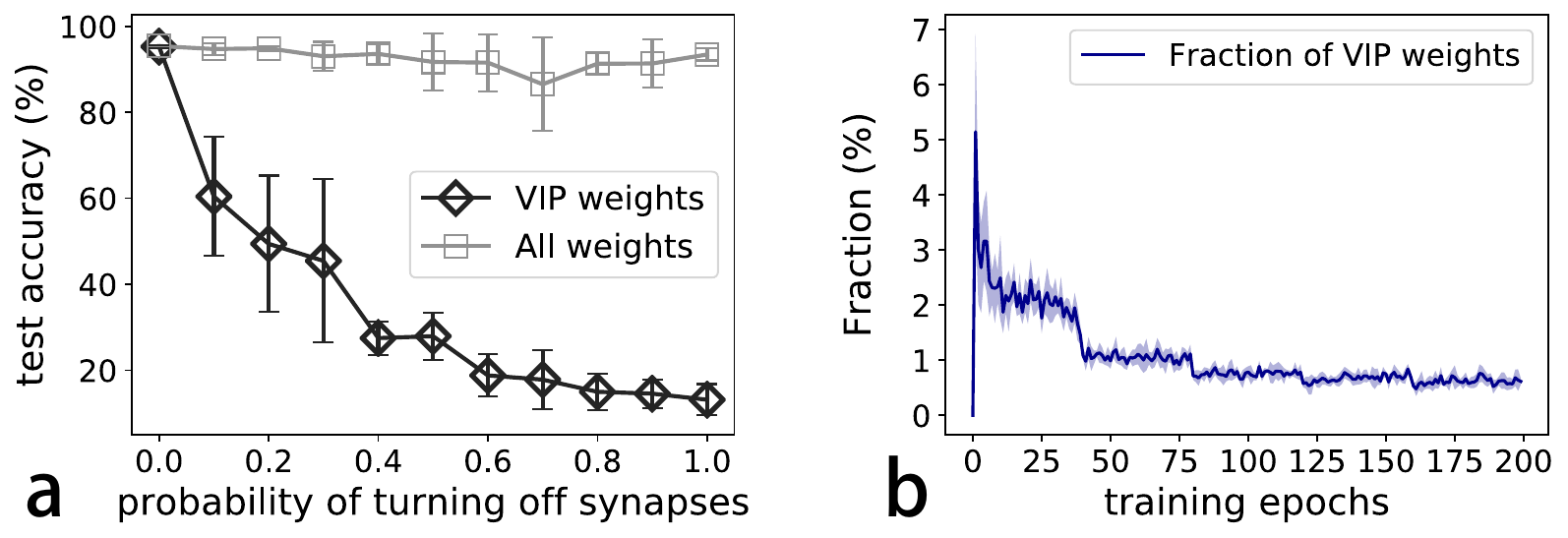}
  \caption{
 Targeted weight perturbation in the recurrent layer of RNNs.  
 All the training conditions are kept the same as  in Fig~\ref{pixel}, and
 the results are averaged over five independent runs. (a) VIP weights are 
 stochastically turn off, and the same number of randomly selected weights
 from the entire weight population are also deleted for comparison of the 
 resultant performance.  (b) The fraction of VIP weights changes during the training process. 
  }\label{target}
\end{figure}

Our method can identify the exact nature of each directed connection in the network. 
Targeted weight perturbation can thus be performed in the recurrent layer (Fig.~\ref{target}) . 
In the course of training, the fraction of VIP connections slowly decreases,
although the fraction of VIP connections is not significant (around $1\%$) [Fig.~\ref{target}(b)]. 
Surprisingly, pruning these VIP weights could strongly deteriorate the test performance
of the network (up to the chance level), whereas pruning the same number of randomly 
selected connections from the entire weight population yields a negligible drop of the
test accuracy. We thus conclude that in a RNN, there exists a minority of VIP weights
carrying key spatio-temporal information essential for decision making of the network.

\begin{figure}
     \includegraphics[bb=9 17 4238 965,width=0.9\textwidth]{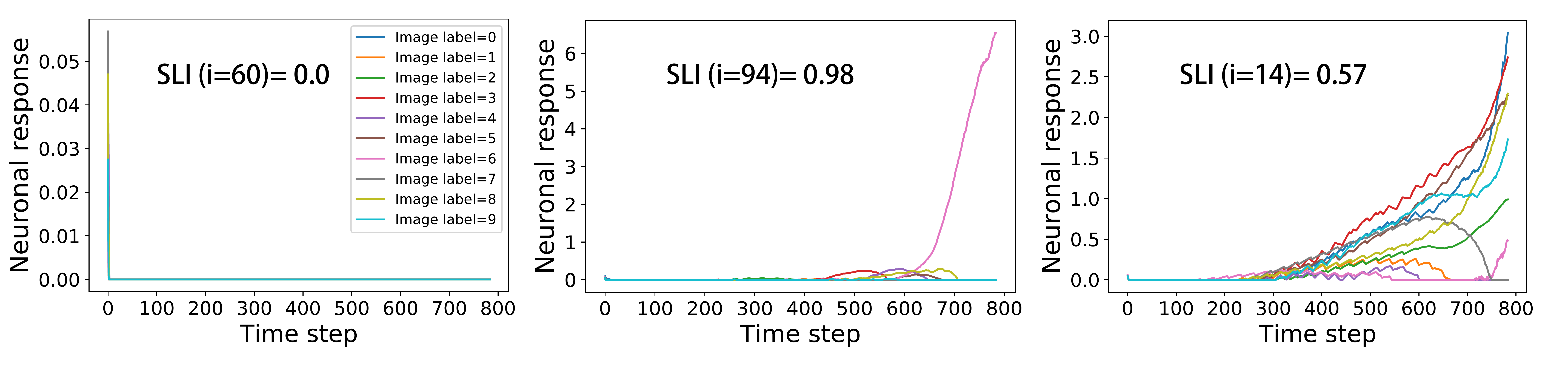}
  \caption{
 Selectivity of representative neurons in RNNs learning MNIST digit classification.
 (Left) the non-selectivity case. (Middle) the uni-selectivity case. (Right) the
 mixed-selectivity case. Time steps indicate the dynamic steps during the test phase,  
 and the labels in the left panel are shared for the other two panels.
  }\label{select}
\end{figure}

We then explore whether the selectivity of neurons for the sensory inputs could emerge
from our training. The degree of selectivity for an individual unit, say neuron $i$, 
can be characterized by an index, namely SLI~\cite{Sel-2014} as follows
\begin{equation}
{\rm SLI} (i) = \frac{1}{1-\frac{1}{N_{s}}}\left[1-\frac{(\frac{1}{N_{s}}\sum_{s=1}^{N_{s}}{r_{i, s}})^{2}}{\frac{1}{N_{s}}\sum_{s=1}^{N_s}{(r_{i, s})^{2}}}\right],
\end{equation}
where $r_{i, s}$ indicates the response of the neuron to the input stimulus $s$ with 
the total number of stimulus-class being $N_s$ ($N_s=10$ in the MNIST experiment). 
Note that $r_{i,s}$ for the RNN depends on the time step as well.  The value of 
SLI ranges from $0$ (when the neuron responds identically to all stimuli) to $1$ 
(when the neuron responds only to a single stimulus), and a higher SLI indicates
a higher degree of selectivity, and vice versa. Interestingly, we find that most 
neurons have mixed selectivity shown in Fig.~\ref{select} (the right panel), whose
SLI lies between $0$ and $1$, and such neurons respond strongly to a few types of images. 
The mixed selectivity was also discovered in complex cognitive tasks~\cite{Mixed-2013}.
We also observe that some neurons are particularly selective to only one type of images
with a high value of SLI [Fig.~\ref{select} (middle)], while the remaining part of 
neurons keeps silent with an SLI near to zero [Fig.~\ref{select} (left)]. As expected, the 
selectivity property emerges slightly before the decision making of the network.

\begin{figure}
     \includegraphics[bb=1 3 779 291,width=0.9\textwidth]{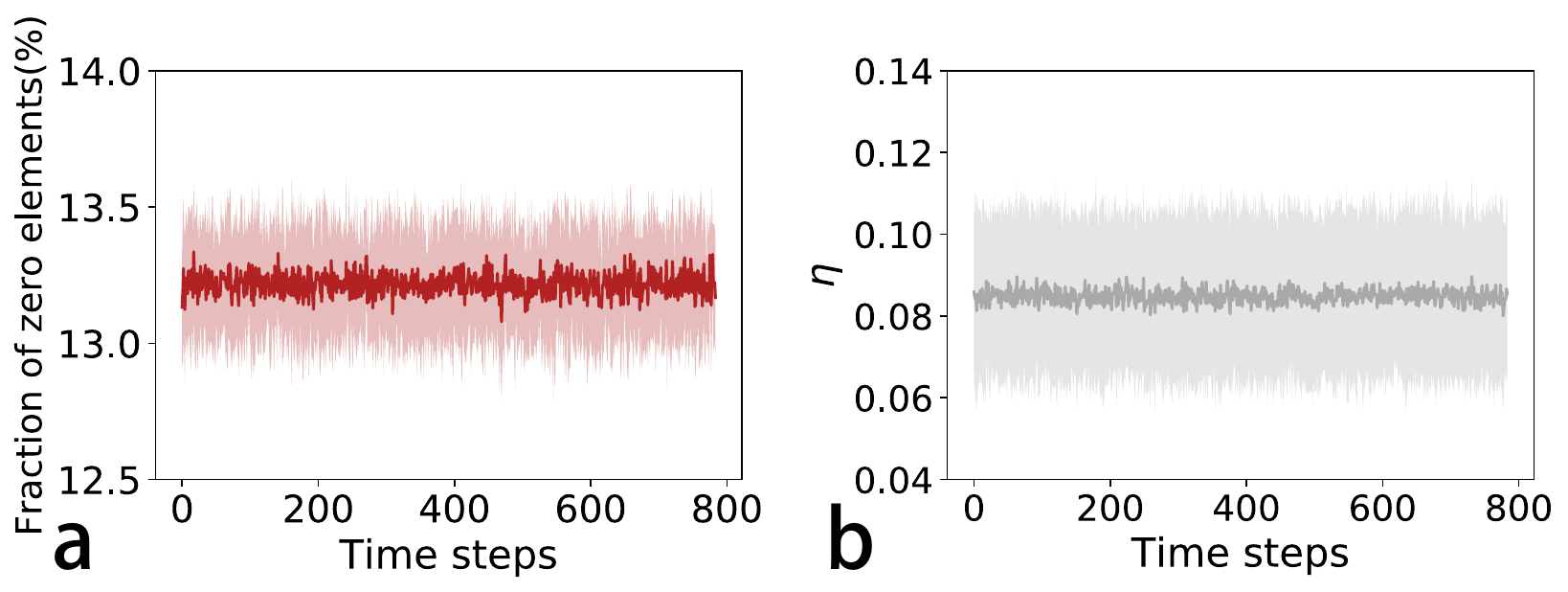}
  \caption{
 The statistics of time-dependent networks does not change significantly. (a) 
 The fraction of zero elements of sampled weights across $784$ time steps.
 The fluctuation is computed from five independent runs. (b) The asymmetry measure $\eta$ defined by
 $\frac{\overline{w_{ij}w_{ji}}}{\overline{w_{ij}^2}}$
 versus time step. The over-bar means the average over all reciprocal connections. The result is averaged over five independent runs.
 }\label{sampling}
\end{figure}

We finally remark that gBPTT produces a network ensemble, characterized by the hyper-parameter set of the SaS
distribution. A concrete network can thus be sampled from this ensemble. During training, we use independently time-dependent
Gaussian noise $\boldsymbol{\epsilon}^{\rm u}(t)$ to approximate the statistics of the ensemble. Accordingly, we find that
a time-dependent concrete network yields the identical test performance with the mean-field propagation (i.e., using the parametrized 
pre-activation, see Eq.~(\ref{MFP})]. The overall statistics of sampled networks does not vary significantly across dynamic steps (Fig.~\ref{sampling}).
This observation is in stark contrast to the traditional RNN training, where a deterministic weight matrix is used in all time steps.
In fact, in a biological neural circuit, changes in synaptic connections are essential for the development and function
of the nervous system~\cite{Kasai-2008,SD-2009}. In other words, the specific details of the connection pattern for a circuit may not be critical to the behavior,
but rather, the parameters underlying the statistics of weight distributions become dominant factors for the behavioral output of the network (see also a recent paper demonstrating
that innate face-selectivity could emerge from statistical variation of the feedforward projections in hierarchical neural networks~\cite{FSN-2019}).
Therefore, our current ensemble perspective of training RNNs offers a promising framework to achieve neural computation with a dynamic network, which is likely 
to be consistent with biologically plausible computations with fluctuating dendritic spines, e.g., dendritic spines can undergo morphological remodeling 
in adaptation to sensory stimuli or in learning~\cite{SD-2009}.

\begin{figure}
\centering
     \includegraphics[bb=18 113 864 720,width=0.8\textwidth]{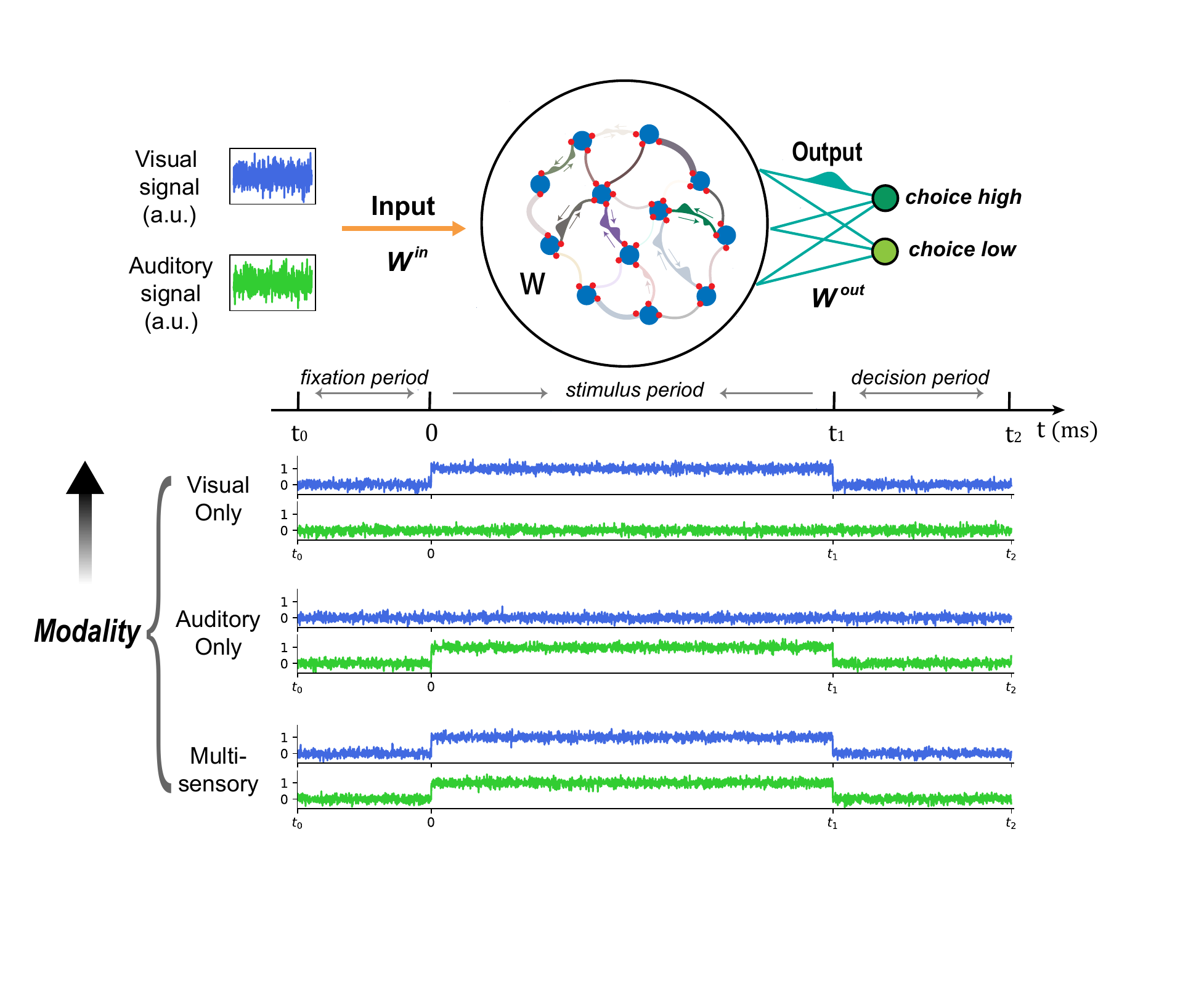}
  \caption{
  Schematic illustration of the RNN performing a
	multi-sensory integration task. In one trial, the task consists of three consecutive stages (of lengths $300$, $900$, and $400$ ms, respectively)
	including a fixation period ($t_0 \sim 0$), 
	a stimulus period ($0\sim t_1$), and a decision period ($t_1\sim t_2$).
	Note that in a test trial, the lengths of three stages are $500$, $1000$, and $300$ ms, respectively.
	 During the fixation period, 
the network must maintain a low output value, ensuring that
the network will be sensitive to the stimulus signals. During the stimulus period, 
either or both (or multisensory) sources of signals
with the same frequency $f$, including visual signal and auditory signal, 
act as the input to the network. The frequency $f$ ranges from
$9$ to $16$. Each type of sensory input contains 
positively tuned and negatively tuned signals (only the positively tuned one is shown in 
the plot). 
	During the decision period, there are no modality input signal, and the network must
	learn to discriminate the even rate of the stimulus period, i.e., the output must hold a
	high or low value correctly. The output at the last time step of the decision 
period is assigned the final decision of one trial. Other parameters for training are as follows: baseline $x_0=0.2$ for all neurons, $\tau=100$ ms,
$\alpha=0.2$, $\sigma=0.15$,
$\sigma_{\rm in}=0.01$, $\ell_2$ regularization strength is $10^{-3}$, and the initial learning rate ${\rm lr}_0=0.001$. $\Delta t=0.5$ ms for testing.
  }\label{model-msi}
\end{figure}

\begin{figure}
\centering
     \includegraphics[bb=4 2 639 398,width=0.6\textwidth]{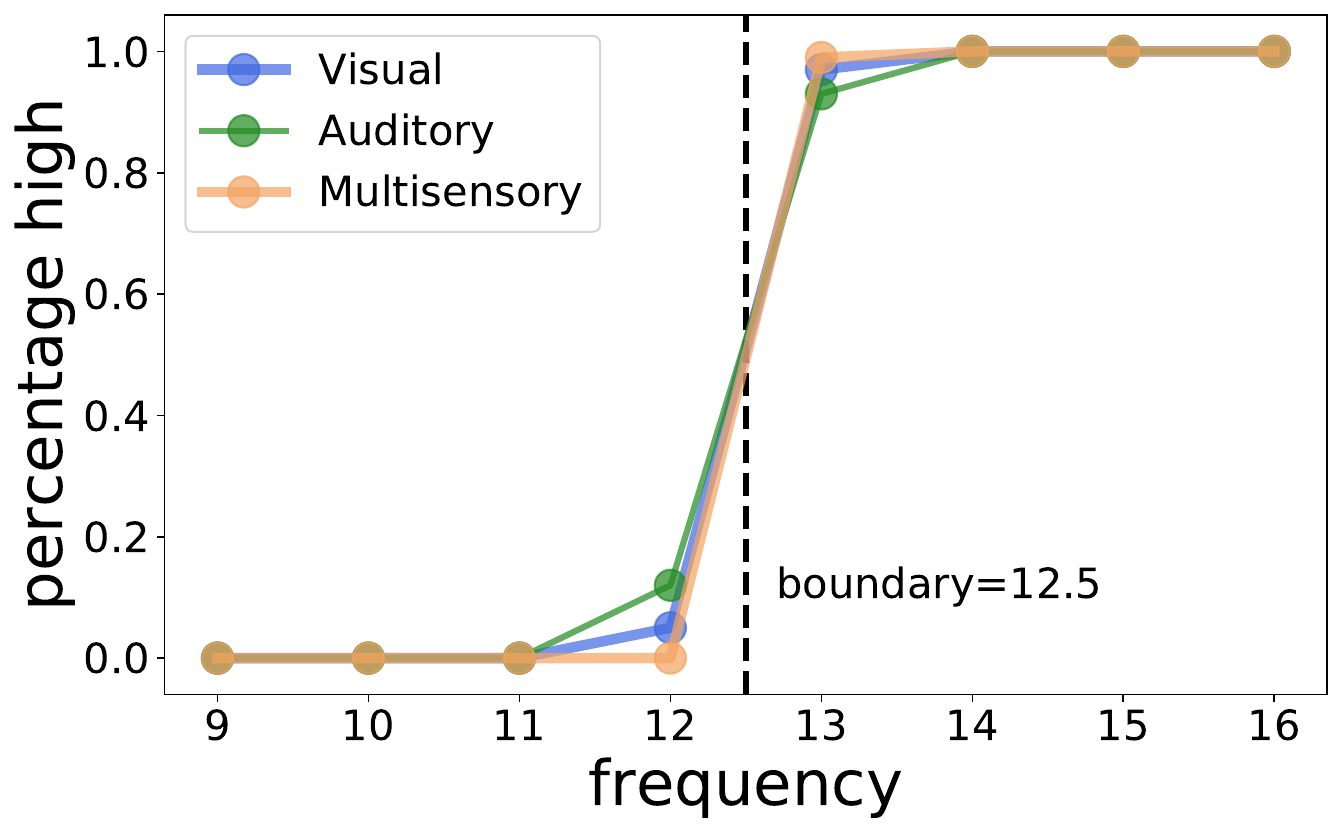}
  \caption{
  Psychometric function for the MSI task. This function shows the percentage of high choices made by the network as 
  a function of the event rate for both unisensory and multisensory trials.
	Each marker is the mean result over $100$ independent test trials. Other training conditions are the same as in 
	Fig.~\ref{model-msi}.
  }\label{psy}
\end{figure}
\subsection{Multisensory integration task}
Multisensory integration is a fundamental ability of the brain to combine cues from multiple senses to form robust perception of signals in the noisy world~\cite{MSI-2009,Shams-2008}.
In typical cognitive experiments, two separate sources of information are provided to animals, e.g., rats, before the animals make a decision.
The source of information can be either auditory clicks or visual flashes, or both~\cite{Cate-2014}. When the task becomes difficult, the
multisensory training is more effective than the unisensory one. Akin to animals trained to perform a behavior task, RNNs can also learn
the input-output mapping through an optimization procedure, which is able to provide a quantitative way to understand the computational principles 
underlying the multisensory integration. In particular, by increasing biological plausible levels of network architectures and even learning dynamics~\cite{Miconi-2017,Murray-2019},
one may generate quantitative hypotheses about the dynamical mechanisms that may be implemented in real neural circuits to solve the same behavior task.
In this section, we restrict the network setting to the simplest case, i.e., a group of neurons are reciprocally connected with continuous firing rate dynamics.
To consider more biological details is straightforward, e.g., taking Dale's principle~\cite{Song-2016}. The goal in this section is to show that our ensemble perspective
also works in modeling cognitive computational tasks.

\begin{figure}
\centering
     \includegraphics[bb=4 4 603 410,width=0.5\textwidth]{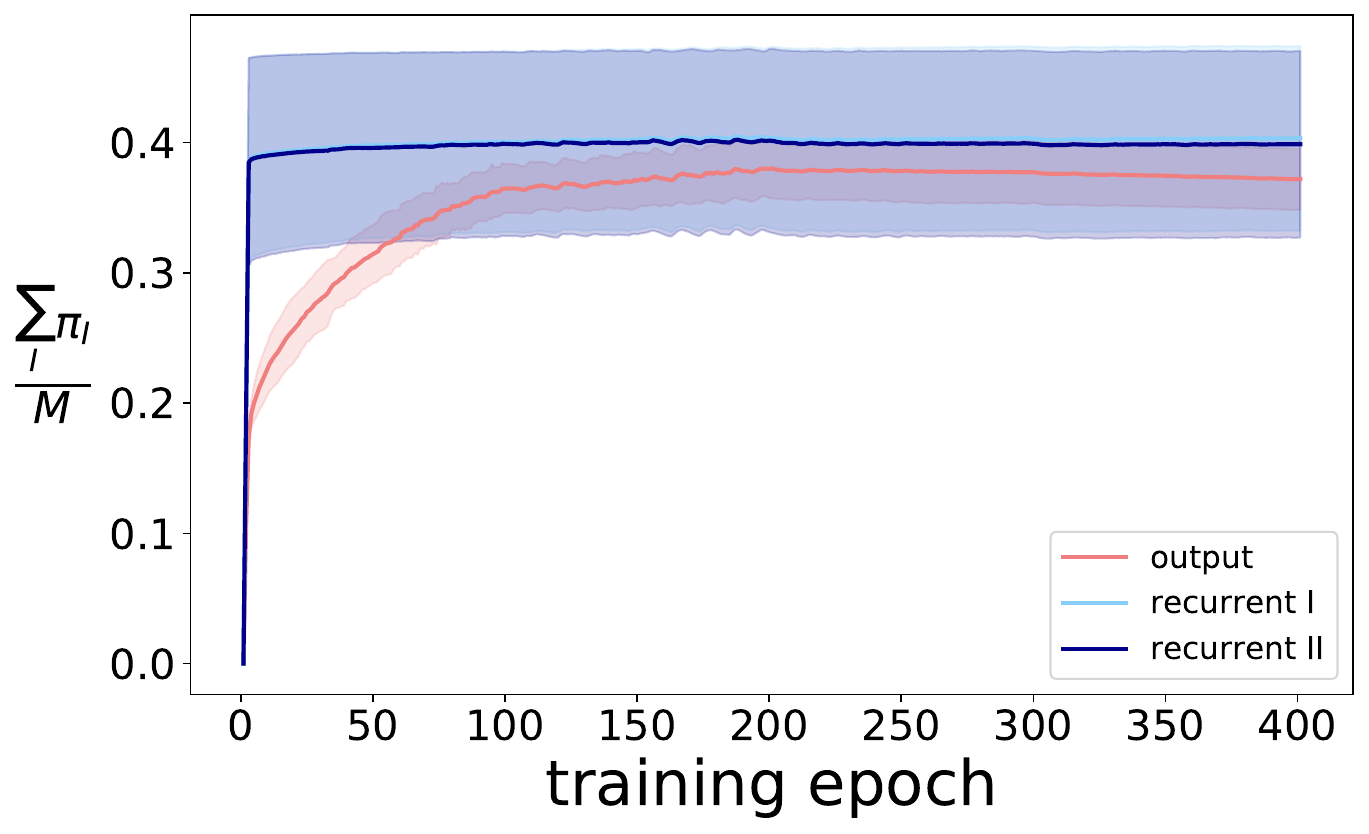}
  \caption{
  Evolution of sparsity densities with training epochs for the MSI task. The lines are the mean results of ten independent training
	trials, and the shadow indicates the fluctuation. Recurrent I and II are the lower and upper 
	triangles of the $\pi$ matrix, respectively. Other training conditions are the same as in 
	Fig.~\ref{model-msi}.
  }\label{traj}
\end{figure}

\begin{figure}
\centering
     \includegraphics[bb=5 6 1307 753,width=0.8\textwidth]{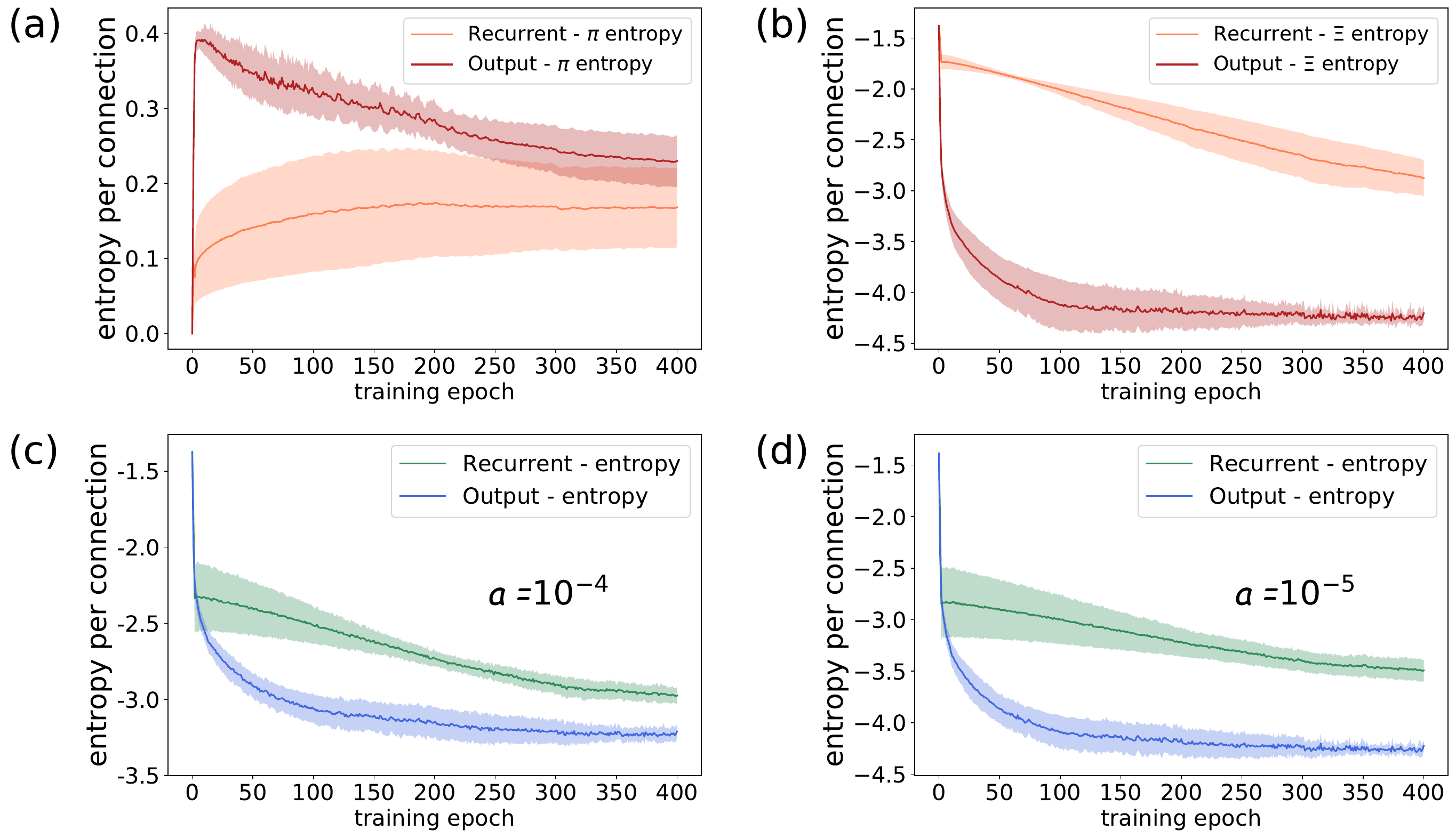}
  \caption{
  Evolution of entropy densities with training epochs for the MSI task. The lines are the mean results of ten independent training
	trials, and the shadow indicates the fluctuation. The definitions of entropies are the same as in the MNIST experiment (see Fig.~\ref{pixel-entro}).
	The value of $a$ does not affect the qualitative behavior of the entropy profile.
	Other training conditions are the same as in 
	Fig.~\ref{model-msi}.
  }\label{msi-entro}
\end{figure}

In the multisensory integration experiment (MSI), either or both of auditory and visual signals are fed into the recurrent
network via deterministic weights (the same reason as in the pixel-by-pixel MNIST task),
which is required, after a stimulus period,
to report whether the frequency of the presented stimulus is above a decision boundary ($12.5$ events per second), as shown in Fig.~\ref{model-msi}.
One third of network units receive only visual input, while another third receive only auditory input, and the remaining third do not 
receive any input.
The continuous variable $\bm{r}(t)$ in our model is an activity vector indicating the firing rates of neurons, obtained through a non-linear transfer function (ReLU here) of the 
synaptic current input [$\bm{h}(t)$]. The current includes both of external input and recurrent feedback.
The output $\bm{z}(t)$ is a weighted readout of the neural responses in the reservoir (a binary choice for the MSI task).
The RNN is used here to model the multisensory event-rate discrimination task for the rats~\cite{Song-2016,Cate-2014}, and is trained 
by our gBPTT to solve the same audiovisual integration task. The visual and auditory inputs can be either positively (increasing function of event rate) or negatively
tuned (decreasing function of event rate). Showing the network both types of tunned inputs could improve the training~\cite{Miller-2003}. The RNN is composed of $150$ neurons, whose recurrent
dynamics is required to hold a high output value if the input event rate (represented by time-dependent noisy inputs)
was above the decision boundary, and hold a low output value otherwise. Neurons are reciprocally connected, and the global statistics of the topology is 
learned from the training trials.

\begin{figure}
\centering
     \includegraphics[bb=6 2 1129 1105,width=0.7\textwidth]{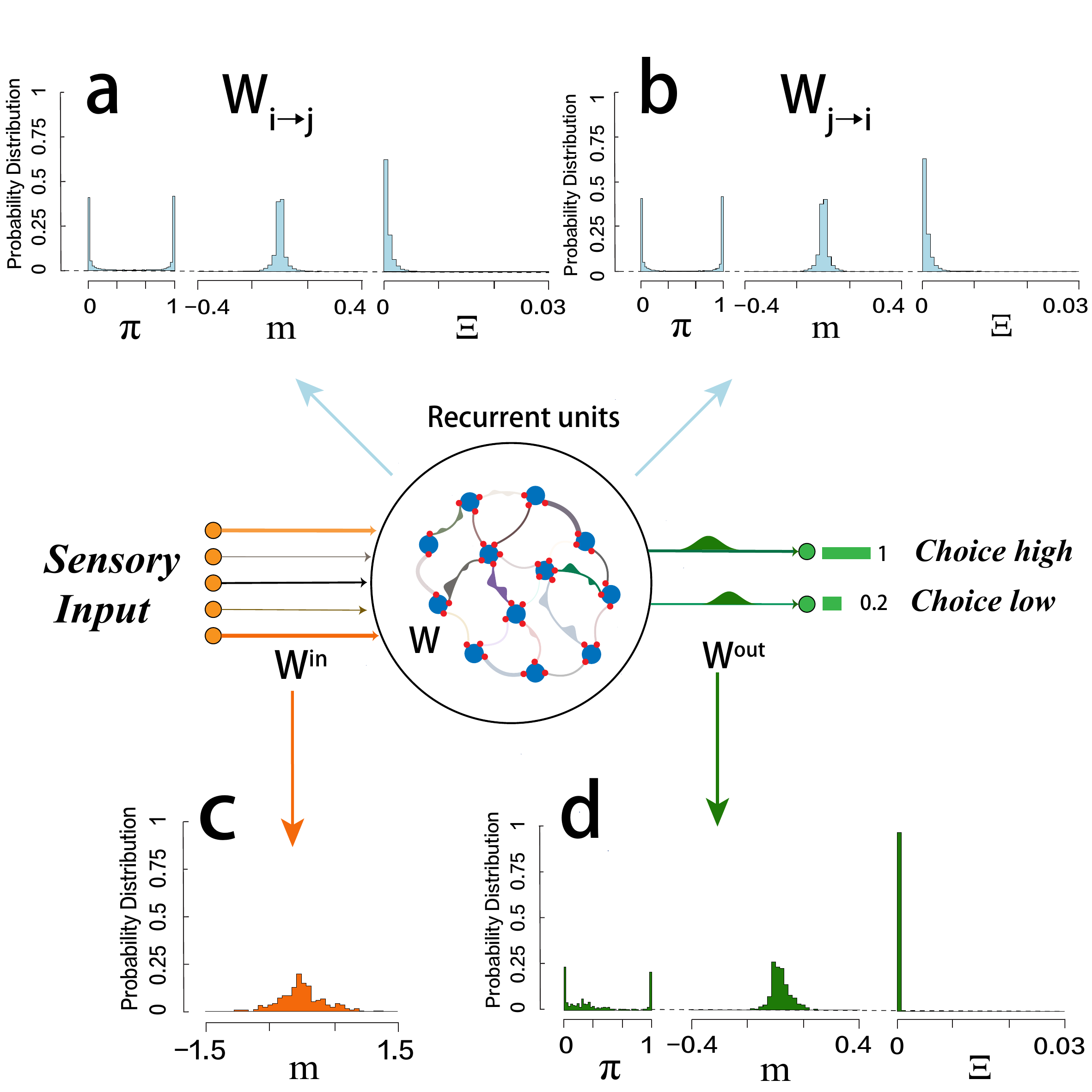}
  \caption{
  Distributions of hyper-parameters $(\pi, m, \Xi)$ in input, recurrent and
 output layers of the RNN model for the MSI task. Training conditions are the same as in 
	Fig.~\ref{model-msi}. In (a,b), $i<j$ is assumed.
  }\label{para-msi}
\end{figure}

\begin{figure}
\centering
     \includegraphics[bb=4 2 1486 741,width=0.8\textwidth]{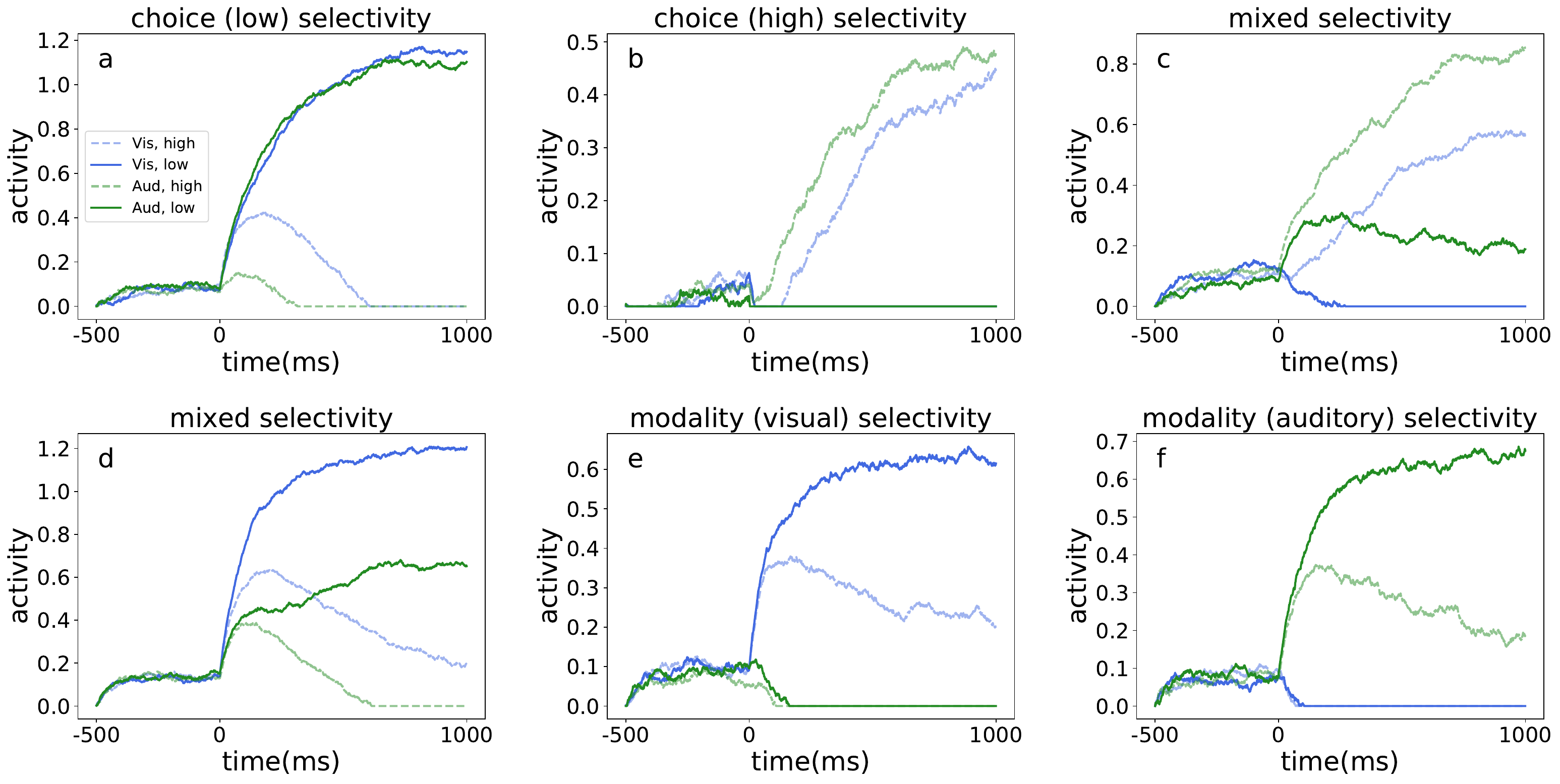}
  \caption{
 Selectivity of constituent neurons for the MSI task. Training conditions are the same as in 
	Fig.~\ref{model-msi}.
 The neurons show selectivity with respect to choice (a, b), modality (e, f) or both (c, d).
  }\label{sel-msi}
\end{figure}
The benefits of multisensory inputs, as known in cognitive science~\cite{Shams-2008}, are reproduced by
our RNN model trained by gBPTT (Fig.~\ref{psy}). Integrating multiple sources of information, rather than unisensory inputs, helps decision making
particularly when the task becomes hard (i.e., around the decision boundary). As the training proceeds, the sparsity of the recurrent layer
grows rapidly until saturation, while the sparsity of the output layer grows in the same manner but finally reaching a lower value yet with a small fluctuation
(Fig.~\ref{traj}).
The recurrent layer becomes sparser with training, demonstrating that the latent dynamics is likely low dimensional, because of existence of 
some unnecessary degrees of freedom in recurrent feedbacks. In contrast, the goal of the output layer is to decode the recurrent dynamics,
and the output layer should therefore keep all relevant dimensions of information integrated, which requires a
densely connected output layer. This behavior is consistent with the evolution of the entropy profile.
The recurrent layer maintains a relatively higher level of variability, compared with
that of the output layer at the end of training (see the continuous entropy computed according to
Eq.~(\ref{enEq}), or the $\Xi$-entropy in Fig.~\ref{msi-entro}). In addition,
the discrete $\pi$-entropy decreases with training in the output layer, in contrast to the increasing behavior of the same type of entropy in the recurrent layer (Fig.~\ref{msi-entro}).

\begin{figure}
\centering
     \includegraphics[bb=4 2 1467 410,width=0.9\textwidth]{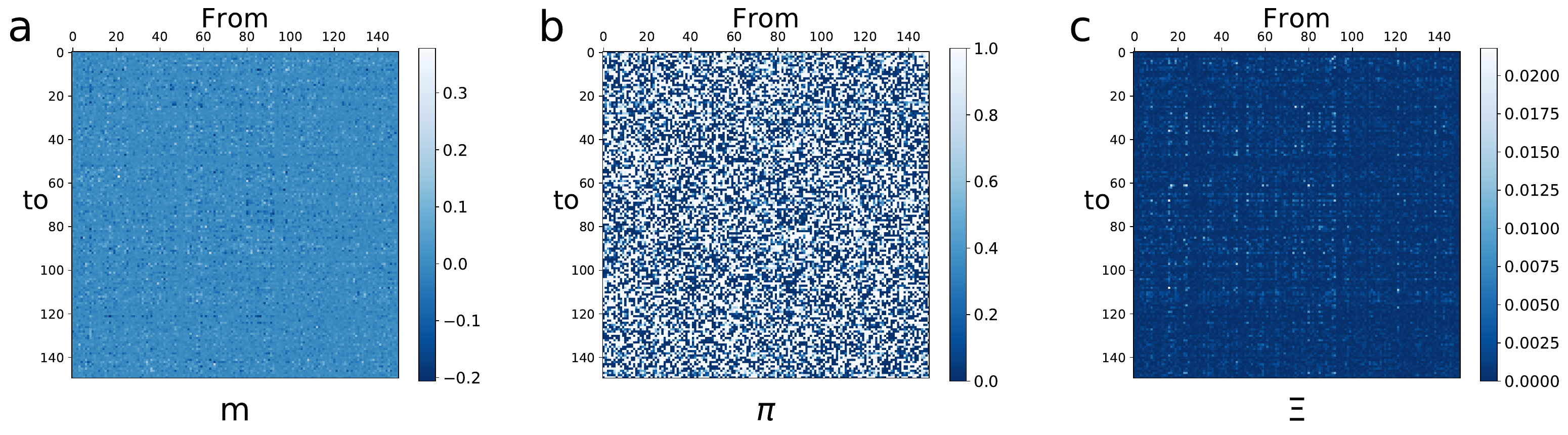}
  \caption{
  Hyper-parameter matrix of the recurrent layer in a trained network for the MSI task.
	Trained conditions are the same as in Fig.~\ref{model-msi}. 
	Hyper-parameters $(\pi, m, \Xi)$ are plotted in the matrix form with 
	the dimension $N\times N$, where $N$ indicates the number of
	neurons in the reservoir.
  }\label{matrix-msi}
\end{figure}

\begin{figure}
\centering
     \includegraphics[bb=2 2 1440 517,width=0.8\textwidth]{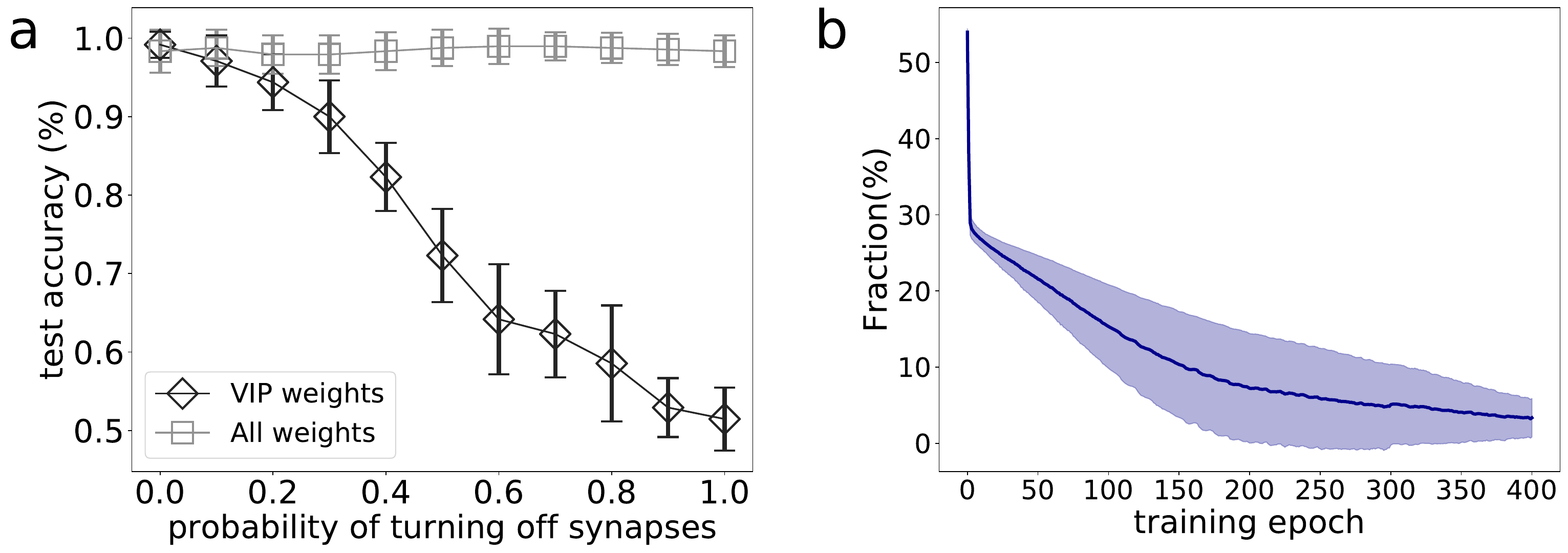}
  \caption{
  Targeted-weight perturbation in the recurrent layer for the MSI task. 
	(a) VIP weights (a relaxed version, i.e., $\pi<0.1,\,|m|>0.05$, note that $\Xi$ is very small) are stochastically turned off, in comparison with randomly selected weights pruned with the 
	same number.
	(b) The fraction of VIP weights changes
	during the training process. Trained conditions are the same as in Fig.~\ref{model-msi}.
  }\label{target-msi}
\end{figure}

Let us then look at the distribution profile of hyper-parameters (Fig.~\ref{para-msi}).
In the output layer, the distribution of $\pi$ is U-shaped, and $\Xi$ shows a
sharp single peak at zero, which demonstrates
that a dominant part of the weight distribution reduces to the Bernoulli distribution
with two peaks at $0$ and $m\neq 0$ respectively. The observed less variability in weight values of 
the output layer is consistent with the decoding stability.
In the recurrent layer, the profile of $\pi$-distribution is U-shaped, and the distribution profile of 
$\Xi$ resembles an L shape. 
This implies that there emerge VIP and UIP connections in the network. Moreover, a certain level of variability is allowed 
for the weight values, making a flexible computation possible in the internal dynamics.

Next, we ask whether our training leads to the emergence of neural selectivity, which indeed exists in
the prefrontal cortex of behaving animals~\cite{Mixed-2013}. The selectivity properties of neurons play 
a critical role in the complex cognitive ability.
 In our trained networks, we also find that neurons in the recurrent 
layer display different types of selectivity (Fig.~\ref{sel-msi}). In other words, neurons become highly active for either of
modality (visual or auditory) and choice (high or low), or 
both (mixed selectivity).

We then explore the detailed patterns of the hyper-parameter matrices, which conveys individual contributions of each connection 
to the behavioral performance. By inspecting the sparsity matrix [Fig.~\ref{matrix-msi} (b)], one can identify both unnecessary ($\pi=1$) and important
connections ($\pi=0$). We also find that, some 
neurons prefer receiving or sending information during the recurrent computation. An interpretation is that,
the spatio-temporal information is divided into relevant and irrelevant parts; the relevant parts are maintained through sending preference, while 
the irrelevant parts are blocked through receiving preference. 

\begin{figure}
\centering
     \includegraphics[bb=2 2 1207 412,width=0.8\textwidth]{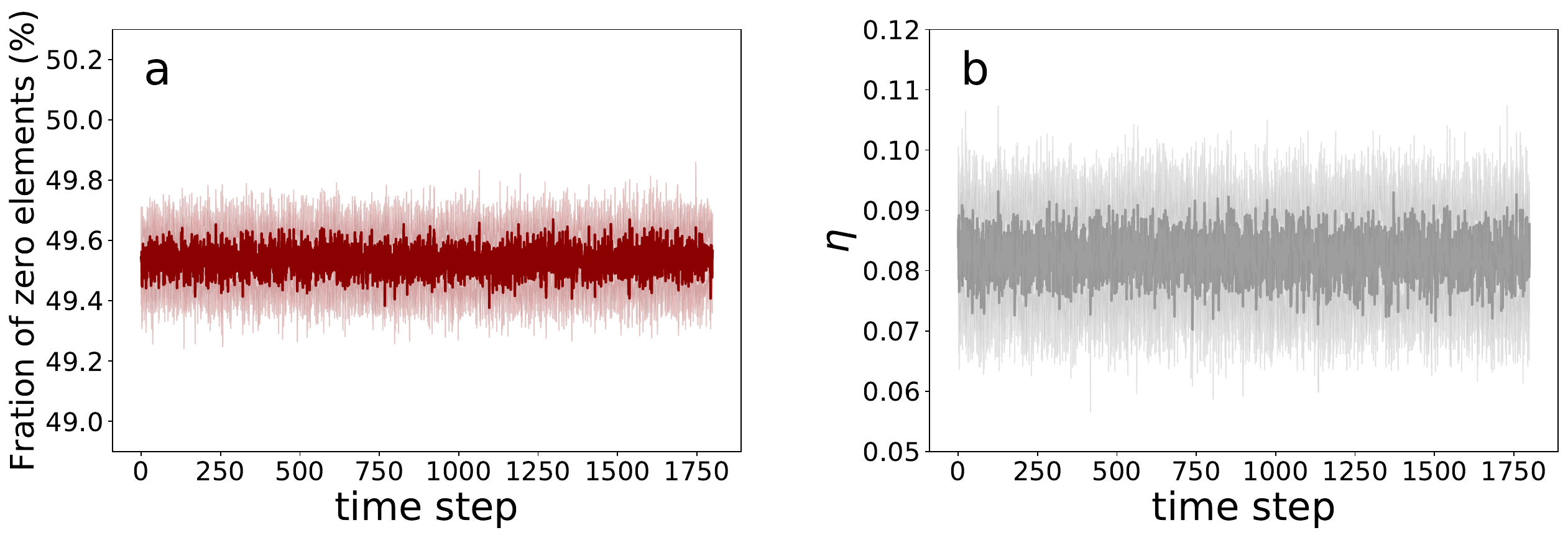}
  \caption{
 The statistics of the dynamic network implementing the MSI task could be preserved.
 (a) The fraction of zero elements of sampled weights across one test trial.
	The fluctuation is computed from ten independent runs. (b) The asymmetry measure $\eta$ defined by
 $\frac{\overline{w_{ij}w_{ji}}}{\overline{w_{ij}^2}}$
 versus time step. The over-bar means the average over all reciprocal connections. The result is averaged over ten independent runs.
  }\label{sampling-msi}
\end{figure}

Target weight perturbation experiments (Fig.~\ref{target-msi}) show that
the VIP
weights play a fundamental role in supporting the task accuracy reached by the recurrent computation.
Our method can thus provide precise temporal credit assignment to the MSI task, which the standard BPTT could not.

Finally, we remark that a dynamic network with time-dependent specified weight values is able to reach an equivalent accuracy with 
the network using the mean-field propagation. Note that the overall statistics of the network does not change significantly (Fig.~\ref{sampling-msi}), suggesting that
the hyper-parameters for the the weight statistics are more important than precise values of weights. In fact, dendritic spines in neural circuits, 
biological substrates for synaptic contacts,
are also subject to 
fluctuation, i.e., a highly dynamic structure~\cite{SD-2009,Kasai-2008}. Future exploration of this interesting connection would be fruitful, as an ensemble perspective
is much more abstract than a concrete topology, while the specified stationary topology is
still widely used in modern machine learning. Therefore,
the ensemble perspective yielding a dynamic network in adaptation of external stimuli could shed light on our understanding of 
adaptive RNNs.

\section{Mechanistic analysis of the recurrent SaS learning}
In this section, we provide a complete analysis of the SaS learning through different angles. First, we assume 
that the learning does not move far away from the random initialization, which is reasonable given that the learning rate is small and the network
is sufficiently large. This assumption leads to the lazy learning regime~\cite{NTK-2018}. Under this assumption, the dynamics can be predicted by calculating a recurrent neural tangent kernel (RNTK)~\cite{RNTK-2021}.  The tangent
kernel does not change (or less formally, the change is not significant) over the course of training. The kernel does not depend on the specific choice of network weights, but
depends only on the feature matrix of input data.
Second, when the kernel changes during learning, the lazy learning setting does not hold. In most practical learning with finite-size networks and arbitrarily-tuned learning rate,
the feature learning regime sets in. It is challenging to have a closed-form theory for this regime. Instead, we investigate our model by looking at the low dimensional projection of 
the neural and synaptic dynamics, which forms the task manifold in the ambient $N$-dimensional state space. Moreover, we find that the weight uncertainty impacts the learning accuracy, supporting 
the key role of stochastic plasticity. Finally, we also explain the feature learning as the emergent symmetry breaking in the hyper-parameter space. 
\subsection{Lazy learning regime}
In this section, we address a question whether there exists a
stationary RNTK in our model. The question is
non-trivial because we should consider not only the randomness of trainable hyper-parameters, but also the randomness of the particular
auxiliary variables---$\boldsymbol{\epsilon}^{t}(\boldsymbol{x})$ for recurrent units and $\epsilon^{\mathrm{out}}(\boldsymbol{x})$ for the single output unit 
in the simplest setting. 
These auxiliary variables capture the fluctuation effect in the learning, which plays a key role in our ensemble learning setting. Note that these variables are absent in the vanilla RNTK~\cite{RNTK-2021}.
\subsubsection{Recurrent neural tangent kernel}
To proceed, we first simplify our model as
\begin{equation}\label{SAS-RNN-simplified}
	\begin{aligned}
		h_{i}^{t+1}(\boldsymbol{x}) &= (1-\alpha)h_{i}^{t}(\boldsymbol{x}) +\alpha u_{i}^{t+1}(\boldsymbol{x}),\\
		u_{i}^{t+1}(\boldsymbol{x}) &= \sum_{j=1}^{N} \mu^{\mathrm{rec}}_{ij} r_{j}^{t}(\boldsymbol{x}) + \sum_{j=1}^{\nin} m^{\mathrm{in}}_{i j} x_{j}^{t+1} + \epsilon_i^{t+1}(\boldsymbol{x})\Delta_i^{t}(\boldsymbol{x}),\\
		r_{i}^{t+1}(\boldsymbol{x}) &=\phi\left(h_{i}^{t+1}(\boldsymbol{x})\right), \\
		f(\boldsymbol{x}) &=\sum_{i=1}^{N} \mu^{\mathrm{out}}_{i} r_{i}^{T}(\boldsymbol{x}) + \epsilon^{\mathrm{out}}(\boldsymbol{x}) \Delta^{\mathrm{out}}(\boldsymbol{x}).
	\end{aligned}
\end{equation}
In the above dynamics, we take the linear readout at the last time step and do not consider the external sensory noise for simplicity. At the random initialization for training,
we assume $h_i^0(\boldsymbol{x})$ is an i.i.d. random variable sampled from the Gaussian distribution
$\mathcal{N}(0,\sigma_h^2)$ for every input $\boldsymbol{x}$ that is denoted as $\boldsymbol{x}=\{\boldsymbol{x}_t\}_{t=1}^{T}$.

We then adopt the parameter initialization scheme as in the previous work~\cite{NTK-2018}
for our model. We use the following rescaled model parameters to replace the original ones in Eq.~(\ref{SAS-RNN-simplified}), and thus
the trainable parameters are still defined by the non-hatted variables, i.e.,
\begin{equation}\label{init-scheme}
	\begin{aligned}
		\tilde{\boldsymbol{m}}^{\mathrm{rec}} = \frac{\sigma_{\mathrm{rec}}}{\sqrt{N}}\boldsymbol{m}^{\mathrm{rec}}, \quad \tilde{\boldsymbol{m}}^{\mathrm{in}} = \frac{\sigma_{\mathrm{in}}}{\sqrt{N_{\mathrm{in}}}}\boldsymbol{m}^{\mathrm{in}}, \quad \tilde{\boldsymbol{m}}^{\mathrm{out}} = \frac{\sigma_{\mathrm{out}}}{\sqrt{N}}\boldsymbol{m}^{\mathrm{out}},\\
		\tilde{\boldsymbol{\pi}}^{\mathrm{rec}} = \boldsymbol{\pi}^{\mathrm{rec}}, \quad 
		\tilde{\boldsymbol{\pi}}^{\mathrm{out}} = \boldsymbol{\pi}^{\mathrm{out}}, \quad
		\tilde{\boldsymbol{\Xi}}^{\mathrm{rec}} = \frac{1}{N}\boldsymbol{\Xi}^{\mathrm{rec}}, \quad  \tilde{\boldsymbol{\Xi}}^{\mathrm{out}} = \frac{1}{N}\boldsymbol{\Xi}^{\mathrm{out}}.
	\end{aligned}
\end{equation}
where in the initialization $m^{\mathrm{rec}}_{ij} \sim \mathcal{N}(0,1)$, $m^{\mathrm{in}}_{ij} \sim \mathcal{N}(0,1)$, $m^{\mathrm{out}}_{i} \sim \mathcal{N}(0,1)$, $\pi_{ij}^{\mathrm{rec}}=0$, $\pi_i^{\mathrm{out}}=0$, $\Xi_{ij}^{\mathrm{rec}}\sim U(0,1)$, and $\Xi_{ij}^{\mathrm{out}}\sim U(0,1)$.
$U(0,1)$ indicates the uniform distribution with the support within $[0,1]$.
Note that $\boldsymbol{\pi}^{\rm in}=0$ and $\boldsymbol{\Xi}^{\rm in}=0$ for considering the case of small $\nin$.

We now derive the RNTK for $\alpha=1$ and leave the more involved derivation details for general case of $\alpha\in[0,1]$ in
Appendix \ref{app-a}. In our model, we find that each
synaptic current $h_i^t(\boldsymbol{x})$ and back-propagation error
$\delta_i^t(\boldsymbol{x}) = \sqrt{N}\frac{\partial f(\boldsymbol{x})}{\partial h_i^t(\boldsymbol{x})}$ follow 
centered Gaussian processes when $N$ tends to infinity. The associated kernels are given by
\begin{equation}
	\begin{aligned}
	\Sigma^{\left(t, t^{\prime}\right)}\left(\boldsymbol{x}, \boldsymbol{x}^{\prime}\right) &=\mathbb{E}_{\boldsymbol{\theta}}\left[h_i^{t}(\boldsymbol{x})h_i^{t^{\prime}}(\boldsymbol{x}^{\prime})\right],\\
	\Pi^{\left(t, t^{\prime}\right)}\left(\boldsymbol{x}, \boldsymbol{x}^{\prime}\right)
	&= \mathbb{E}_{\boldsymbol{\theta}}\left[\delta_i^t(\boldsymbol{x}) \delta_i^{t^{\prime}}(\boldsymbol{x}^{\prime})\right],
	\end{aligned}
\end{equation}
where the expectation is done over the trainable parameter vector $\boldsymbol{\theta}$.
These kernels can be computed recursively across time in the forward pass by
\begin{equation}
	\begin{aligned}\label{kernel-sigma}
		\Sigma^{\left(0, 0\right)}\left(\boldsymbol{x}, \boldsymbol{x}^{\prime}\right) & = \delta_{\boldsymbol{x}=\boldsymbol{x^{\prime}}}\sigma_h^2,\\
		\Sigma^{\left(t, t\right)}\left(\boldsymbol{x}, \boldsymbol{x}^{\prime}\right) &= \left(\sigma_{\mathrm{rec}}^{2}+\frac{1}{2}\delta_{\boldsymbol{x}=\boldsymbol{x^{\prime}}}\right) \mathrm{F}_{\phi}\left[\boldsymbol{K}^{\left(t, t\right)}\left(\boldsymbol{x}, \boldsymbol{x}^{\prime}\right)\right]+\frac{\sigma_{\mathrm{in}}^{2}}{\nin}\left\langle\boldsymbol{x}_{t}, \boldsymbol{x}^{\prime}_{t}\right\rangle,
	\end{aligned}
\end{equation}
where we define an operator $\mathrm{F}_{\phi}[\boldsymbol{K}]$ for an arbitrary function $\phi(\cdot)$ and
a semi-positive definite matrix $\boldsymbol{K}\in \mathbb{R}^{2\times2}$ as follows 
\begin{equation}
	\mathrm{F}_{\phi}[\boldsymbol{K}]=\mathbb{E}\left[\phi\left(\mathrm{z}_{1}\right) \cdot \phi\left(\mathrm{z}_{2}\right)\right], \quad\left(\mathrm{z}_{1}, \mathrm{z}_{2}\right) \sim \mathcal{N}(0, \boldsymbol{K}).
\end{equation}
The matrix $\boldsymbol{K}^{\left(t, t\right)}\left(\boldsymbol{x}, \boldsymbol{x}^{\prime}\right)$ in Eq.(\ref{kernel-sigma}) is explicitly defined below,
\begin{equation}
	\boldsymbol{K}^{\left(t, t\right)}\left(\boldsymbol{x}, \boldsymbol{x}^{\prime}\right)=\left[\begin{array}{cc}
		\Sigma^{(t-1, t-1)}(\boldsymbol{x}, \boldsymbol{x}) & \Sigma^{\left(t-1, t-1\right)}\left(\boldsymbol{x}, \boldsymbol{x}^{\prime}\right) \\
		\Sigma^{\left(t-1, t-1\right)}\left(\boldsymbol{x}, \boldsymbol{x}^{\prime}\right) & \Sigma^{\left(t-1, t-1\right)}\left(\boldsymbol{x}^{\prime}, \boldsymbol{x}^{\prime}\right)
	\end{array}\right].
\end{equation}
During the back-propagation pass, the kernels are computed in an analogous way,
\begin{equation}\label{kernels-pi}
	\begin{aligned}
		\Pi^{\left(T, T\right)}\left(\boldsymbol{x}, \boldsymbol{x}^{\prime}\right) 
		&=\sigma_{\mathrm{out}}^{2}\mathrm{F}_{\phi^{\prime}}\left[\boldsymbol{K}^{\left(T+1, T+1\right)}\left(\boldsymbol{x}, \boldsymbol{x}^{\prime}\right)\right] ,\\
		\Pi^{\left(t, t\right)}\left(\boldsymbol{x}, \boldsymbol{x}^{\prime}\right)&=\sigma_{\mathrm{rec}}^{2}\mathrm{F}_{\phi^{\prime}}\left[\boldsymbol{K}^{\left(t+1, t+1\right)}\left(\boldsymbol{x}, \boldsymbol{x}^{\prime}\right)\right]\Pi^{\left(t+1, t+1\right)}\left(\boldsymbol{x}, \boldsymbol{x}^{\prime}\right).\\
	\end{aligned}
\end{equation}

The RNTK is defined as $\Theta\left(\boldsymbol{x}, \boldsymbol{x}^{\prime}\right) = \sum_{p=1}^{|\boldsymbol{\theta}|} \frac{\partial f(\boldsymbol{x})}{\partial \mathbf{\theta}_p} \frac{\partial f(\boldsymbol{x}^{\prime})}{\partial \mathbf{\theta}_p}$, 
where trainable parameters $\boldsymbol\theta\equiv\{\boldsymbol{m}^{\mathrm{rec}},\boldsymbol{m}^{\mathrm{in}},\boldsymbol{m}^{\mathrm{out}},\boldsymbol{\pi}^{\mathrm{rec}},\boldsymbol{\Xi}^{\mathrm{rec}},\boldsymbol{\pi}^{\mathrm{out}},\boldsymbol{\Xi}^{\mathrm{out}}\}$. 
When $N$ goes to infinity, $\Theta\left(\boldsymbol{x}, \boldsymbol{x}^{\prime}\right)$ can be calculated
by using the Gaussian process kernels [Eq.~(\ref{kernel-sigma}), and Eq.~(\ref{kernels-pi})]. More precisely,
\begin{equation}
	\begin{aligned}
		\Theta\left(\boldsymbol{x}, \boldsymbol{x}^{\prime}\right)
		&= \sum_{t=1}^{T}  \Pi^{\left(t, t\right)}\left(\boldsymbol{x}, \boldsymbol{x}^{\prime}\right) \left(2 \sigma_{\mathrm{rec}}^2\mathrm{F}_{\phi}\left[\boldsymbol{K}^{\left(t, t\right)}\left(\boldsymbol{x}, \boldsymbol{x}^{\prime}\right)\right] + \frac{\alpha^2 \sigma_{\mathrm{in}}^2\langle \boldsymbol{x}_{t}, \boldsymbol{x}^{\prime}_{t}\rangle}{N_{\mathrm{in}}}\right)\\
		&\quad + 2\sigma_{\mathrm{out}}^2\mathrm{F}_{\phi}\left[\boldsymbol{K}^{\left(T+1, T+1\right)}\left(\boldsymbol{x}, \boldsymbol{x}^{\prime}\right)\right],
	\end{aligned}
\end{equation}
where $\langle \boldsymbol{x}_{t}, \boldsymbol{x}^{\prime}_{t}\rangle$ denotes the inner product between two inputs.

\begin{figure}
	\centering
	\includegraphics[bb=3 1 1083 394,width=0.9\textwidth]{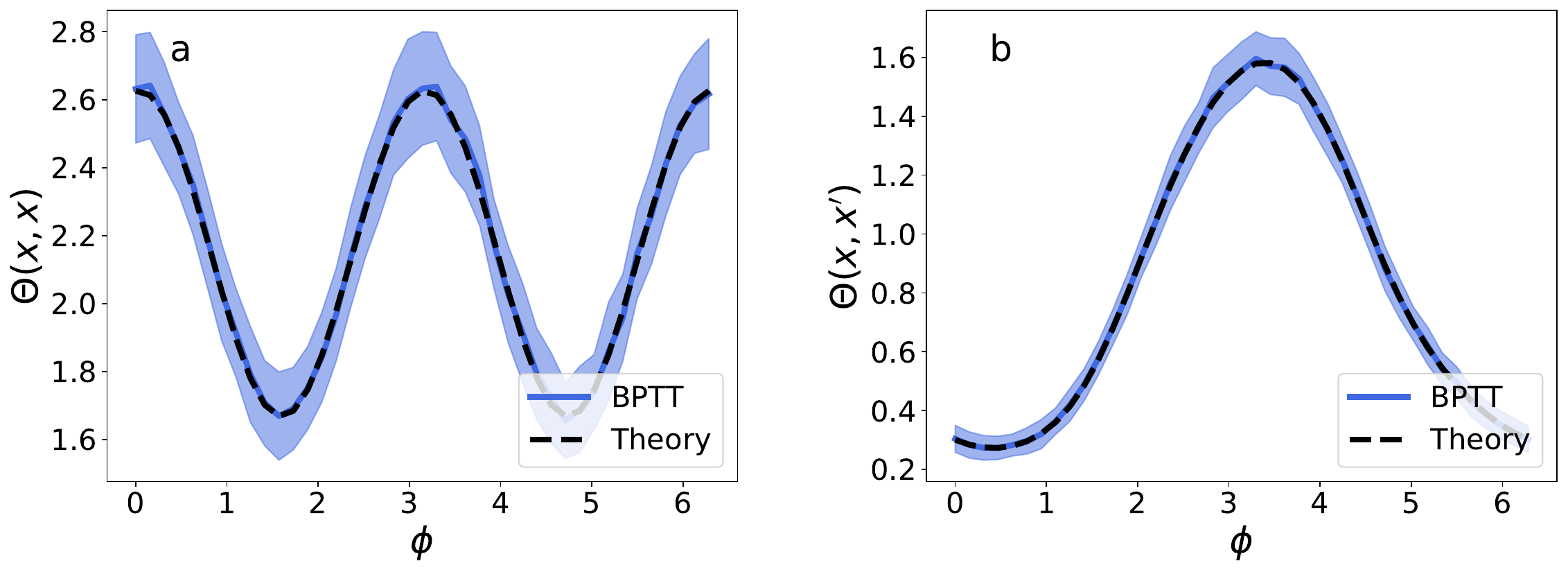}
	\caption{Comparison between analytical and empirical results for RNTK when $\alpha=0.9$. 
	Empirical results are obtained by running BPTT on recurrent neural networks where the recurrent size $N=2000$.
	The fluctuations are computed from 100 independent trials. (a) RNTK for a pair of identical inputs $\boldsymbol{x}$ and
	$\boldsymbol{x}$. (b) RNTK for a pair of different inputs $\boldsymbol{x}$ and $\boldsymbol{x}^{\prime}$.}
	\label{comprntk}
\end{figure}

To show the behavior of RNTK evaluated at initialization, we compare the 
empirical RNTK computed by BPTT with our analytical
result given two input sequences $\boldsymbol{x} = \{\cos{\phi},\sin{\phi},-\cos{\phi}\}$ 
and $\boldsymbol{x}^{\prime} = \{1,-1,1\}$ in Fig.~\ref{comprntk}, which shows an excellent agreement.

\subsubsection{Predicting dynamics of the network output}
In the lazy learning regime, the neural tangent kernel stays constant throughout the training. Hence,
the network output $f$ has a closed form dynamics equation~\cite{Lee-2019}. 
The gradient flow of the parameter vector $\boldsymbol{\theta}$ and network output $f$ can be derived by the chain rule,
\begin{equation}
\label{ode}
	\begin{aligned}
		\dot{\boldsymbol{\theta}}_t&=-\eta \nabla_{\boldsymbol{\theta}} f_{t}(\mathcal{X})^{T} \nabla_{f_{t}(\mathcal{X})} \mathcal{L}, \\
		\dot{f}_{t}(\mathcal{X})&=\nabla_{\boldsymbol{\theta}} f_{t}(\mathcal{X}) \dot{\boldsymbol{\theta}}_t=-\eta \hat{\Theta}_{t}(\mathcal{X}, \mathcal{X}) \nabla_{f_{t}(\mathcal{X})} \mathcal{L},
	\end{aligned}
\end{equation}
where $\dot{f}$ denotes the time derivative of $f$, $\eta$ is learning rate, and $\mathcal{X}\equiv\{\boldsymbol{x}^{\mu}\}_{\mu=1}^{P}$
denotes the training dataset. $f_{t}(\mathcal{X})$ is thus a $P\times 1$ vector consisting of all 
the network outputs. The loss function is given by the squared error
 $\mathcal{L} = \frac{1}{2}||f_t(\mathcal{X}) - \mathcal{Y} ||_2^2$, where the target output
$\mathcal{Y}\equiv\{y^{\mu}\}_{\mu=1}^{P}$. The empirical neural tangent 
kernel $\hat{\Theta}_{t} \equiv \hat{\Theta}_{t}(\mathcal{X}, \mathcal{X})$ at
time $t$ is a $P\times P$ matrix given by
\begin{equation}
	\hat{\Theta}_{t}=\nabla_{\boldsymbol{\theta}} f_{t}(\mathcal{X}) \nabla_{\boldsymbol{\theta}} f_{t}(\mathcal{X})^{T}=\sum_{p=1}^{|\boldsymbol{\theta}|} \nabla_{\theta_{p}} f_{t}(\mathcal{X}) \nabla_{\theta_{p}} f_{t}(\mathcal{X})^{T}.
\end{equation}

As assumed in the lazy regime, $\hat{\Theta}_{t}$ is a constant matrix $\Theta_0$ which is the
analytical neural tangent kernel calculated at the initialization. As a result, the exact solution of the 
ordinary differential equation [Eq.~(\ref{ode})] for the output $f_t(\mathcal{X})$ is given by
\begin{equation}{}
	f_{t}(\mathcal{X})=e^{-\eta \Theta_{0}t}\left(f_{0}(\mathcal{X})-\mathcal{Y}\right)+\mathcal{Y}.
\end{equation}

To obtain the time evolution of the network output $f$ for an arbitrary $\boldsymbol{x}$ (e.g., a test input),
we carry out a first-order Taylor expansion for $f_t(\boldsymbol{x})$ around 
the initial parameters $\boldsymbol{\theta}_0$, 
\begin{equation}\label{linear-model}
	f_{t}^{\mathrm{lin}}(\boldsymbol{x}) \equiv f_{0}(\boldsymbol{x})+\left.\nabla_{\theta} f_{0}(\boldsymbol{x})\right|_{\boldsymbol{\theta}=\boldsymbol{\theta}_{0}} \boldsymbol{\omega}_{t}
\end{equation}
where $\boldsymbol{\omega}_{t} = \boldsymbol{\theta}_t - \boldsymbol{\theta}_0$ is
 the relative change of parameters with respect to the initial point. Using Eq.~(\ref{ode}), $\boldsymbol{\omega}$ obeys the following update rule,
\begin{equation}
	\dot{\boldsymbol{\omega}}_{t}=-\eta \nabla_{\boldsymbol{\theta}} f_{0}(\mathcal{X})^{T} \nabla_{f_{t}^{\mathrm{lin}}(\mathcal{X})} \mathcal{L}.
\end{equation}
As $\nabla_{\boldsymbol{\theta}} f_{0}(\mathcal{X})$ stays constant during training, the dynamics of $\boldsymbol{\omega}$ has an analytic solution
as
\begin{equation}\label{omega-solution}
	\omega_{t}=-\nabla_{\boldsymbol{\theta}} f_{0}(\mathcal{X})^{T} \hat{\Theta}_{0}^{-1}\left(I-e^{-\eta \hat{\Theta}_{0} t}\right)\left(f_{0}(\mathcal{X})-\mathcal{Y}\right),
\end{equation}
where $I$ denotes the identity matrix.
Inserting Eq.~(\ref{omega-solution}) into the linear output Eq.~(\ref{linear-model}), we can obtain the final result
\begin{equation}\label{fx}
	f_{t}^{\mathrm{lin}}(\boldsymbol{x})=f_{0}(\boldsymbol{x})-\Theta_{0}(\boldsymbol{x}, \mathcal{X})\Theta_{0}^{-1}\left(I-e^{-\eta {\Theta}_{0}t}\right)\left(f_{0}(\mathcal{X})-\mathcal{Y}\right),
\end{equation}
where we have used $\hat{\Theta}_0=\Theta_0$.

Finally, we arrive at the analytic form of the output dynamics as
\begin{equation}\label{NTK-output}
	\begin{aligned}
		f_{t}(\mathcal{X})&=e^{-\eta\Theta_{0}t}\left(f_{0}(\mathcal{X})-\mathcal{Y}\right)+\mathcal{Y},\\
		f_{t}(\mathcal{X}^{\prime})&=f_{0}(\mathcal{X}^{\prime})-\Theta_{0}(\mathcal{X}^{\prime}, \mathcal{X})\Theta_{0}^{-1}\left(I-e^{-\eta {\Theta}_{0} t}\right)\left(f_{0}(\mathcal{X})-\mathcal{Y}\right),
	\end{aligned}
\end{equation}
where we denote the test dataset as $\mathcal{X}^{\prime}$ and omit the superscript `lin' for the following analysis. 

\begin{figure}
	\centering
	\includegraphics[bb=191 87 2353 1294, width=1.0\textwidth]{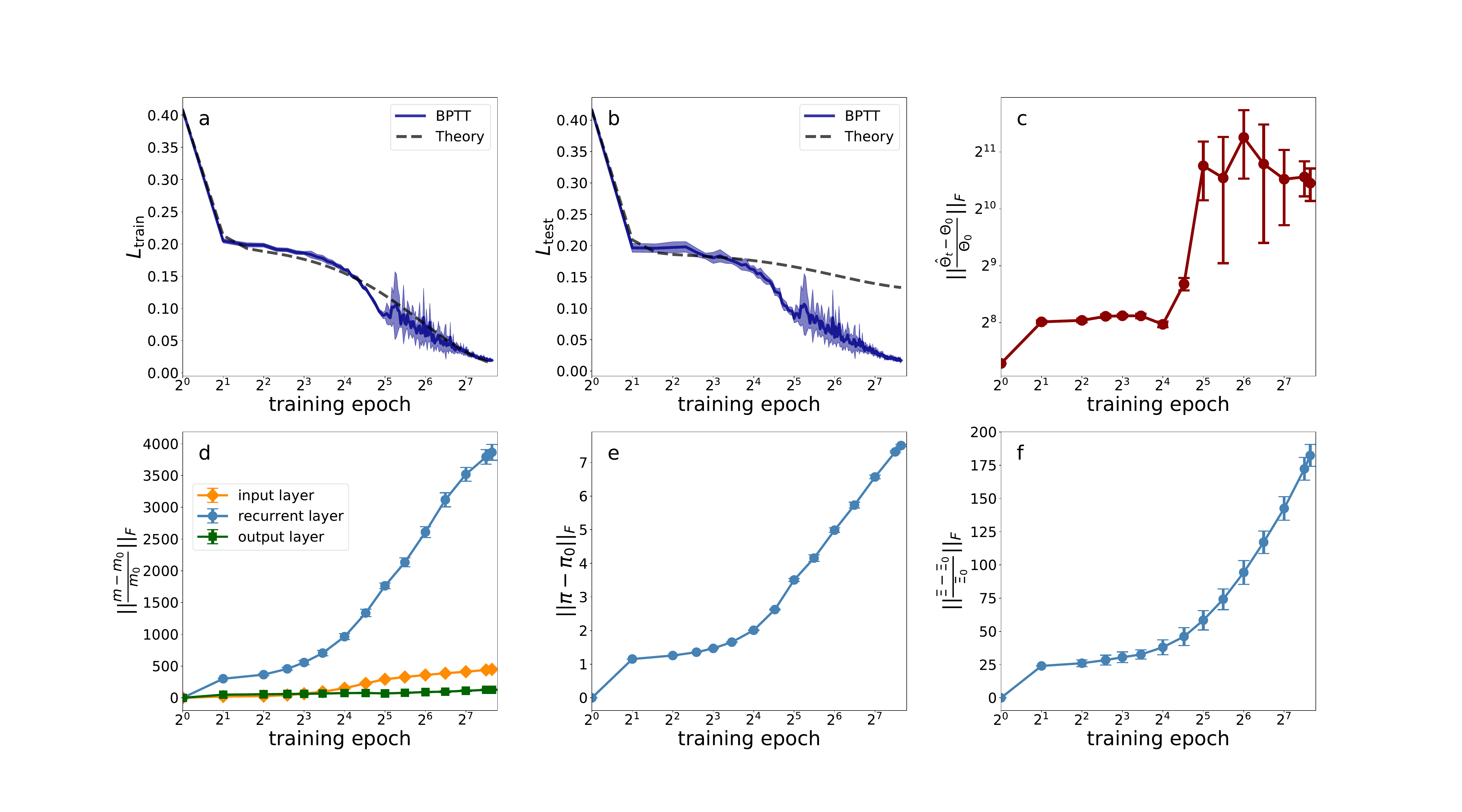}
	\caption{Comparison between BPTT and analytical learning dynamics when $\alpha=1$ and $N=512$.
	The hyper-parameters $\sigma_{\mathrm{in}}=\sigma_{\mathrm{rec}}=\sigma_{\mathrm{out}}=\sigma_h=1$.
	We set learning rate $\eta$ to $0.02$. The fluctuation is computed from $10$ independent runs. (a) 
	Training error per data sample $\mathcal{L}_{\mathrm{train}}$ versus epoch. The training sample size is equal to $512$. (b) 
	Test error per sample $\mathcal{L}_{\mathrm{test}}$ versus epoch. The test sample size is equal to $128$. (c)
	Frobenius norm of the relative difference between $\hat{\Theta}_t$ and $\Theta_{0}$ during training. (d) Frobenius norm of the 
	relative difference between 
	$\boldsymbol{m}_t$ and $\boldsymbol{m}_{0}$ for all layers during training. (e) Frobenius norm of the difference
	between $\boldsymbol{\pi}^{\mathrm{rec}}_t$ and $\boldsymbol{\pi}^{\mathrm{rec}}_{0}$ during training.
	(f) Frobenius norm of the relative difference between $\boldsymbol{\Xi}^{\mathrm{rec}}_t$ and $\boldsymbol{\Xi}^{\mathrm{rec}}_{0}$ during training. }
	\label{ntkdyn}
\end{figure}

To investigate whether our model performs the lazy learning when $N$ is large.
We design a binary classification task that the RNN learns to classify the
handwritten digits $0$ and $1$ from the MNIST dataset. The pixel sequence in an image is fed to the RNN in a manner of
row by row (i.e., $\nin=28$), and the corresponding label is set
to $0$ or $1$ for the digits $0$ or $1$ respectively.

We compare the analytical learning curves computed from Eq.~(\ref{NTK-output}) and 
the BPTT learning curves obtained from gradient descent training by back propagation
through time in Fig.~(\ref{ntkdyn}). As training progresses, two learning curves 
show a good agreement at an early stage and then deviate at a later stage.
The deviation is much more stronger for the test error curve. To explain this phenomenon, we plot the Frobenius 
norm $\|\frac{\hat\Theta_t(\mathcal{X}_{\mathrm{total}},\mathcal{X}_{\mathrm{total}})- \Theta_0(\mathcal{X}_{\mathrm{total}},\mathcal{X}_{\mathrm{total}})}{\Theta_0(\mathcal{X}_{\mathrm{total}},\mathcal{X}_{\mathrm{total}})}\|_{F}$ 
of the relative difference between the empirical tangent kernel $\hat\Theta_t(\mathcal{X}_{\mathrm{total}},\mathcal{X}_{\mathrm{total}})$
and analytical tangent kernel $\Theta_0(\mathcal{X}_{\mathrm{total}},\mathcal{X}_{\mathrm{total}})$ during 
training, where $\mathcal{X}_{\mathrm{total}} = \{\mathcal{X}, \mathcal{X}^{\prime}\}$ contains both training and test dataset.
Frobenius norms of model parameters are also plotted. 
Interestingly, we find that the RNTK Frobenius norm stays constant at the early stage of learning, and then increases rapidly.
This observation implies that the early stage can be described by the RNTK theory, while the later stage escapes from the lazy regime, and the learning becomes active.
The active or feature learning is also supported by the dynamics of parameter Frobenious norm [Fig.~\ref{ntkdyn} (d-f)].
The evident deviation for the test output also implies that the linear approximation used to derive Eq.~(\ref{linear-model}) needs to be corrected by taking into account
higher-order non-linear terms.

\begin{figure}
	\centering
	\includegraphics[bb=0 0 526 391,width=0.6\textwidth]{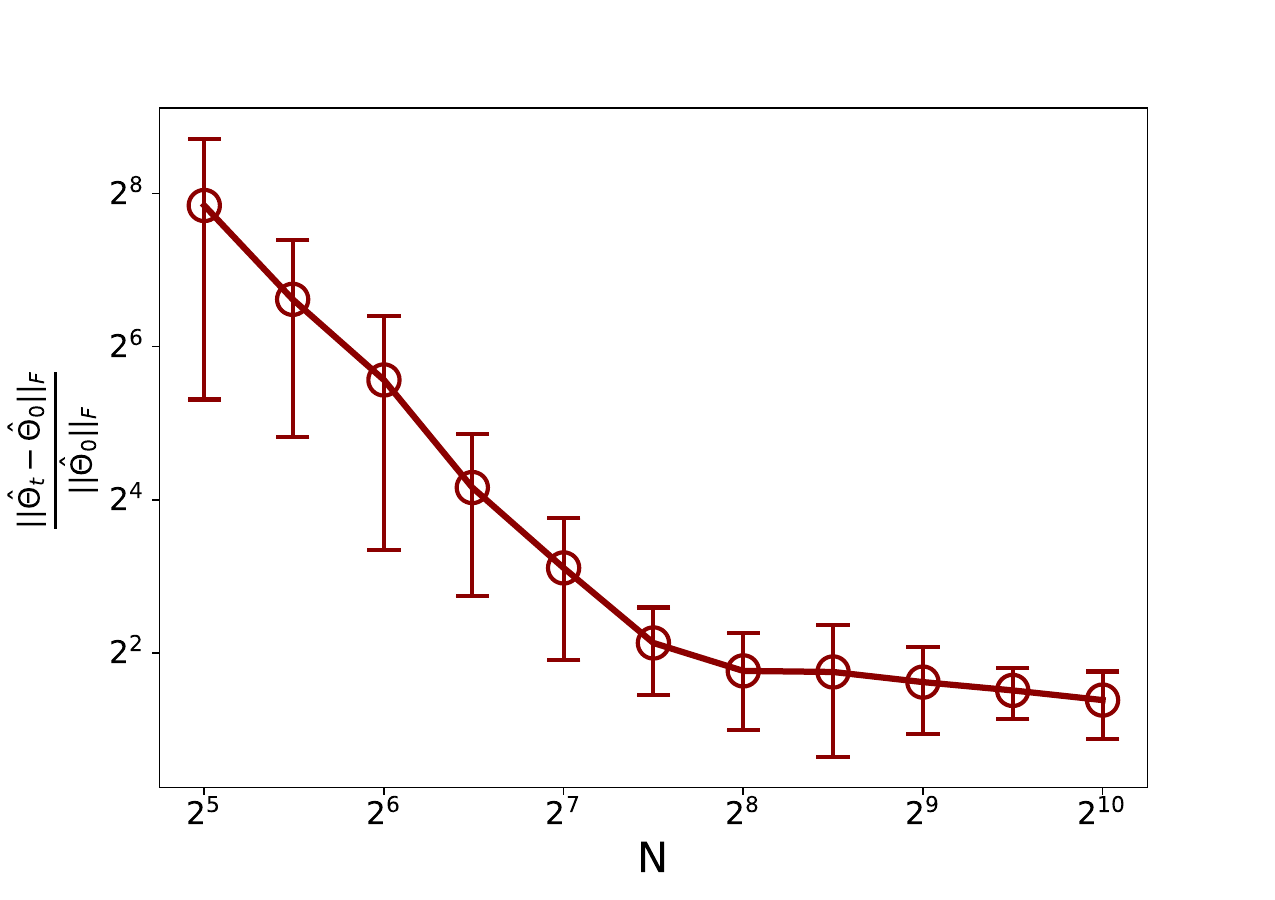}
	\caption{Kernel variation versus network size $N$. The learning rate $\eta=0.002$ and the total number of training epochs is $1000$.
	Other training conditions are the same as Fig.~\ref{ntkdyn}.}
	\label{ntkN}
\end{figure}

We finally ask whether the training will always stay in the lazy regime for infinite-width networks. 
To address this question, we measure the variation of the kernel between each pair of test 
data samples as $\frac{\left\|\hat\Theta_t(\mathcal{X}^{\prime},\mathcal{X}^{\prime})-\hat\Theta_0\left(\mathcal{X}^{\prime},\mathcal{X}^{\prime}\right)\right\|_{F} }{\left\|\hat\Theta_0\left(\mathcal{X}^{\prime},\mathcal{X}^{\prime}\right)\right\|_{F}}$
for increasing network size (Fig.~\ref{ntkN}). 
The kernel variation decreases as $N$ increases, which suggests that the training 
will be trapped in the lazy regime when $N$ is sufficiently large, and in this case the learning
dynamics becomes tractable. For a computational task using networks of practical sizes, it is necessary to analyze the feature learning regime,
which will be done in the following sections.

\subsection{Feature learning regime}
\subsubsection{Low-dimensional synaptic dynamics}
\label{LDsyn}
To explore the underlying picture of the feature learning, we first perform a low-dimensional projection of the synaptic dynamics along specific
directions which explain most of variances in the noisy synaptic dynamics.
The learning dynamics will experience two phases:
the  initial fast learning phase and the later slow exploration phase. 
In the initial phase, the training and test loss decreases 
rapidly, while in the exploration phase, 
the training loss gets close to zero, but the test loss is still decreasing albeit much more slowly.

\begin{figure}
	\centering
	\includegraphics[bb=2 1 805 485,width=0.8\textwidth]{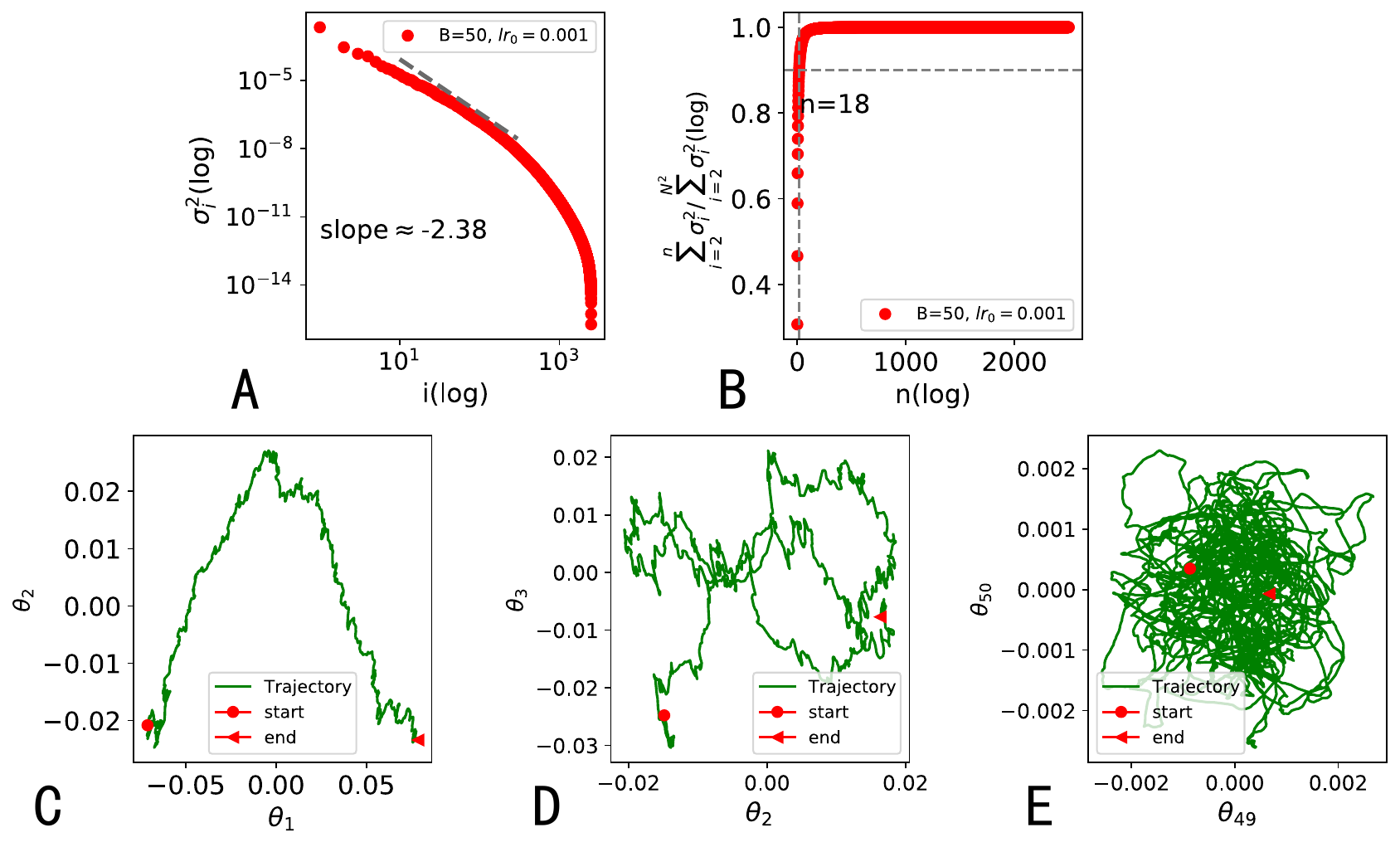}
	\caption{PCA results and the drift-diffusion motion for the parameter $\bm{m}$ in the pixel-by-pixel MNIST task.
	(A) The rank-ordered variance $\sigma_{i}^{2}$ versus different principal component (PC) directions $i$. 
	For $10 \leq i \leq 300$, $\sigma_{i}^{2}$ decreases with $i$ as a power law $i^{-\gamma}$ where
	$\gamma \sim 2.38$. (B) The normalized accumulative variance of the top $(n - 1)$ PCs excluding $i = 1$. 
	It reaches $\sim 90\%$ at $n = 18$ which is much smaller than the ambient dimension $N^2 = 2\,500$. 
	(C) The projected synaptic dynamics in the $(\vartheta_1,\vartheta_2)$ plane. 
	The persistent drift motion in $\vartheta_1$ and the diffusive random motion in $\vartheta_2$ are clearly shown.
	(D) The diffusive motion in the $(\vartheta_3,\vartheta_4)$ plane. (E) The diffusive motion in the $(\vartheta_{49},\vartheta_{50})$ plane.
	}
	\label{pixelm}
\end{figure}

In our model, there are $N$ recurrent units, and thus the recurrent parameter
matrix has the shape of $N \times N$, which is then flattened to a vector of the shape $1\times N^2$.
We take one minibatch as a single time unit, and thus for a learning process composed of $M$ minibatches, 
we obtain a parameter matrix of the shape $M\times N^2$. The size of $M$ depends on the time window, and 
we choose a large time window $t\in [t_0,t_0+T]$ where $T = 10$ 
epochs and $t_0$ is some moment in the exploration phase of the learning. From the perspective of the principal component analysis (PCA),
the synaptic dynamics
can be decomposed into their variations in different principal components as follows~\cite{Tu-2021},
\begin{equation}
\label{pcasyn}
	\begin{aligned}
		\bm{m}(t)&=\langle\bm{m}\rangle_{T}+\sum_{i=1}^{N^2} \vartheta_{i}^m(t) \bm{\psi}_{i}^m,\\
		\bm{\pi}(t)&=\langle\bm{\pi}\rangle_{T}+\sum_{i=1}^{N^2} \vartheta_{i}^{\pi}(t) \bm{\psi}_{i}^{\pi},\\
		\bm{\Xi}(t)&=\langle\bm{\Xi}\rangle_{T}+\sum_{i=1}^{N^2} \vartheta_{i}^{\Xi}(t) \bm{\psi}_{i}^{\Xi},\\
	\end{aligned}
\end{equation}
where $\bm{\psi}_i$ is the $i$-th principal component basis satisfying $\bm{\psi}_i\cdot \bm{\psi}_j = \delta_{ij}$,
and $\vartheta_i(t)$ is the value of the parameter matrix projected along the PCA direction $\bm{\psi}_i$.
The temporal average $\langle\bm{m}\rangle_{T}=T^{-1} \int_{t_{0}}^{t_{0}+T} \bm{m}(t) d t$ denotes
the mean of the parameter $\bm{m}$ in the time window, and so are the other parameters ($\langle\bm{\pi}\rangle_{T}$ and $\langle\bm{\Xi}\rangle_{T}$).
In other words, $\bm{\vartheta}$ is the projected parameter vector along the PCA coordinate.

\begin{figure}
	\centering
	\includegraphics[bb=2 1 805 485,width=0.8\textwidth]{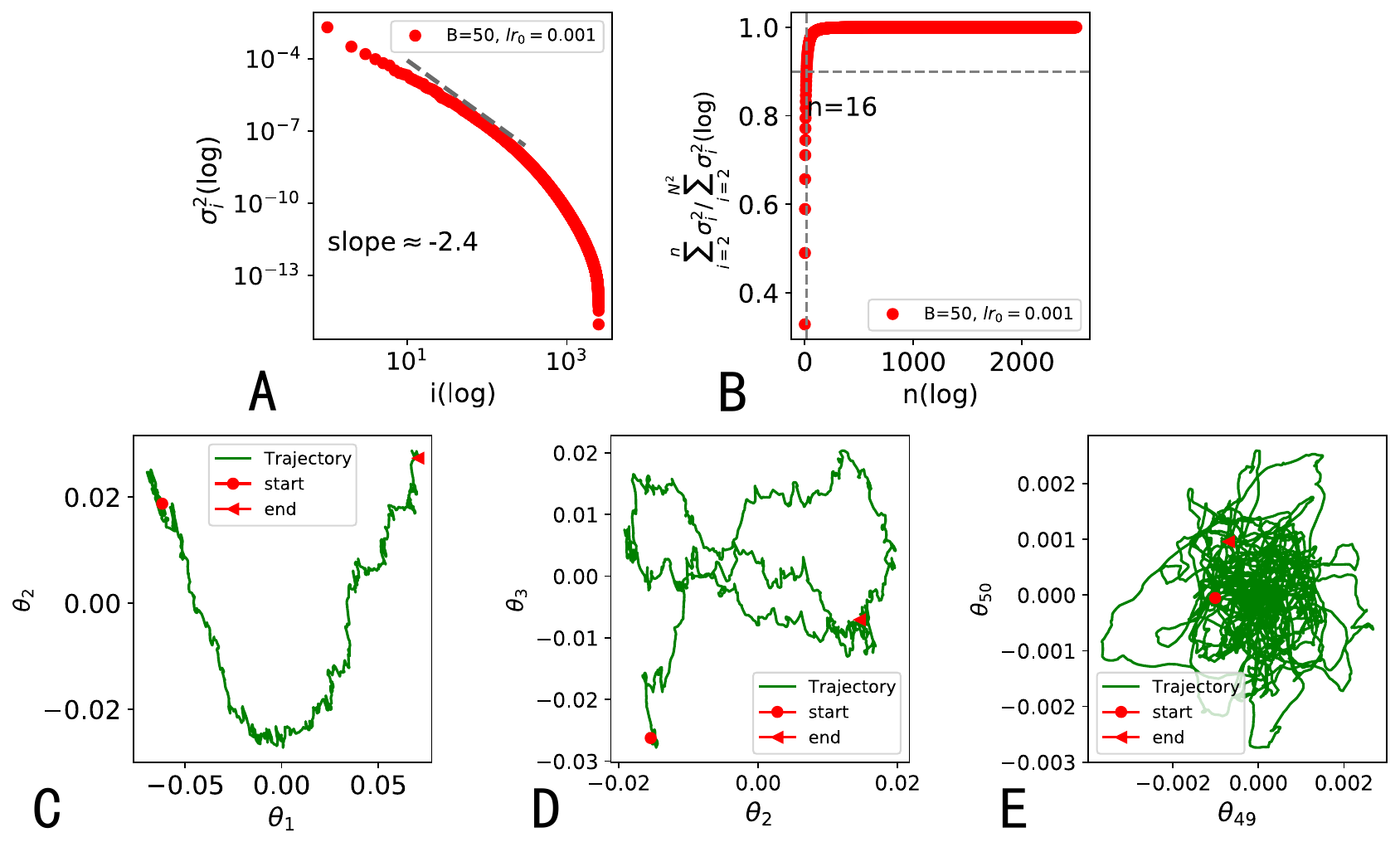}
	\caption{The PCA results and the drift-diffusion motion for the parameter $\bm{\pi}$ in the pixel-by-pixel MNIST task.
	Other conditions are the same as in Fig.~\ref{pixelm}.
	}
	\label{pixelpi}
\end{figure}

\begin{figure}
	\centering
	\includegraphics[bb=2 1 805 485,width=0.8\textwidth]{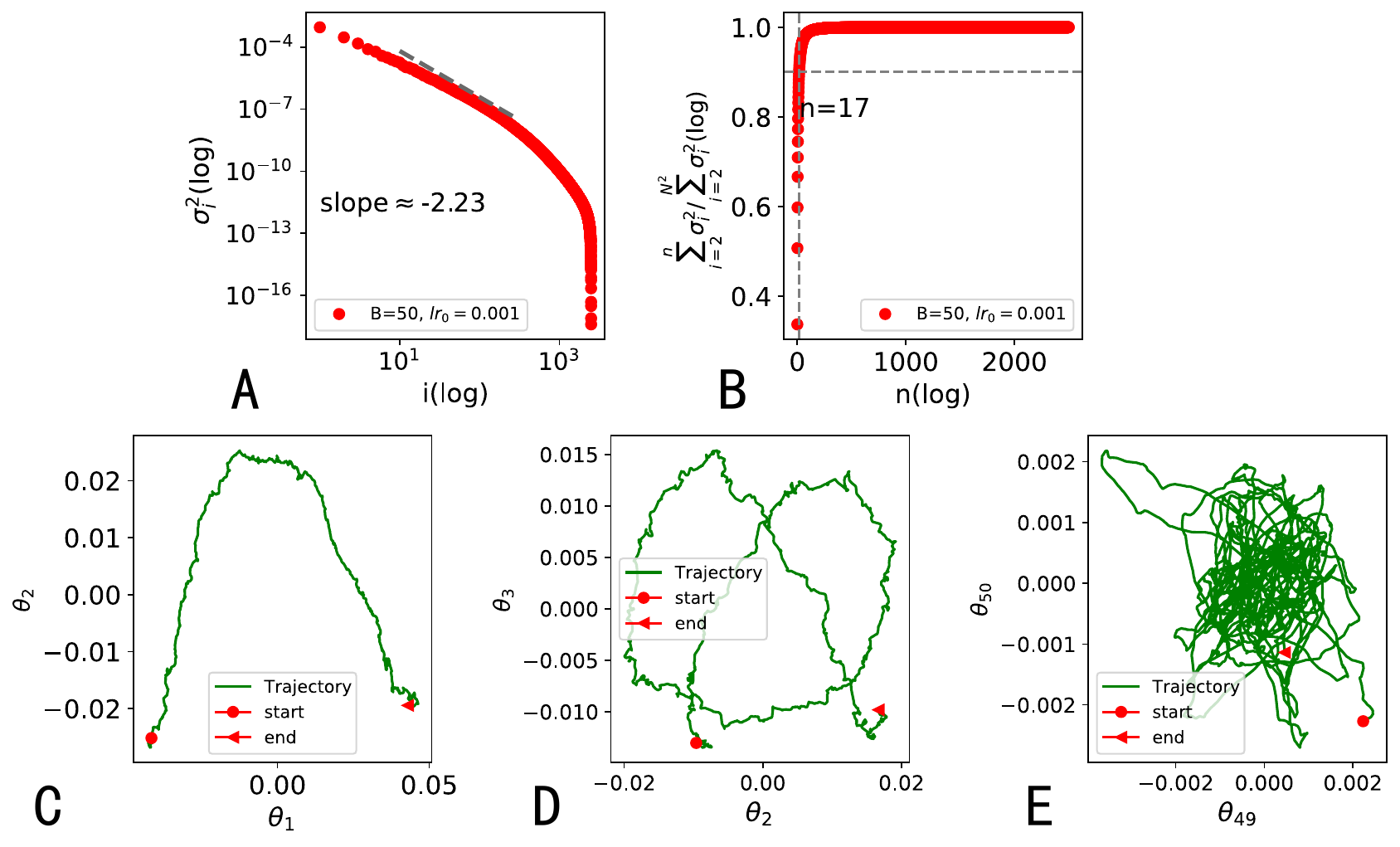}
	\caption{The PCA results and the drift-diffusion motion for the parameter $\bm{\Xi}$ in the pixel-by-pixel MNIST task.
	Other conditions are the same as in Fig.~\ref{pixelm}.
	}
	\label{pixelxi}
\end{figure}

In the pixel-by-pixel MNIST digit classification task, the number of
recurrent units are set to be $50$ for the analysis here.
The minibatch size $B$ is chosen to be $50$, which implies that
the network experiences $1200$ minibatch-updating during one epoch (a full training dataset is used). Therefore, 
$M=12000$ for $T = 10$ epochs. Taking $\bm{m}$ as an example, we show in Fig. \ref{pixelm} (A) the PCA spectrum, in which 
the variance is defined as $\sigma_i^2\equiv T^{-1} \int_{t_{0}}^{t_{0}+T} \vartheta_{i}^{2}(t) d t$ where 
the rank $i$ is arranged in the descending order (i.e., $\sigma_{i+1} < \sigma_i$).
With increasing rank, the variance first decreases exponentially with the rank
and then reduces more rapidly to a value of the magnitude $10^{-14}$, which implies that
most of the variations (for the learning dynamics) are captured by a relatively small number of PCA directions (see 
Fig. \ref{pixelm} (B) for a more precise estimation). 
The number of PCA dimensions explaining the variation of synaptic dynamics
is much smaller than the dimension of the ambient space ($N^2=2500$). 
These results show clearly that the SGD dynamics of the parameter $\bm{m}$ is embedded in a low-dimensional manifold.

\begin{figure}
	\centering
	\includegraphics[bb=2 1 805 485,width=0.8\textwidth]{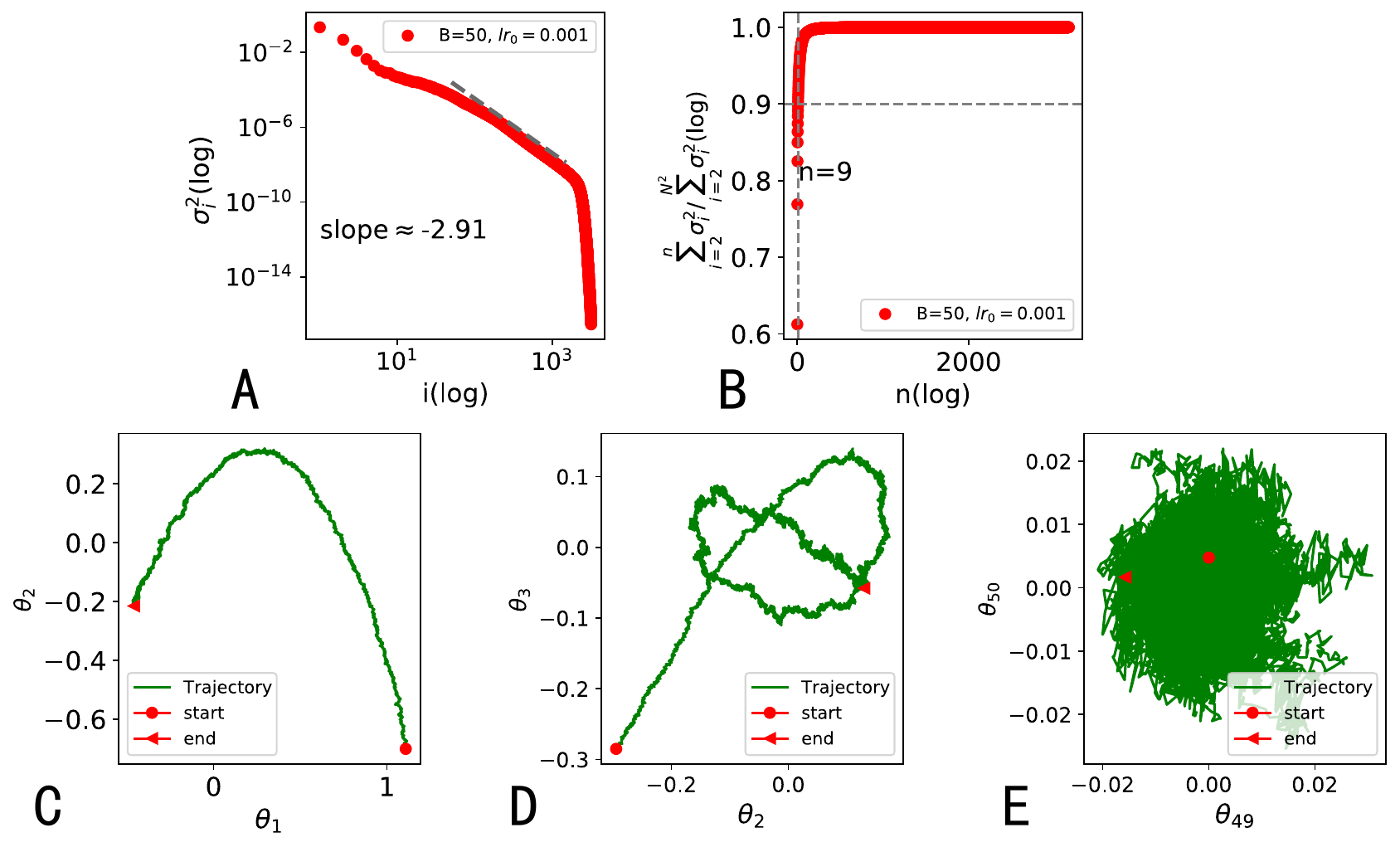}
	\caption{The PCA results and the drift-diffusion motion for the parameter $\bm{m}$ in the MSI task.
	(A) The rank-ordered variance $\sigma_{i}^{2}$ versus different principal component (PC) directions $i$. 
	For $50 \leq i \leq 1500$, $\sigma_{i}^{2}$ decreases with $i$ as a power law $i^{-\gamma}$ with $\gamma \sim2.91$.
	(B) The normalized accumulative variance of the top $(n - 1)$ PCs excluding $i = 1$. 
	It reaches $\sim 90\%$ at $n = 9$ much smaller than the ambient dimension $N^2 = 3\,600$.
	(C) The projected synaptic dynamics in the $(\vartheta_1,\vartheta_2)$ plane.
	The persistent drift motion in $\vartheta_1$ and the diffusive random motion in $\vartheta_2$ are clearly shown. 
	(D) The diffusive motion in the $(\vartheta_3,\vartheta_4)$ plane. (E) The diffusive motion in the $(\vartheta_{49},\vartheta_{50})$ plane.
	}
	\label{cognim}
\end{figure}
\begin{figure}
	\centering
	\includegraphics[bb=2 1 805 485,width=0.8\textwidth]{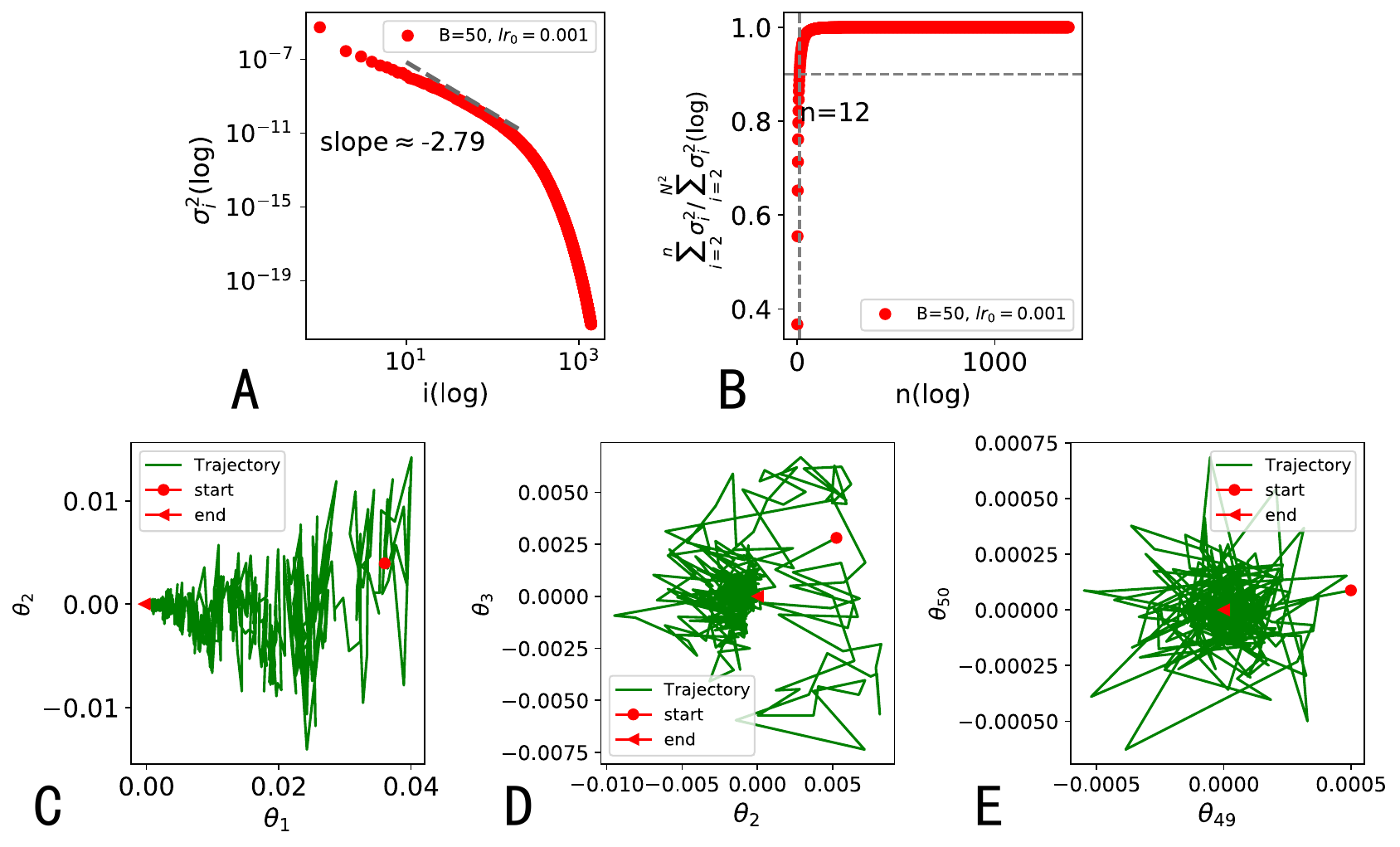}
	\caption{The PCA results and the drift-diffusion motion for the parameter $\bm{\pi}$ in the MSI task. The power law fitting is done for the range $10 \leq i \leq 200$.
	Other conditions are the same as in Fig.~\ref{cognim}. 
	}
	\label{cognipi}
\end{figure}
A salient feature in Fig. \ref{pixelm} (A) is that the variance along the first PCA direction $\bm{\psi}_1$ is
much larger than the other directions. To reveal the underlying picture, we analyze the
synaptic dynamics along specific directions. In Fig. \ref{pixelm} (C), we display the synaptic 
dynamics projected onto the $(\vartheta_1, \vartheta_2)$ space. We observe that along the
first PCA direction $\bm{\psi}_1$, there is a net drift velocity $\frac{d\vartheta}{dt}$.
For other PCA directions, the low dimensional dynamics becomes noisier (like random walks) with increasing
rank [Fig.~\ref{pixelm} (D) and (E)]. The same qualitative behavior is also observed for the other two hyper-parameters (see 
Fig.~\ref{pixelpi} and Fig.~\ref{pixelxi}).

 We also carry out the same analysis for the MSI task which uses the mean squared error as the loss function. In this analysis, we choose $N=60$,
 and collect $20\,000$ epochs in the exploration phase (no minibatch is used).
 We can see similarly that,
the dynamics of the parameters is embedded in the low-dimensional space with a persistent
net drift velocity along the direction $\bm{\psi}_1$, as shown in Fig. \ref{cognim} (for the parameter $\bm{m}$) and Fig. \ref{cognipi} (for the parameter $\bm{\pi}$).
The dynamics of $\bm{\Xi}$ shows a similar behavior.

\begin{figure}
	\centering
	\includegraphics[bb=4 5 930 260,width=0.9\textwidth]{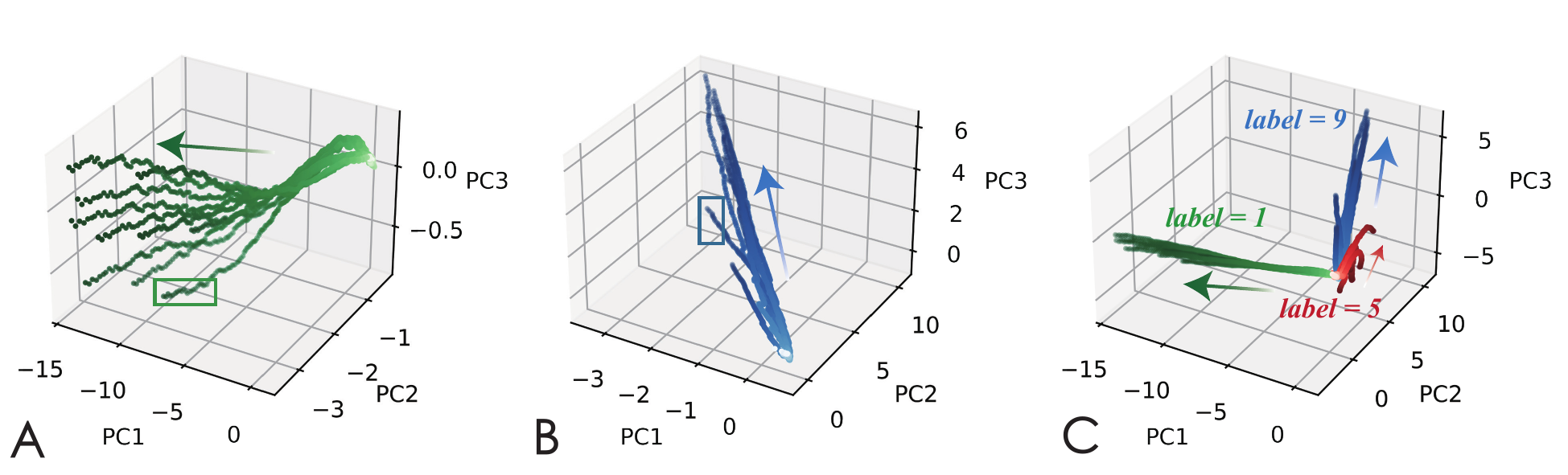}
	\caption{The three-dimensional projection of the neural dynamics with different inputs. 
	The arrow direction indicates the evolution of the dynamics, and the intensity of the color is gradually 
	increased from $t=0$ to $t=T$ for each sample.  
	All the trajectories start from the same point. Trajectories produced by a network sampled from the well-trained ensemble 
	are shown for the inputs of the label 1 (A) and the label 9 (B). The trajectories marked by rectangles denote 
	samples with a high loss or misclassified.  (C) Trajectories of different inputs shown in the same neural space. 
	If the input samples of the same label are correctly classified, the trajectories converge to the same subspace. 
	Simulation parameters: $N=100$ , $T=785$ and $P = 3\times 10$ (ten pictures for each label).}
	\label{pixelact}
\end{figure}

\subsubsection{Low-dimensional neural dynamics}
The internal representation for the task learning can be constructed by the ensemble algorithm. The geometric organization of the 
internal representation is also a crucial factor determining the success of the algorithm. Therefore, we investigate the low-dimensional
projection of the neural activity in response to time-dependent inputs in this section. The ambient state space is described by $N$ coordinates, and one point in this space indicates
an $N$-dimensional firing rate activity. When the neural network adapts to the input sequence, the firing rate point 
draws a trajectory in this ambient space. To check whether the trajectory is embedded in a low-dimensional subspace, 
we can use the PCA method. 
For a given input sequence $k$, we extract the states of all $N$ neurons across all
discrete time steps of the total length $T$ to construct an $N\times T$ matrix called the dynamics matrix for the input $k$.
If the network receives $P$ input sequences, we stack the dynamics matrices from different inputs
horizontally to construct a $PT \times N$ full dynamics matrix $\mathbf{\hat{r}}$. The PCA is applied to this full dynamics matrix. 
To implement PCA, we first compute the equal-time cross-correlation matrix
\begin{equation}
	D_{i j}=\langle\left(\hat{r}_{i}(\hat{t})-\langle \hat{r}_{i}(\hat{t})\rangle\right)\left(\hat{r}_{j}(\hat{t})-\langle \hat{r}_{{j}}(\hat{t})\rangle \right)\rangle, \quad \hat{t} = 0 ,\ldots, PT,
\end{equation}
where the average $\langle \cdot \rangle$ denotes the temporal average. We then perform the spectral
decomposition of the $N\times N$ cross-correlation matrix as $\mathbf{D} = \Phi\Sigma\Phi^{-1}$, 
and the dynamics of $\mathbf{\hat{r}}$ can be decomposed into its variations in different principal components as follows,
\begin{equation}
\begin{aligned}
	\mathbf{\hat{r}}(\hat{t}) &= \langle \mathbf{\hat{r}}(\hat{t})\rangle+ \sum_{i=1}^{N}\vartheta_i(\hat{t})\boldsymbol{\phi}_i,\\
	\vartheta_i(t) & = [\Phi\cdot (\mathbf{\hat{r}}(\hat{t}) -  \langle \mathbf{\hat{r}}(\hat{t})\rangle)]_i
\end{aligned}
\end{equation}
where $\vartheta_i (\hat{t}) $ denotes the projection of $\mathbf{\hat{r}}(\hat{t})$ along the $i$-th PC direction $\bm{\phi}_i$
with $\bm{\phi}_i \cdot \bm{\phi}_j = \delta_{ij}$ (orthogonal bases).

We then analyze the neural dynamics of pixel-by-pixel MNIST task. Surprisingly, we find that the first three PCA modes explain more than $85\%$ of the total variance, 
which clearly shows that the dynamics of network activity is embedded in
a low-dimensional subspace. We thus keep the first three PCA modes [$\vartheta_1(t), \vartheta_2(t)$ and $\vartheta_3(t)$],
and plot the three-dimensional trajectory of the network dynamics in a subspace with coordinates
$(\vartheta_1, \vartheta_2, \vartheta_3)$. The result is shown in Fig. \ref{pixelact}, which indicates that the ensemble learning can 
segregate the information for different types of digits. The learning would fail if the neural activity does not move towards the specific manifold.

\begin{figure} 
	\centering
	\includegraphics[bb=5 5 836 713,width=0.8\textwidth]{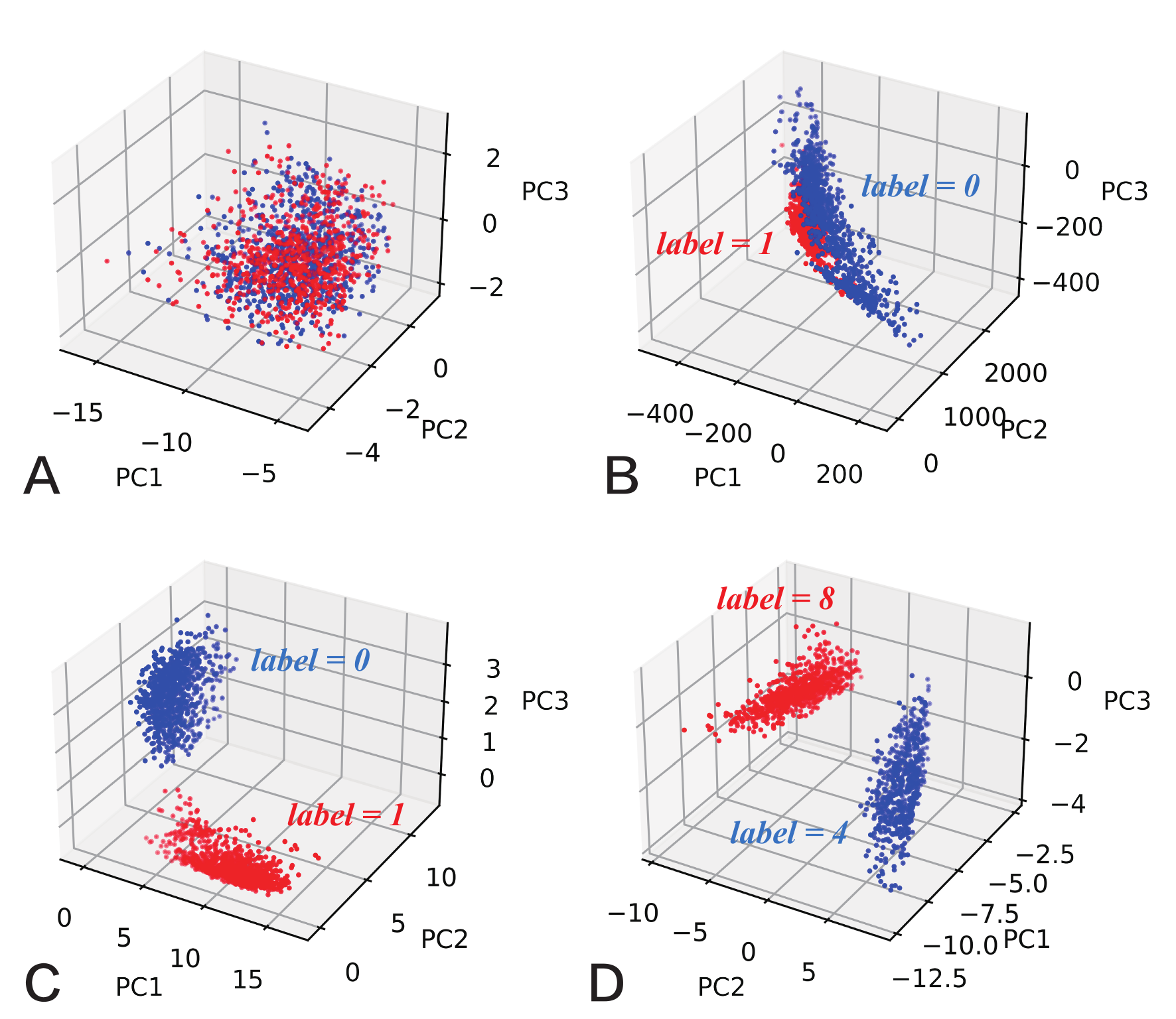}
	\caption{The low-dimensional projection of the neural activity at the last time step of the dynamics. 
	(A) Trajectories corresponding to one label (e.g., digit 8 here, the samples are taken from test (blue points) or training (red points) dataset)
	converge to the same subspace.  All the samples are perfectly classified
	(i.e., the softmax output for the right class is more than $0.95$). (B) For a random (untrained) network, 
	the low dimensional projections of the last-time-step neural states are entangled for labels $0$ and $1$ (other pairs of labels show a similar behavior). 
	(C, D) For trained networks, the projections are well separated (an example of label pair (0,1) or (4,8) is shown).}
	\label{piactFT}
\end{figure} 

In particular, for inputs with the same label, the trajectories of network dynamics converge to 
the same subspace in the low-dimensional neural space, provided that the inputs are all perfectly classified. 
It can be clearly recognized that if one input sample is misclassified (or with a high loss), the induced trajectory
will deviate from the category manifold, marked by the rectangles in Fig. \ref{pixelact} (A) and (B). 
Our ensemble training thus leads to disentangled category manifolds (the subspace the dynamics of correctly classified inputs flow to),
as shown in Fig. \ref{pixelact} (C). This segregation supports the success of our algorithm.  
As the pixel-by-pixel MNIST classification task depends on the decision made at the last time step when the input sequence is completed,
we also plot the projected neural state at the last time step in Fig. \ref{piactFT}. 
For input samples from test and training dataset, the corresponding trajectories converge to the 
same subspace, as illustrated in Fig. \ref{piactFT} (A).
Hence, we choose all the samples with label 0 and 1 in Fig. \ref{piactFT} (C), label 4 and 8 in Fig. \ref{piactFT} (D)
from the test dataset, and project the last-time-step neural activity for each sample in the three-dimensional space. The category manifolds are well separated
for trained networks [Fig.~\ref{piactFT} (C) and (D)], but entangled for untrained networks [Fig.~\ref{piactFT} (B)].
This picture explains how the ensemble algorithm drives the segregation of the input information into distinct category manifolds.

\begin{figure}
	\centering
	\includegraphics[bb=5 7 1663 648,width=0.8\textwidth]{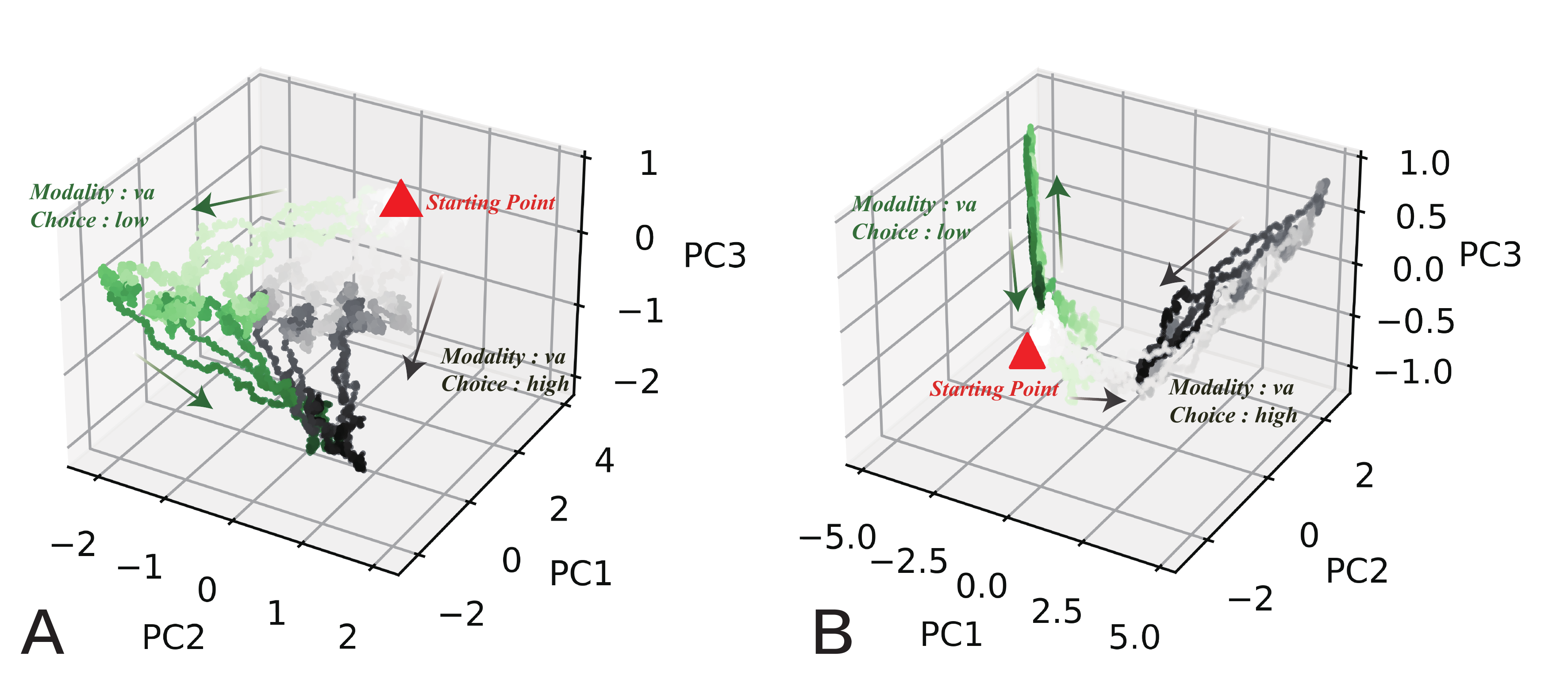}
	\caption{The low dimensional projections of the neural activity for the MSI task. 
	All the trajectories start from the same point indicated by a triangle. Two kinds of input 
	samples are considered: the modality is `va' (multisensory), and the frequency choice is `low' (green dots); the modality is `va' 
	and the frequency choice is `high' (gray dots). The arrow direction indicates the evolution of the dynamics, and the intensity of the color is gradually 
	increased from $t=0$ to $t=T$ for each sample.  (A) The projected trajectories produced by an untrained network. 
	(B) The projected trajectories produced by the network sampled from the well-trained ensemble. We use $N=150$ and $T = 3601$ in this task.}
	\label{sasva}
\end{figure}

We next analyze the low-dimensional neural dynamics of the MSI task.
In Fig. \ref{sasva}, we randomly choose two kinds of input samples with the same modality but 
different choices (four samples for each choice are considered). We then make a low-dimensional projection of the neural activity, and 
find that the first three PCA modes explain more than $85\%$ of the total variance. For a well-trained network, the dynamics for inputs of 
different frequencies are well separated [Fig.~\ref{sasva} (B)]. In contrast, the random untrained network does not have this nice property [Fig.~\ref{sasva} (A)].

\subsubsection{Symmetry breaking in the hyper-parameter space}
Symmetry breaking is an important concept in understanding the feature learning~\cite{Huang-2019,Huang-2020}.
In this section, we relate the hyper-parameter symmetry breaking to the learning performance of the ensemble training.
First, we assume a symmetric initialization, because we have no prior knowledge about the true solution of the network connectivity with proper weights on each link.
This symmetry is formulated as follows,
\begin{equation}
	\begin{aligned}
		P(w_{ij})& = \frac{1}{2}\delta(w_{ij}) + \frac{1}{2}\mathcal{N}(w_{ij}|0,\frac{1}{N}).
	\end{aligned}
\end{equation}
This initialization means that all entries of the vector are identical, i.e., $\mathbf{m} = 0,\boldsymbol{\pi} = \frac{1}{2}, \mathbf{\Xi} = \frac{1}{N}$. 
We then study when and how this symmetry is broken during learning. For simplicity, we focus on the task of 28-by-28 MNIST digit classification, where 
the network reads 28 pixels at each time step. We also analyze the MSI task.

\begin{figure}
	\centering
	\includegraphics[bb=4 7 1400 422,width=0.8\textwidth]{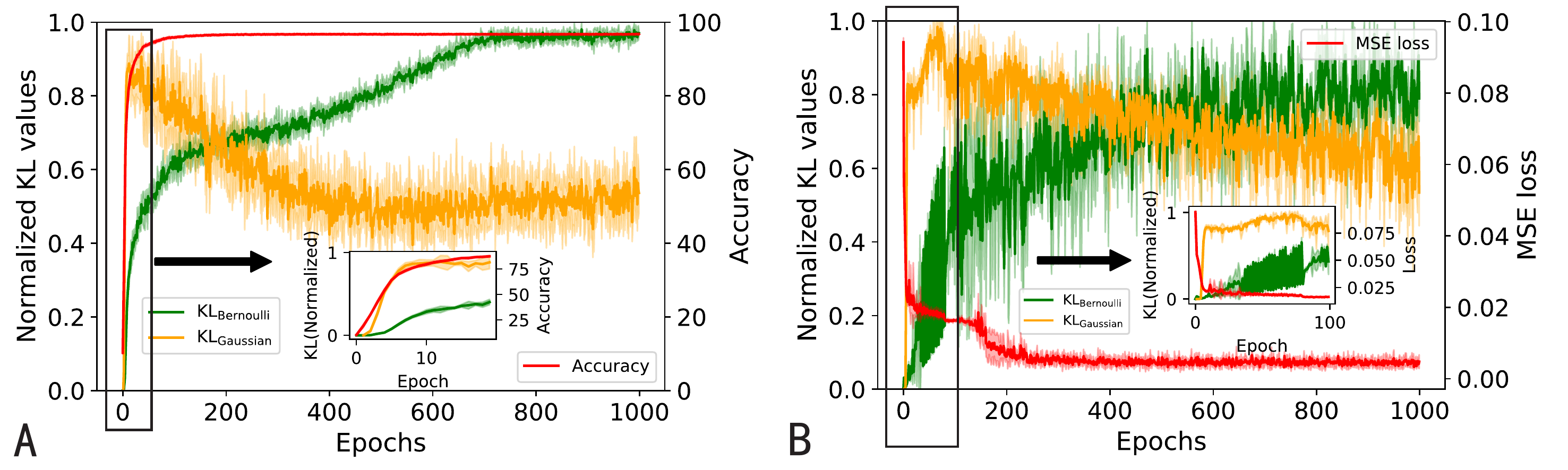}
	\caption{Symmetry breaking of hyper-parameters in the 28-by-28 MNIST digit classification and MSI task. 
	The fluctuations are computed over five independent runs. (A) Results of the 28-by-28 MNIST digit classification task. 
	The KL values at two levels (Bernoulli and Gaussian) are shown in green and yellow lines respectively. 
	These KL values are both normalized by their maxima and averaged over all the connections. 
	The test accuracy over $1\,000$ epochs of training is shown in red. Inset: The two normalized KL values over the first 20 epochs. 
	(B) Results of the MSI task.
	}
	\label{KL}
\end{figure}

To quantify the degree of the symmetry breaking, we use the Kullback-Leibler (KL)
divergence to measure the distance between the weight distribution at the 
initialization and the distribution at the epoch $t$. Specifically, we analyze the KL divergence at two levels: the discrete Bernoulli
and the continuous Gaussian levels. The Bernoulli KL distance is evaluated as
\begin{equation}
	\begin{aligned}
		 \mathrm{KL}_{\rm Bernoulli}(\pi_0||\pi_t)&= -\pi_0\ln\pi_t-(1-\pi_0)\ln(1-\pi_t)+\pi_0\ln \pi_0+(1-\pi_0)\ln(1-\pi_0),\\
		 & =  -0.5\ln\pi_t-0.5\ln(1-\pi_t)-\ln 2,
	\end{aligned}
\end{equation}
where $\pi_0=0.5$, and $\pi_t$ is truncated to $[a,1-a]$ for which we set $a =10^{-10}$ in simulations to avoid numerical divergence. 
The Gaussian KL distance is computed as follows,
\begin{equation}
	\begin{aligned}
		\mathrm{KL}_{\rm Gaussian}(\mathcal{N}_0||\mathcal{N}_t)
		& = \ln (N^{\frac{1}{2}}{\sqrt{\Xi_{t}}})+\frac{N^{-1}+\left(m_{t}\right)^{2}}{2 \Xi_{t}}-\frac{1}{2}.
	\end{aligned}
\end{equation}
Similarly, we also truncate $\Xi_t \in [a,\infty]$ for which we set $a = 10^{-10}$ in simulations to avoid divergence.
As shown in Fig. \ref{KL}, there exist two stages. In the first stage, the test accuracy increases rapidly, as the symmetry starts to break at an epoch value
less than five. The continuous symmetry is first broken, followed by the discrete symmetry. The Gaussian KL distance grows faster than the Bernoulli one.
In the second stage, the test accuracy reaches a steady value, while both KL distances are still changing, implying that the learning explores
the low-dimensional manifold of synaptic activity (see Sec~\ref{LDsyn}). This qualitative behavior also holds in the MSI task.

\subsubsection{Stochastic plasticity impacts the learning accuracy}
\label{stoc}
\begin{figure}
	\centering
	\includegraphics[bb=8 29 806 551,width=0.7\textwidth]{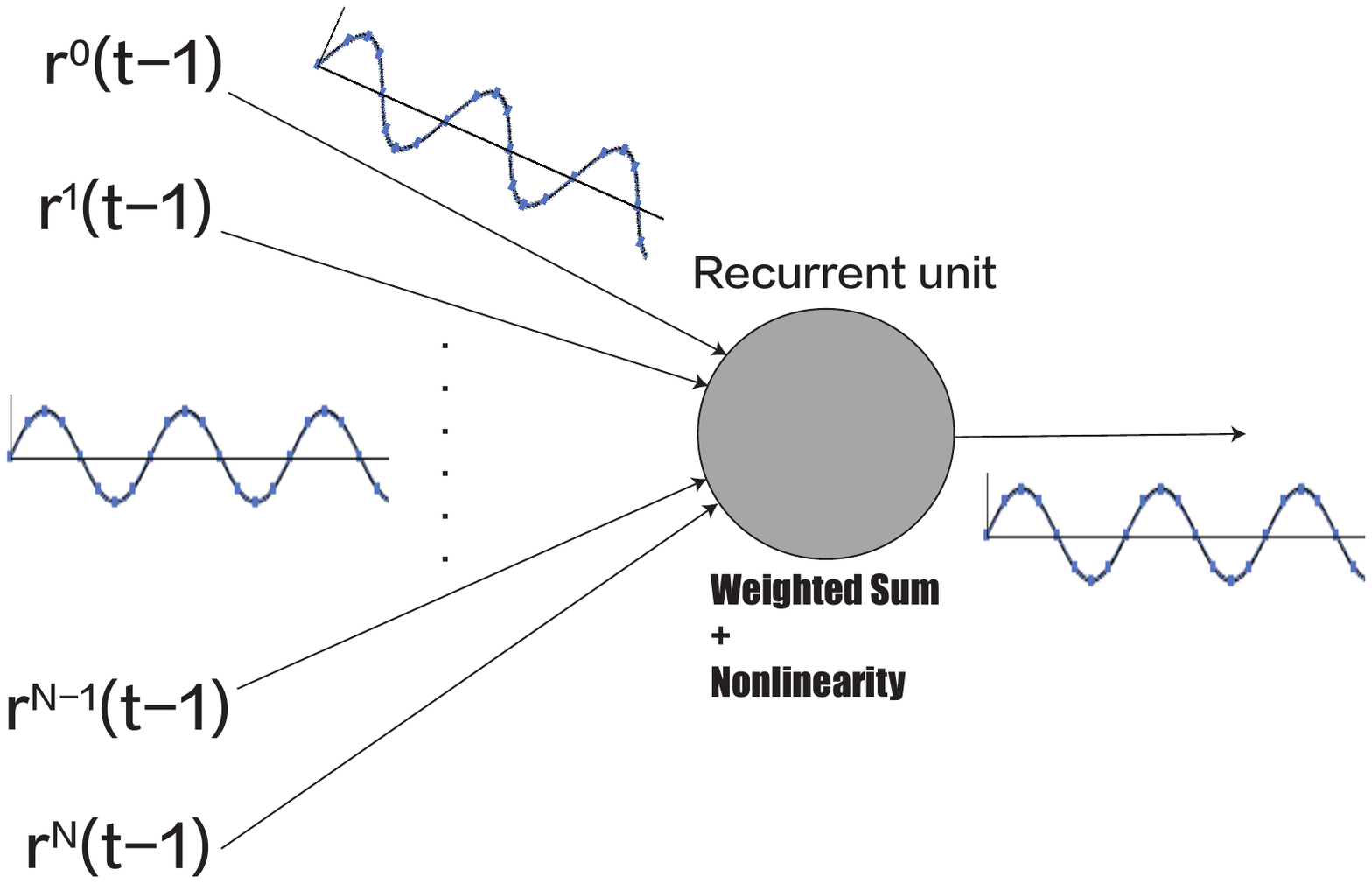}
	\caption{An illustration of the toy model. 
	The network receives $N$ identical input sequences, i.e., $r^1(t) = r^{2}(t)\ldots = r^{N}(t)$, where the sequence is denoted by
	$r(t) = \sin (t)$ where $t\in[0,21]$ in the simulation. The goal of the network is to output the same sine wave dynamics as the input.
	}
	\label{toymodel}
\end{figure}

In this section, we design a toy model in a simplest setting to see the nature of the ensemble learning rule. As shown in Fig.~\ref{toymodel}, the single output unit mimics its 
rhythmic inputs. The output unit can be thought of as a typical unit in a recurrent network.

The dynamics of this toy network can be described as
\begin{equation}
	\begin{aligned}
		h(t)& = (1-\alpha)h(t-1)+\frac{\alpha}{\sqrt{N}} \sum_{j}W_{j}r^{j}(t),\\
		y(t)& = f(h(t)),\\
		\mathcal{L} &= \frac{1}{2}\sum_t (y(t) - \hat{y}(t))^2,
	\end{aligned}
\end{equation}
where $r^j(t)$ indicates the dynamic input from node $j$, $\mathcal{L}$ is the training loss, $\hat{y}(t)$ denotes the target output at time $t$, and $f(\cdot)$ is the nonlinear ReLU function.
The gradient of the parameter $\mathbf{W}$ in a vanilla RNN training can be computed as follows,
\begin{equation}
	\begin{aligned}
		\Delta W_{j}& = \frac{1}{\sqrt{N}}(y(t)-\hat{y}(t))f^{\prime}(h(t))\alpha r^j(t).
	\end{aligned}
\end{equation}
In our ensemble training, the distribution of weights $\mathbf{W}$ and its statistics are given by
\begin{equation}
\begin{aligned}
	P(W_i) &= \pi_i \delta(W_i) + (1-\pi_i)\mathcal{N}(W_i|m_i,\Xi_i),\\
	\mu_i & = (1-\pi_i)m_i,\\
	\varrho_{i}&=\left(1-\pi_{i}\right)\left(\left(m_{i}\right)^{2}+\Xi_{i}\right).
\end{aligned}
\end{equation} 
Using the central-limit theorem, the dynamics can be recast as follows,
\begin{equation}
	\begin{aligned}
		h(t+1) &=(1-\alpha) h(t)+\alpha \left(G(t+1)+\epsilon(t+1)\sqrt{\Delta(t+1)}\right),\\ 
		y(t)& =\phi\left(h(t)\right), \\ 
	\end{aligned}
\end{equation}
where $G(t) = \frac{1}{\sqrt{N}}\sum_{j=1}^{N}\mu_j r^j(t)$ and $\Delta(t)  =  \frac{1}{{N}}\sum_{j=1}^{N}(\varrho_{j} - \mu^{2}_j) (r^j(t))^2$. 
Then the gradients of the three set of parameters $(\mathbf{m},\boldsymbol{\pi},\boldsymbol{\Xi})$ are calculated as
\begin{equation}
	\begin{aligned}
		\Delta m_i(t) & = \frac{\partial \mathcal{L}}{\partial m_i} = \frac{\partial \mathcal{L}}{\partial y(t)}\frac{\partial y(t)}{\partial m_i},\\
		& = ( y(t)-\hat{y}(t))\frac{\partial y(t)}{\partial h(t)}\frac{\partial h(t)}{\partial m_i},\\
		& = ( y(t)-\hat{y}(t))\phi^{\prime}(h(t))\times\alpha\left(\frac{\partial G}{\partial m_i}+\epsilon(t)\frac{\partial \sqrt{\Delta}}{\partial m_i}\right),\\
		& = ( y(t)-\hat{y}(t))\phi^{\prime}(h(t))\times\alpha\left(\frac{1}{\sqrt{N}}(1-\pi_i)r^i(t)+ \frac{\epsilon(t)}{N}\frac{\mu_i\pi_i(r^i(t))^2}{\sqrt{\Delta(t)}}\right),
	\end{aligned}
\end{equation}
and
\begin{equation}
	\begin{aligned}
		\Delta \pi_i(t) & = \frac{\partial \mathcal{L}}{\partial \pi_i} = \frac{\partial \mathcal{L}}{\partial y(t)}\frac{\partial y(t)}{\partial \pi_i},\\
		& = ( y(t)-\hat{y}(t))\frac{\partial y(t)}{\partial h(t)}\frac{\partial h(t)}{\partial \pi_i},\\
		& = ( y(t)-\hat{y}(t))\phi^{\prime}(h(t))\times\alpha\left(\frac{\partial G}{\partial \pi_i}+\epsilon(t)\frac{\partial \sqrt{\Delta}}{\partial \pi_i}\right),\\
		& = ( y(t)-\hat{y}(t))\phi^{\prime}(h(t))\times\alpha\left(-\frac{m_ir^i(t)}{\sqrt{N}}+ \frac{\epsilon(t)}{N}\frac{\left(m_i^2(1-2\pi_i)-\Xi_i\right)(r^i(t))^2}{2\sqrt{\Delta(t)}}\right),
	\end{aligned}
\end{equation}
and finally
\begin{equation}
	\begin{aligned}
		\Delta \Xi_i(t) & = \frac{\partial \mathcal{L}}{\partial \Xi_i} = \frac{\partial \mathcal{L}}{\partial y(t)}\frac{\partial y(t)}{\partial \Xi_i},\\
		& = ( y(t)-\hat{y}(t+1))\frac{\partial y(t)}{\partial h(t)}\frac{\partial h(t)}{\partial \Xi_i},\\
		& = (y(t)-\hat{y}(t))\phi^{\prime}(h(t))\times\alpha\left(\frac{\partial G}{\partial \Xi_i}+\epsilon(t)\frac{\partial \sqrt{\Delta}}{\partial \Xi_i}\right),\\
		& = ( y(t)-\hat{y}(t))\phi^{\prime}(h(t))\times\alpha\frac{\epsilon(t)}{N}\frac{(1-\pi_i)(r^i(t))^2}{2\sqrt{\Delta(t)}}.
	\end{aligned}
\end{equation}

The above learning rule for the toy model has a direct physics interpretation. The mean of the Gaussian distribution ($\mathbf{m}$) is updated by two terms: one is determined by the product of 
input and output activities with the learning rate regulated by two factors---nonlinearity of the transfer function and the spike mass; the other term is proportional to the stochastic noise capturing effects of 
weight uncertainty, which plays an important role in biological computation~\cite{Alex-2021}. We can see later the role of this term in the neural computation.
The spike mass update is also driven by a similar two-term form being a function of Gaussian mean $\mathbf{m}$ and variance $\mathbf{\Xi}$. The motion of the
Gaussian variance during learning has no drift term, and is purely driven by the stochastic noise. Hence, our ensemble learning rule is one form of stochastic plasticity, and our equation shows precisely 
how the local neural activity, connection probability and weight uncertainty affect the synaptic plasticity.

\begin{figure}
	\centering
	\includegraphics[bb=2 1 380 271,width=0.75\textwidth]{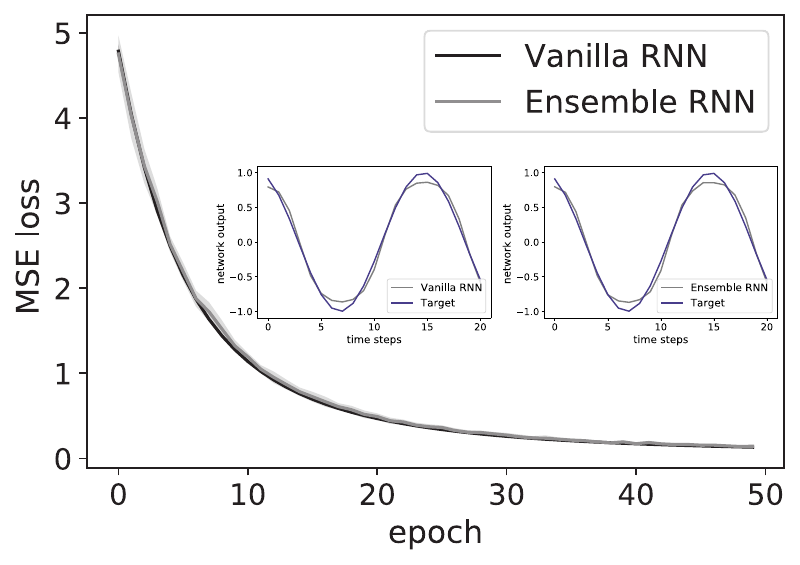}
	\caption{The MSE loss of the vanilla RNN with the parameter $\mathbf{W}$ and the ensemble RNN with parameters
	$(\mathbf{m},\boldsymbol{\pi},\boldsymbol{\Xi})$. We use $\alpha = 0.9$ and $N=10000$ for both training scenarios. 
	The fluctuations are computed over ten independent runs. The insets show the input-reconstruction performance.
	}
	\label{toyloss}
\end{figure}

\begin{figure}
	\centering
	\includegraphics[bb=5 2 820 270,width=0.85\textwidth]{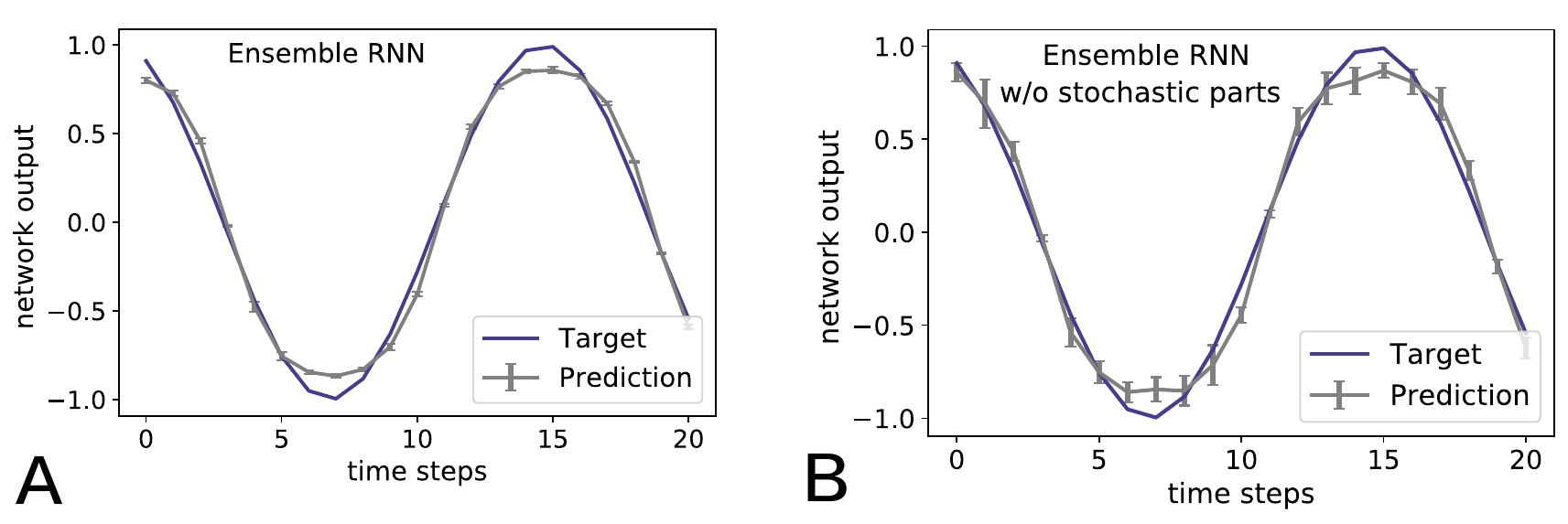}
	\caption{The reconstruction performance of the full ensemble training (A) and the ensemble training without the stochastic part (B).
	We use $N=100$ and $\alpha = 0.9$.  Each marker is an average over ten independent runs.}
	\label{plast}
\end{figure}

The ensemble training works in this simplest setting, with a decreasing training error (Fig.~\ref{toyloss}). The input can be reconstructed using the ensemble training, similar to performance of training directly the weights in
vanilla RNNs. The shape of the hyper-parameter distribution looks also similar to the more complex case of MNIST classification (data not shown here).
To verify the role of the second term in the hyper-parameter gradients, we remove the second term, i.e., no stochastic plasticity for the parameter $\mathbf{m}$
and $\boldsymbol{\pi}$, but keep the stochasticity for $\mathbf{\Xi}$. Note that the second term is a fluctuation term, smaller than the first drift term. The fluctuation term
is nevertheless significant in a finite size network.
We find that the reconstruction accuracy is sacrificed in the absence of this stochastic term (Fig.~\ref{plast}). In other words, the stochastic plasticity guarantees
the accuracy of the neural computation, which is quite important in motor control through the brain's cortical computation~\cite{Solla-2018,Suss-2009}.

\section{Conclusion}
In this study, we propose an ensemble perspective for understanding temporal credit assignment, 
which yields a generalized BPTT algorithm for training RNNs in practice. Our training protocol 
produces the hyper-parameters underlying the weight distribution, or the statistics of the RNN ensemble. 
In contrast to the standard BPTT, gBPTT highlights the importance of network statistics, which can particularly 
make a dynamic network changing its specified network weights a potential substrate for recurrent computation in 
adaptation to sensory inputs. It is thus interesting in future works to explore the biological counterpart 
of neural computation in brain circuits, e.g., in terms of dendritic spine dynamics~\cite{Kasai-2008}. 

Our SaS model has three types of hyper-parameters with distinct roles. $\bm{m}$ tells us the mean of
the continuous Gaussian slab, while $\bm{\Xi}$ controls the variance (fluctuation) of the slab. 
In both computational tasks we are interested in, we find that the $\Xi$-distribution profile is L-shaped. 
In other words, the peak at zero turns the SaS distribution into a Bernoulli distribution, while the 
tail at finite values of variance endows a computational flexibility to each connection, which may be critical to 
recode the high-dimensional sensory inputs into a low-dimensional latent dynamics.
It is thus interesting to establish this hypothesis, by addressing precisely how
the ensemble perspective helps to clarify the mechanism of low-dimensional latent 
dynamics encoding relevant spatio-temporal features of inputs. Our low-dimensional projections of synaptic and neural dynamics
supports this intuitive picture, and could inspire future theoretical and experimental studies of the connections between synaptic dynamics and neural dynamics,
particularly from a geometry viewpoint.

The spike probability $\bm{\pi}$ tells us the sparsity level of the network, which is a salient feature of the model offering a key to unlock the black box of the 
network function.
We find that a sparse RNN emerges after training. In particular, the recurrent
layer is sparser than the output layer, reflecting different roles of both layers.
The sparseness allows the recurrent layer to remove some unnecessary degrees of 
freedom in the recurrent dynamics, making recurrent feedbacks maintain only relevant
information for decision making. In contrast, the output layer is much denser, 
being thus able to read out relevant information both completely and robustly
(because most of $\Xi$ values are zero). It is worth noticing that the $\pi$-distribution profile 
implies the existence of VIP weights, which we show has a critical contribution to the overall 
performance of the network. Our method thus provides a practical way to identify the critical 
elements of the network topology, which could contribute to understanding temporal credit assignment
underlying the behavior of the RNN.

Working at the ensemble level, the constituent neurons of our model also display
neural selectivity of different nature in response to the task parameters. Some 
neurons show uni-selectivity, while the others show mixed selectivity. The selectivity 
property of neurons, shaped by the recurrent connections, is also a key factor impacting 
the behavioral output of the network, and thus deserves future studies particularly on the structure basis of the neural
selectivity~\cite{Sel-2020}.

Another promising direction is how to derive a biological learning rule inspired by our ensemble perspective, as already discussed in Sec.~\ref{stoc}.
In particular, the rule takes local neural activity, connection probability and weight uncertainty together to guide the global behavior of the system, whose
mechanism can be compared with neurophysiological experiments on the synapse level~\cite{Goto-2021}.

Our mechanistic analysis shows that the ensemble learning is divided into two stages: the first is the lazy regime, where the output dynamics can be predicted by the 
derived RNTK theory; the second is the active regime, where both synaptic and neural dynamics have low-dimensional manifold structure, and the symmetry breaking of hyper-parameters
drives the feature learning and exploration on the task manifold.

Taken together, our model and the associated theoretical and numerical analysis are able to provide insights 
toward a mechanistic understanding of temporal credit assignment, not only in engineering applications 
(e.g., MNIST digit classification), but also in modeling brain dynamics (e.g., multisensory integration task).


\begin{acknowledgments}
This research was supported by the National Natural Science Foundation of China for
Grant numbers 12122515 and 11805284.  
\end{acknowledgments}

\appendix
\section{Derivation of RNTK}\label{app-a}
In this section, we derive the RNTK formula. In essence,
computing the RNTK relies on calculating the corresponding Gaussian process (GP) kernels, which are 
the forward pass kernel $\Sigma^{\left(t, t^{\prime}\right)}\left(\boldsymbol{x}, \boldsymbol{x}^{\prime}\right) =\mathbb{E}_{\boldsymbol{\theta}}\left[h_i^{t}(\boldsymbol{x})h_i^{t^{\prime}}(\boldsymbol{x}^{\prime})\right]$ 
and backward pass kernel $\Pi^{\left(t, t^{\prime}\right)}\left(\boldsymbol{x}, \boldsymbol{x}^{\prime}\right)
= \mathbb{E}_{\boldsymbol{\theta}}\left[\delta_i^t(\boldsymbol{x}) \delta_i^{t^{\prime}}(\boldsymbol{x}^{\prime})\right]$ in our model.
The case of $\alpha\neq1$ is more involved than the case of $\alpha=1$. Our analysis shows that the case of $\alpha=1$ is a special example of the general $\alpha$ case.
We finally give two types of algorithms to compute the RNTK corresponding to these two cases, where the $\alpha=1$ case takes a much lower computation cost.

\subsection{Preprocessing the Fluctuation Term}
Due to the application of re-parameterization trick, the mean field dynamics includes fluctuation
terms like $\Delta_{i}^{t}(\boldsymbol{x})$ for recurrent units and $\Delta^{\mathrm{out}}(\boldsymbol{x})$ for the output unit, which leads to the
appearance of $\left(\Delta_{i}^{t}(\boldsymbol{x})\right)^2$ and $\left(\Delta^{\mathrm{out}}(\boldsymbol{x})\right)^2$ in the
GP kernels that seem difficult to handle. Thanks to the law of large number, we can use the mean of a random quantity to replace the quantity itself. 
Taking $\left(\Delta_{i}^{t}(\boldsymbol{x})\right)^2$ as an example, we have
\begin{equation}
	\left(\Delta_{i}^{t}(\boldsymbol{x})\right)^{2} =\sum_{j}\left(\tilde\varrho_{i j}^{\text {rec}}-\left(\tilde\mu_{i j}^{\text {rec}}\right)^{2}\right)\left(r_{j}^{t}(\boldsymbol{x})\right)^{2}.
\end{equation}
Note that the r.h.s is the sum of a large number of i.i.d. random variables. Therefore, when $N$ is large, 
we have $\left(\Delta_{i}^{t}(\boldsymbol{x})\right)^{2}\approx\mathbb{E}\left(\Delta_{i}^{t}(\boldsymbol{x})\right)^{2}$ where
\begin{equation}
	\begin{aligned}
		\mathbb{E}\left(\Delta_{i}^{t}(\boldsymbol{x})\right)^{2} &= \sum_{j}\mathbb{E}\left[\tilde\varrho_{i j}^{\text {rec}}-\left(\tilde\mu_{i j}^{\text {rec}}\right)^{2}\right]\mathbb{E}\left[\left(r_{j}^{t}(\boldsymbol{x})\right)^{2}\right]\\
		&= \frac{1}{2}\mathrm{F}_{\phi}\left[\boldsymbol{K}^{\left(t+1, t+1\right)}\left(\boldsymbol{x}, \boldsymbol{x}\right)\right],
	\end{aligned}
\end{equation}
where we apply the initialization scheme [Eq.~(\ref{init-scheme})]. 
$\left(\Delta^{\mathrm{out}}(\boldsymbol{x})\right)^2$ is approximated in the same way. We thus obtain the following results as
\begin{equation}
	\begin{aligned}		
		\left(\Delta_{i}^{t}(\boldsymbol{x})\right)^{2}&\approx\mathbb{E}\left[\left(\Delta_{i}^{t}(\boldsymbol{x})\right)^{2}\right] = \frac{1}{2}\mathrm{F}_{\phi}\left[\boldsymbol{K}^{\left(t+1, t+1\right)}\left(\boldsymbol{x}, \boldsymbol{x}\right)\right],\\
		\left(\Delta^{\mathrm{out}}(\boldsymbol{x})\right)^2&\approx\mathbb{E}\left[\left(\Delta^{\mathrm{out}}(\boldsymbol{x})\right)^{2}\right] = \frac{1}{2}\mathrm{F}_{\phi}\left[\boldsymbol{K}^{\left(T+1, T+1\right)}\left(\boldsymbol{x}, \boldsymbol{x}\right)\right].	
	\end{aligned}
\end{equation}

\subsection{Forward Pass Kernel $\Sigma^{\left(t, t^{\prime}\right)}\left(\boldsymbol{x}, \boldsymbol{x}^{\prime}\right)$}

We first define an auxiliary kernel $\Omega^{\left(t, t^{\prime}\right)}\left(\boldsymbol{x}, \boldsymbol{x}^{\prime}\right) = \mathbb{E}\left[u_i^{t}(\boldsymbol{x})u_i^{t^{\prime}}(\boldsymbol{x}^{\prime})\right]$ to make the recursive process more clearly,
\begin{equation}
	\begin{aligned}\label{OmegafromSigma}
		&\Omega^{\left(t, t^{\prime}\right)}\left(\boldsymbol{x}, \boldsymbol{x}^{\prime}\right) =\mathbb{E}\left[u_i^{t}(\boldsymbol{x})u_i^{t^{\prime}}(\boldsymbol{x}^{\prime})\right]\\
		&=\mathbb{E}\left[\left(\frac{\sigma_{\mathrm{rec}}}{\sqrt{N}}\sum_{j=1}^N(1-\tilde\pi_{ij}^{\mathrm{rec}})^2 m_{ij}^{\mathrm{rec}} r_j^{t-1}(\boldsymbol{x}) +\frac{\sigma_{\mathrm{in}}}{\sqrt{N_{\mathrm{in}}}}\sum_{j=1}^{\nin} m_{ij}^{\mathrm{in}}x_{t,j} + \epsilon_i^t(\boldsymbol{x})\Delta_i^{t-1}(\boldsymbol{x})\right)\right.\\
		&\left.\qquad\left(\frac{\sigma_{\mathrm{rec}}}{\sqrt{N}}\sum_{j=1}^N(1-\tilde\pi_{ij}^{\mathrm{rec}})^2 m_{ij}^{\mathrm{rec}}r_j^{t^{\prime}-1}(\boldsymbol{x}^{\prime}) +\frac{\sigma_{\mathrm{in}}}{\sqrt{N_{\mathrm{in}}}}\sum_{j=1}^{\nin} m_{ij}^{\mathrm{in}}x^{\prime}_{t^{\prime},j}+ \epsilon_i^{t^{\prime}}(\boldsymbol{x}^{\prime})\Delta_i^{t^{\prime}-1}(\boldsymbol{x}^{\prime}) \right)  \right] \\
		&=\frac{\sigma_{\mathrm{rec}}^2}{N}\sum_{j=1}^N\mathbb{E}\left[(1-\tilde\pi_{ij}^{\mathrm{rec}})^2	\right]\mathbb{E}\left[(m_{ij}^{\mathrm{rec}})^2\right]\mathbb{E}\left[r_j^{t-1}(\boldsymbol{x})r_j^{t^{\prime}-1}(\boldsymbol{x}^{\prime})\right] + \frac{\sigma_{\mathrm{in}}^2}{N_{\mathrm{in}}}\sum_{j=1}^{\nin}\mathbb{E}\left[(m_{ij}^{\mathrm{in}})^2\right]x_{t,j} x^{\prime}_{t^{\prime},j} \\
		&\qquad+\mathbb{E}\left[\epsilon_i^t(\boldsymbol{x})\epsilon_i^{t^{\prime}}(\boldsymbol{x}^{\prime})\right]\mathbb{E}\left[\Delta_i^{t-1}(\boldsymbol{x})\Delta_i^{t^{\prime}-1}(\boldsymbol{x}^{\prime})\right]\\
		&=\left(\sigma_{\mathrm{rec}}^{2}+\frac{1}{2}\delta_{\boldsymbol{x}=\boldsymbol{x}^{\prime}}\delta_{t=t^{\prime}}\right) \mathrm{F}_{\phi}\left[\boldsymbol{K}^{\left(t, t^{\prime}\right)}\left(\boldsymbol{x}, \boldsymbol{x}^{\prime}\right)\right]+\frac{\sigma_{\mathrm{in}}^{2}}{N_{\mathrm{in}}}\left\langle\boldsymbol{x}_{t}, \boldsymbol{x}^{\prime}_{t^{\prime}}\right\rangle,
	\end{aligned}
\end{equation}
where $\boldsymbol{K}^{\left(t, t^{\prime}\right)}\left(\boldsymbol{x}, \boldsymbol{x}^{\prime}\right)$ is the covariance matrix related to the forward pass kernel at a previous time step,
\begin{equation}\label{covar}
	\boldsymbol{K}^{\left(t, t^{\prime}\right)}\left(\boldsymbol{x}, \boldsymbol{x}^{\prime}\right)=\left[\begin{array}{cc}
		\Sigma^{(t-1, t-1)}(\boldsymbol{x}, \boldsymbol{x}) & \Sigma^{\left(t-1, t^{\prime}-1\right)}\left(\boldsymbol{x}, \boldsymbol{x}^{\prime}\right) \\
		\Sigma^{\left(t-1, t^{\prime}-1\right)}\left(\boldsymbol{x}, \boldsymbol{x}^{\prime}\right) & \Sigma^{\left(t^{\prime}-1, t^{\prime}-1\right)}\left(\boldsymbol{x}^{\prime}, \boldsymbol{x}^{\prime}\right)
	\end{array}\right].
\end{equation}
If $t\neq0$ and $t^{\prime}\neq0$, the forward pass kernel at the current time step can be written as 
\begin{equation}
	\begin{aligned}\label{SigmafromOmega}
		&\Sigma^{\left(t, t^{\prime}\right)}\left(\boldsymbol{x}, \boldsymbol{x}^{\prime}\right)
		= \mathbb{E}\left[h_i^t(\boldsymbol{x}) h_i^{t^{\prime}}(\boldsymbol{x}^{\prime})\right]\\
		&= (1-\alpha)^2\mathbb{E}\left[ h_i^{t-1}(\boldsymbol{x}) h_i^{t^{\prime}-1}(\boldsymbol{x}^{\prime})\right] + \alpha^2\mathbb{E}\left[u_i^t(\boldsymbol{x}) u_i^{t^{\prime}}(\boldsymbol{x}^{\prime})\right] \\
		& \qquad + (1-\alpha)\alpha\mathbb{E}\left[u_i^t(\boldsymbol{x})h_i^{t^{\prime}-1}(\boldsymbol{x}^{\prime})\right] + (1-\alpha)\alpha\mathbb{E}\left[u_i^{t^{\prime}}(\boldsymbol{x}^{\prime})h_i^{t-1}(\boldsymbol{x})\right]\\
		&= (1-\alpha)^2\mathbb{E}\left[ h_i^{t-1}(\boldsymbol{x}) h_i^{t^{\prime}-1}(\boldsymbol{x}^{\prime})\right] + \alpha^2\mathbb{E}\left[u_i^t(\boldsymbol{x}) u_i^{t^{\prime}}(\boldsymbol{x}^{\prime})\right] \\
		& \qquad + \sum_{\Delta t^{\prime} = 1}^{t^{\prime}-1} (1-\alpha)^{\Delta t^{\prime}} \alpha^2 \mathbb{E}\left[u_i^t(\boldsymbol{x})u_i^{t^{\prime}-\Delta t^{\prime}}(\boldsymbol{x}^{\prime})\right]  +  \sum_{\Delta t = 1}^{t-1} (1-\alpha)^{\Delta t} \alpha^2 \mathbb{E}\left[u_i^{t^{\prime}}(\boldsymbol{x}^{\prime})u_i^{t-\Delta t}(\boldsymbol{x})\right] \\
		&= (1-\alpha)^2\Sigma^{\left(t-1, t^{\prime}-1\right)}\left(\boldsymbol{x}, \boldsymbol{x}^{\prime}\right) + \alpha^2\Omega^{\left(t, t^{\prime}\right)}\left(\boldsymbol{x}, \boldsymbol{x}^{\prime}\right) \\
		& \qquad + \sum_{\Delta t^{\prime} = 1}^{t^{\prime}-1} (1-\alpha)^{\Delta t^{\prime}} \alpha^2 \Omega^{\left(t, t^{\prime}-\Delta t^{\prime}\right)}\left(\boldsymbol{x}, \boldsymbol{x}^{\prime}\right)  +  \sum_{\Delta t = 1}^{t-1} (1-\alpha)^{\Delta t} \alpha^2 \Omega^{\left(t-\Delta t, t^{\prime}\right)}\left(\boldsymbol{x}, \boldsymbol{x}^{\prime}\right). \\
	\end{aligned}
\end{equation}
To derive the last equality in Eq.~(\ref{SigmafromOmega}), we expand $h_{i}^{t}(\boldsymbol{x})$ as
\begin{equation}\label{h-expand}
	\begin{aligned}
		h_{i}^{t}(\boldsymbol{x})
		&= \alpha u_{i}^{t}(\boldsymbol{x}) + (1-\alpha)h_{i}^{t-1}(\boldsymbol{x})\\
		&= \alpha u_{i}^{t}(\boldsymbol{x}) + (1-\alpha)\alpha u_{i}^{t-1}(\boldsymbol{x}) + (1-\alpha)^2 h_{i}^{t-2}(\boldsymbol{x})\\
		&= \sum_{\Delta t = 0}^{t-1} (1-\alpha)^{\Delta t}\alpha u_{i}^{t-\Delta t}(\boldsymbol{x}) + (1-\alpha)^{t} h_{i}^{0}(\boldsymbol{x}),
	\end{aligned}
\end{equation}
and we also use the fact that $\mathbb{E}\left[u_{i}^{t}(\boldsymbol{x}) h_{i}^{0}(\boldsymbol{x})\right] = 0,~t=1,2,\ldots,T$. 
Thus, we obtain the recursive formula of $\Sigma^{\left(t, t^{\prime}\right)}\left(\boldsymbol{x}, \boldsymbol{x}^{\prime}\right)$
by introducing an auxiliary kernel $\Omega^{\left(t, t^{\prime}\right)}\left(\boldsymbol{x}, \boldsymbol{x}^{\prime}\right)$. Next,
we consider the initial time step ($t=0$) to complete the derivation. According to the expansion of $h_{i}^{t}(\boldsymbol{x})$ [Eq.~(\ref{h-expand}))], we have
\begin{equation}
	\begin{aligned}\label{Sigma0}
		\Sigma^{\left(0, 0\right)}\left(\boldsymbol{x}, \boldsymbol{x}^{\prime}\right) &= \mathbb{E}\left[h_i^0(\boldsymbol{x}) h_i^{0}(\boldsymbol{x}^{\prime})\right] = \delta_{\boldsymbol{x}=\boldsymbol{x^{\prime}}}\sigma_h^2,\\
		\Sigma^{\left(0, t^{\prime}\right)}\left(\boldsymbol{x}, \boldsymbol{x}^{\prime}\right) &= \mathbb{E}\left[h_i^0(\boldsymbol{x}) h_i^{t^{\prime}}(\boldsymbol{x}^{\prime})\right] = (1-\alpha)^{t^{\prime}} \delta_{\boldsymbol{x}=\boldsymbol{x^{\prime}}}\sigma_h^2,\\
		\Sigma^{\left(t, 0\right)}\left(\boldsymbol{x}, \boldsymbol{x}^{\prime}\right) & = \mathbb{E}\left[h_i^t(\boldsymbol{x}) h_i^{0}(\boldsymbol{x}^{\prime})\right] = (1-\alpha)^{t} \delta_{\boldsymbol{x}=\boldsymbol{x^{\prime}}}\sigma_h^2.
	\end{aligned}
\end{equation}

\subsection{Backward Pass Kernel $\Pi^{\left(t, t^{\prime}\right)}\left(\boldsymbol{x}, \boldsymbol{x}^{\prime}\right)$}
We first define the backpropagation error $\delta_i^t(\boldsymbol{x}) = \sqrt{N}\frac{\partial f(\boldsymbol{x})}{\partial h_i^t(\boldsymbol{x})}$. We then 
compute the error at the last time step and intermediate step separately,
\begin{equation}
	\begin{aligned}
		\delta_i^T(\boldsymbol{x}) 
		&= \sqrt{N}\frac{\partial f(\boldsymbol{x})}{\partial h_i^T(\boldsymbol{x})}\\
		&= \sqrt{N}\frac{\partial f(\boldsymbol{x})}{\partial r_i^T(\boldsymbol{x})}\frac{\partial r_i^T(\boldsymbol{x})}{\partial h_i^T(\boldsymbol{x})}\\
		&= \sqrt{N}\phi^{\prime}( h_i^T(\boldsymbol{x})) \left[\tilde{\mu}_{i}^{\mathrm{out}} + \epsilon^{\mathrm{out}}(\boldsymbol{x}) \frac{\left(\tilde{\varrho}_{i}^{\text {out}}-\left(\tilde{\mu}_{i}^{\text {out }}\right)^{2}\right) r_{i}^{T}(\boldsymbol{x})}{\Delta^{\mathrm{out}}(\boldsymbol{x})}  \right],\\
		\delta_i^t(\boldsymbol{x})
		&= \sqrt{N}\frac{\partial f(\boldsymbol{x})}{\partial h_i^t(\boldsymbol{x})}\\
		&= \sqrt{N}\sum_j \frac{\partial f(\boldsymbol{x})}{\partial h_j^{t+1}(\boldsymbol{x})} \frac{\partial h_j^{t+1}(\boldsymbol{x})}{\partial h_i^t(\boldsymbol{x})}\\
		&= (1-\alpha)\delta_i^{t+1}(\boldsymbol{x}) + \alpha\phi^{\prime}(h_i^{t}(\boldsymbol{x}))\sum_{j}\left[\tilde{\mu}_{ji}^{\mathrm{rec}} + \epsilon_{j}^{t}(\boldsymbol{x}) \frac{\left(\tilde{\varrho}_{ji}^{\text {rec }}-\left(\tilde{\mu}_{ji}^{\text {rec }}\right)^{2}\right) r_{i}^{t}(\boldsymbol{x})}{\Delta_{j}^{t}(\boldsymbol{x})} \right] \delta_j^{t+1}(\boldsymbol{x})\\
		&= (1-\alpha)\delta_i^{t+1}(\boldsymbol{x}) + \alpha v_i^t(\boldsymbol{x}),
	\end{aligned}
\end{equation}
where we define a new variable $v_i^t(\boldsymbol{x})$ and introduce an auxiliary kernel
$\Gamma^{t,t^{\prime}}(\boldsymbol{x},\boldsymbol{x}^{\prime}) = \mathbb{E}\left[ v_i^t(\boldsymbol{x}) v_i^{t^{\prime}}(\boldsymbol{x}^{\prime}) \right]$, which is computed as follows,
\begin{equation}
	\begin{aligned}\label{GammafromPi}
		&\Gamma^{\left(t, t^{\prime}\right)}\left(\boldsymbol{x}, \boldsymbol{x}^{\prime}\right) 
		=\mathbb{E}\left[v_i^{t}(\boldsymbol{x}) v_i^{t^{\prime}}(\boldsymbol{x}^{\prime})\right] \\
		&=\mathbb{E}\left[\phi^{\prime}(h^t_i(\boldsymbol{x}))\phi^{\prime}(h^{t^{\prime}}_i(\boldsymbol{x}^{\prime}))\right]\sum_j\Biggl(\frac{\sigma_{\mathrm{rec}}^{2}}{N}\mathbb{E}\left[(1-\tilde\pi_{ji}^{\mathrm{rec}})^2\right]\mathbb{E}\left[(m_{ji}^{\mathrm{rec}})^2 \right]\\
		&\qquad+ \mathbb{E}\left[\epsilon_{j}^{t}(\boldsymbol{x})\epsilon_{j}^{t^{\prime}}(\boldsymbol{x}^{\prime})\right]\frac{\mathbb{E}\left[\left(\tilde{\varrho}_{ji}^{\text {rec }}-\left(\tilde{\mu}_{ji}^{\text {rec }}\right)^{2}\right)^2\right]\mathbb{E}\left[r_{i}^{t}(\boldsymbol{x})r_{i}^{t^{\prime}}(\boldsymbol{x}^{\prime})\right]}{\mathbb{E}\left[\Delta_{j}^{t}(\boldsymbol{x})\Delta_{j}^{t^{\prime}}(\boldsymbol{x}^{\prime})\right]}\Biggl)\mathbb{E}\left[\delta_j^{t+1}(\boldsymbol{x}) \delta_j^{t^{\prime}+1}(\boldsymbol{x}^{\prime})\right] \\
		&=\mathrm{F}_{\phi^{\prime}}\left[\boldsymbol{K}^{\left(t+1, t^{\prime}+1\right)}\left(\boldsymbol{x}, \boldsymbol{x}^{\prime}\right)\right]\sum_j\left(\frac{\sigma_{\mathrm{rec}}^{2}}{N}+\delta_{\boldsymbol{x}=\boldsymbol{x}^{\prime}}\delta_{t=t^{\prime}} \frac{\frac{1}{3N^2}\mathrm{F}_{\phi}\left[\boldsymbol{K}^{\left(t, t\right)}\left(\boldsymbol{x}, \boldsymbol{x}\right)\right]}{\frac{1}{2}\mathrm{F}_{\phi}\left[\boldsymbol{K}^{\left(t, t\right)}\left(\boldsymbol{x}, \boldsymbol{x}\right)\right]}\right)\Pi^{\left(t+1, t^{\prime}+1\right)}\left(\boldsymbol{x}, \boldsymbol{x}^{\prime}\right) \\
		&=\left(\sigma_{\mathrm{rec}}^{2}+\frac{2}{3N}\delta_{\boldsymbol{x}=\boldsymbol{x}^{\prime}}\delta_{t=t^{\prime}}\right)\mathrm{F}_{\phi^{\prime}}\left[\boldsymbol{K}^{\left(t+1, t^{\prime}+1\right)}\left(\boldsymbol{x}, \boldsymbol{x}^{\prime}\right)\right]\Pi^{\left(t+1, t^{\prime}+1\right)}\left(\boldsymbol{x}, \boldsymbol{x}^{\prime}\right),
	\end{aligned}
\end{equation}
which relies on the backward pass kernel of the next time step $\Pi^{\left(t+1, t^{\prime}+1\right)}\left(\boldsymbol{x}, \boldsymbol{x}^{\prime}\right)$.

Similarly, if $t\neq T$ and $t^{\prime}\neq T$, the backward pass kernel at the current time step can also be expanded as
\begin{equation}
	\begin{aligned}\label{PifromGamma}
		\Pi^{\left(t, t^{\prime}\right)}\left(\boldsymbol{x}, \boldsymbol{x}^{\prime}\right)
		&= \mathbb{E}\left[\delta_i^t(\boldsymbol{x}) \delta_i^{t^{\prime}}(\boldsymbol{x}^{\prime})\right]\\
		&= (1-\alpha)^2\mathbb{E}\left[ \delta_i^{t+1}(\boldsymbol{x}) \delta_i^{t^{\prime}+1}(\boldsymbol{x}^{\prime})\right] + \alpha^2\mathbb{E}\left[v_i^t(\boldsymbol{x}) v_i^{t^{\prime}}(\boldsymbol{x}^{\prime})\right] \\
		& \quad + (1-\alpha)\alpha\mathbb{E}\left[v_i^t(\boldsymbol{x})\delta_i^{t^{\prime}+1}(\boldsymbol{x}^{\prime})\right] + (1-\alpha)\alpha\mathbb{E}\left[v_i^{t^{\prime}}(\boldsymbol{x}^{\prime})\delta_i^{t+1}(\boldsymbol{x})\right]\\
		&= (1-\alpha)^2\mathbb{E}\left[ \delta_i^{t+1}(\boldsymbol{x}) \delta_i^{t^{\prime}+1}(\boldsymbol{x}^{\prime})\right] + \alpha^2\mathbb{E}\left[v_i^t(\boldsymbol{x}) v_i^{t^{\prime}}(\boldsymbol{x}^{\prime})\right] \\
		& \quad + \sum_{\Delta t^{\prime} = 1}^{T-t^{\prime}-1} (1-\alpha)^{\Delta t^{\prime}} \alpha^2 \mathbb{E}\left[v_i^t(\boldsymbol{x})v_i^{t^{\prime}+\Delta t^{\prime}}(\boldsymbol{x}^{\prime})\right]  +  \sum_{\Delta t = 1}^{T-t+1} (1-\alpha)^{\Delta t} \alpha^2 \mathbb{E}\left[v_i^{t^{\prime}}(\boldsymbol{x}^{\prime})v_i^{t+\Delta t}(\boldsymbol{x})\right] \\
		&= (1-\alpha)^2\Pi^{\left(t+1, t^{\prime}+1\right)}\left(\boldsymbol{x}, \boldsymbol{x}^{\prime}\right) + \alpha^2\Gamma^{\left(t, t^{\prime}\right)}\left(\boldsymbol{x}, \boldsymbol{x}^{\prime}\right) \\
		& \quad + \sum_{\Delta t^{\prime} = 1}^{T-t^{\prime}-1} (1-\alpha)^{\Delta t^{\prime}} \alpha^2 \Gamma^{\left(t, t^{\prime}+\Delta t^{\prime}\right)}\left(\boldsymbol{x}, \boldsymbol{x}^{\prime}\right)  +  \sum_{\Delta t = 1}^{T-t-1} (1-\alpha)^{\Delta t} \alpha^2 \Gamma^{\left(t+\Delta t, t^{\prime}\right)}\left(\boldsymbol{x}, \boldsymbol{x}^{\prime}\right), \\
	\end{aligned}
\end{equation}
where to derive the last equality, we expand $\delta_{i}^{t}(\boldsymbol{x})$ as
	\begin{equation}\label{delta-expand}
	\begin{aligned}
		\delta_{i}^{t}(\boldsymbol{x})
		&= \alpha v_{i}^{t}(\boldsymbol{x}) + (1-\alpha)\delta_{i}^{t+1}(\boldsymbol{x})\\
		&= \alpha v_{i}^{t}(\boldsymbol{x}) + (1-\alpha)\alpha v_{i}^{t+1}(\boldsymbol{x}) + (1-\alpha)^2 \delta_{i}^{t+2}(\boldsymbol{x})\\
		&= \sum_{\Delta t = 0}^{T-t-1} (1-\alpha)^{\Delta t}\alpha v_{i}^{t+\Delta t}(\boldsymbol{x}) + (1-\alpha)^{T-t} \delta_{i}^{T}(\boldsymbol{x}),
	\end{aligned}
\end{equation}
 and we also use the fact that $\mathbb{E}\left[v_{i}^{t}(\boldsymbol{x}) \delta_{i}^{T}(\boldsymbol{x})\right] = 0,~t=1,2,\ldots,T$.
After we obtain the recursive formula of $\Pi^{\left(t, t^{\prime}\right)}\left(\boldsymbol{x}, \boldsymbol{x}^{\prime}\right)$ by 
introducing the auxiliary kernel $\Gamma^{\left(t, t^{\prime}\right)}\left(\boldsymbol{x}, \boldsymbol{x}^{\prime}\right)$, 
we can consider the last time step ($t=T$) to complete the derivation.  When $t=t^{\prime}=T$, we have
\begin{equation}
	\begin{aligned}\label{PiTT}
		&\Pi^{\left(T, T\right)}\left(\boldsymbol{x}, \boldsymbol{x}^{\prime}\right) 
		=\mathbb{E}\left[\delta_i^{T}(\boldsymbol{x}) \delta_i^{T}(\boldsymbol{x}^{\prime})\right] \\
		&=\mathbb{E}\left[\phi^{\prime}(h^T_i(\boldsymbol{x}))\phi^{\prime}(h^{T}_i(\boldsymbol{x}^{\prime}))\right]\cdot N\Biggl(\mathbb{E}\left[(1-\tilde\pi_{i}^{\mathrm{out}})^2\right]\mathbb{E}\left[(\tilde{m}_{i}^{\mathrm{out}})^2\right]\\
		&\qquad +\epsilon^{\mathrm{out}}(\boldsymbol{x})\epsilon^{\mathrm{out}}(\boldsymbol{x}^{\prime})\frac{\mathbb{E}\left[\left(\tilde\varrho_{i}^{\text {out }}-\left(\tilde\mu_{i}^{\text {out }}\right)^{2}\right)^2\right]\mathbb{E}\left[ r_{i}^{T}(\boldsymbol{x}) r_{i}^{T}(\boldsymbol{x}^{\prime}) \right] }{\mathbb{E}\left[\Delta^{\text {out}}(\boldsymbol{x})\Delta^{\text {out}}(\boldsymbol{x}^{\prime})\right]}\Biggl)\\
		&=\mathrm{F}_{\phi^{\prime}}\left[\boldsymbol{K}^{\left(T+1, T+1\right)}\left(\boldsymbol{x}, \boldsymbol{x}'\right)\right] N\left(\frac{\sigma_{\mathrm{out}}^{2}}{N}+\epsilon^{\mathrm{out}}(\boldsymbol{x})\epsilon^{\mathrm{out}}(\boldsymbol{x}^{\prime})\frac{\frac{1}{3N^2}\mathrm{F}_{\phi}\left[\boldsymbol{K}^{\left(T+1, T+1\right)}\left(\boldsymbol{x}, \boldsymbol{x}^{\prime}\right)\right]}{\frac{1}{2}\sqrt{\mathrm{F}_{\phi}\left[\boldsymbol{K}^{\left(T+1, T+1\right)}\left(\boldsymbol{x}, \boldsymbol{x}\right)\right]\mathrm{F}_{\phi}\left[\boldsymbol{K}^{\left(T+1, T+1\right)}\left(\boldsymbol{x}^{\prime}, \boldsymbol{x}^{\prime}\right)\right]} }\right)\\
		&=\left(\sigma_{\mathrm{out}}^{2}+\frac{2}{3N}\epsilon^{\mathrm{out}}(\boldsymbol{x})\epsilon^{\mathrm{out}}(\boldsymbol{x}^{\prime})\frac{\mathrm{F}_{\phi}\left[\boldsymbol{K}^{\left(T+1, T+1\right)}\left(\boldsymbol{x}, \boldsymbol{x}^{\prime}\right)\right]}{\sqrt{\mathrm{F}_{\phi}\left[\boldsymbol{K}^{\left(T+1, T+1\right)}\left(\boldsymbol{x}, \boldsymbol{x}\right)\right]\mathrm{F}_{\phi}\left[\boldsymbol{K}^{\left(T+1, T+1\right)}\left(\boldsymbol{x}^{\prime}, \boldsymbol{x}^{\prime}\right)\right]} }\right) \mathrm{F}_{\phi^{\prime}}\left[\boldsymbol{K}^{\left(T+1, T+1\right)}\left(\boldsymbol{x}, \boldsymbol{x}'\right)\right] .
	\end{aligned}
\end{equation}
Note that $\epsilon^{\mathrm{out}}(\boldsymbol{x})$ is kept.
When $t$ and $t^{\prime}$ are not both equal to $T$, according to the expansion of $\delta_{i}^{t}(\boldsymbol{x})$ [Eq.~(\ref{delta-expand})], we have
\begin{equation}\label{PiTt}
	\begin{aligned}
		\Pi^{\left(T, t^{\prime}\right)}\left(\boldsymbol{x}, \boldsymbol{x}^{\prime}\right) &=\mathbb{E}\left[\delta_i^{T}(\boldsymbol{x}) \delta_i^{t^{\prime}}(\boldsymbol{x}^{\prime})\right] =(1-\alpha)^{T-t^{\prime}}\Pi^{\left(T, T\right)}\left(\boldsymbol{x}, \boldsymbol{x}^{\prime}\right),\\
		\Pi^{\left(t, T\right)}\left(\boldsymbol{x}, \boldsymbol{x}^{\prime}\right) &=\mathbb{E}\left[\delta_i^{t}(\boldsymbol{x}) \delta_i^{T}(\boldsymbol{x}^{\prime})\right] =(1-\alpha)^{T-t}\Pi^{\left(T, T\right)}\left(\boldsymbol{x}, \boldsymbol{x}^{\prime}\right).
	\end{aligned}
\end{equation}

\subsection{Computation of the RNTK $\Theta(\boldsymbol{x},\boldsymbol{x}^{\prime})$}
Now we can derive the RNTK $\Theta(\boldsymbol{x},\boldsymbol{x}^{\prime})$ based on the results of GP kernels. 
Trainable parameters of our model are $\boldsymbol\theta=\{\boldsymbol{m}^{\mathrm{rec}},\boldsymbol{m}^{\mathrm{in}},\boldsymbol{m}^{\mathrm{out}},\boldsymbol{\pi}^{\mathrm{rec}},\boldsymbol{\Xi}^{\mathrm{rec}},\boldsymbol{\pi}^{\mathrm{out}},\boldsymbol{\Xi}^{\mathrm{out}}\}$. 
The RNTK $\Theta(\boldsymbol{x},\boldsymbol{x}^{\prime})$ is defined as
\begin{equation}
	\begin{aligned}
		\Theta\left(\boldsymbol{x}, \boldsymbol{x}^{\prime}\right) &= \sum_{p=1}^{|\boldsymbol{\theta}|} \frac{\partial f(\boldsymbol{x})}{\partial \theta_p} \frac{\partial f(\boldsymbol{x}^{\prime})}{\partial \theta_p}\\
		& = \sum_{ij} \frac{\partial f(\boldsymbol{x})}{\partial m_{ij}^{\mathrm{rec}}} \frac{\partial f(\boldsymbol{x}^{\prime})}{\partial m_{ij}^{\mathrm{rec}}} + \sum_{ij} \frac{\partial f(\boldsymbol{x})}{\partial m_{ij}^{\mathrm{in}}} \frac{\partial f(\boldsymbol{x}^{\prime})}{\partial m_{ij}^{\mathrm{in}}} + \sum_{i} \frac{\partial f(\boldsymbol{x})}{\partial m_{i}^{\mathrm{out}}} \frac{\partial f(\boldsymbol{x}^{\prime})}{\partial m_{i}^{\mathrm{out}}}\\
		& \quad+ \sum_{ij} \frac{\partial f(\boldsymbol{x})}{\partial \pi_{ij}^{\mathrm{rec}}} \frac{\partial f(\boldsymbol{x}^{\prime})}{\partial \pi_{ij}^{\mathrm{rec}}}  + \sum_{ij} \frac{\partial f(\boldsymbol{x})}{\partial \Xi_{ij}^{\mathrm{rec}}} \frac{\partial f(\boldsymbol{x}^{\prime})}{\partial \Xi_{ij}^{\mathrm{rec}}}\\
		& \quad+ \sum_{i} \frac{\partial f(\boldsymbol{x})}{\partial \pi_{i}^{\mathrm{out}}} \frac{\partial f(\boldsymbol{x}^{\prime})}{\partial \pi_{i}^{\mathrm{out}}}  + \sum_{i} \frac{\partial f(\boldsymbol{x})}{\partial \Xi_{i}^{\mathrm{out}}} \frac{\partial f(\boldsymbol{x}^{\prime})}{\partial \Xi_{i}^{\mathrm{out}}}.
	\end{aligned}
\end{equation}

For the recurrent and input layers, the gradient of $f(\boldsymbol{x})$ with respect to
$\theta_{ij}\equiv\{\theta_{ij}^{\mathrm{in}},\theta_{ij}^{\mathrm{rec}}\}$ is given by
\begin{equation}
	\frac{\partial f(\boldsymbol{x})}{\partial \theta_{ij}} = \sum_{t=1}^{T} \frac{\partial f(\boldsymbol{x})}{\partial h_i^t(\boldsymbol{x})}\frac{\partial h_i^t(\boldsymbol{x})}{\partial u_i^t(\boldsymbol{x})}\frac{\partial u_i^t(\boldsymbol{x})}{\partial \tilde{\theta}_{ij}}\frac{\partial \tilde{\theta}_{ij}}{\partial \theta_{ij}} = \frac{\alpha}{\sqrt{N}}\frac{\partial \tilde{\theta}_{ij}}{\partial \theta_{ij}} \sum_{t=1}^{T} \delta_i^t(\boldsymbol{x}) \frac{\partial u_i^t(\boldsymbol{x})}{\partial \tilde{\theta}_{ij}},
\end{equation}
where $\frac{\partial u_i^t(\boldsymbol{x})}{\partial \tilde{\theta}_{ij}}$ can be explicitly calculated as follows,
\begin{equation}
	\begin{aligned}
		&\frac{\partial u_{i}^t(\boldsymbol{x})}{\partial \tilde m_{i j}^{\mathrm{in}}}=x_{j}^t, \\
		&\frac{\partial u_{i}^t(\boldsymbol{x})}{\partial \tilde  m_{i j}^{\mathrm{rec}}}=\left(1-\tilde\pi_{i j}^{\mathrm{rec}}\right) r_j^{t-1}(\boldsymbol{x})+\epsilon_{i}^{t}(\boldsymbol{x}) \frac{\tilde \mu_{i j}^{\mathrm{rec}}  \tilde\pi_{i j}^{\mathrm{rec}}\left(r_j^{t-1}(\boldsymbol{x})\right)^{2}}{\Delta_i^{t-1}(\boldsymbol{x})}, \\
		&\frac{\partial u_{i}^t(\boldsymbol{x})}{\partial \tilde \pi_{i j}^{\mathrm{rec}}}=-\tilde m_{i j}^{\mathrm{rec}} r_j^{t-1}(\boldsymbol{x})+\epsilon_{i}^{t}(\boldsymbol{x}) \frac{\left(\left(\tilde m_{i j}^{\mathrm{rec}}\right)^{2}\left(1-2 \tilde \pi_{i j}^{\mathrm{rec}}\right)-\tilde \Xi_{i j}^{\mathrm{rec}}\right)\left(r_j^{t-1}(\boldsymbol{x})\right)^{2}}{2 \Delta_i^{t-1}(\boldsymbol{x})}, \\
		&\frac{\partial u_{i}^t(\boldsymbol{x})}{\partial \tilde \Xi_{i j}^{\mathrm{rec}}}=\epsilon_{i}^{t}(\boldsymbol{x}) \frac{\left(1-\tilde \pi_{i j}^{\mathrm{rec}}\right)\left(r_j^{t-1}(\boldsymbol{x})\right)^{2}}{2 \Delta_i^{t-1}(\boldsymbol{x})}.
	\end{aligned}
\end{equation}

For the output layer, the gradient of $f(\boldsymbol{x})$ respect to $\theta_{i}\equiv\{\theta_{i}^{\mathrm{out}}\}$ is given by
\begin{equation}
	\frac{\partial f(\boldsymbol{x})}{\partial \theta_{i}} = \frac{\partial f(\boldsymbol{x})}{\partial \tilde \theta_{i}}\frac{\partial \tilde \theta_{i}}{\partial \theta_{i}},
\end{equation}
where $\frac{\partial f(\boldsymbol{x})}{\partial \tilde{\theta}_{i}}$ can be explicitly derived as follows,
\begin{equation}
	\begin{aligned}
		\frac{\partial f(\boldsymbol{x})}{\partial \tilde m_{i}^{\text {out }}}&=\left(1-\tilde\pi_{i}^{\text {out }}\right) r_{i}^T(\boldsymbol{x})+\epsilon^{\text {out}}(\boldsymbol{x})\frac{\left(\tilde\mu_{i}^{\text {out }}\tilde \pi_{i}^{\text {out }}\right)\left(r_{i}^T(\boldsymbol{x})\right)^{2}}{\Delta^{\text {out }}}, \\
		\frac{\partial f(\boldsymbol{x})}{\partial \tilde\pi_{i}^{\text {out }}}&=-\tilde m_{i}^{\text {out }} r_{i}^T(\boldsymbol{x})+\epsilon^{\text {out}}(\boldsymbol{x})\frac{\left(\left(\tilde m_{i}^{\text {out }}\right)^{2}\left(1-2\tilde \pi_{i}^{\text {out }}\right)-\tilde\Xi_{i}^{\text {out }}\right)\left(r_{i}^T(\boldsymbol{x})\right)^{2}}{2 \Delta^{\text {out }}}, \\
		\frac{\partial f(\boldsymbol{x})}{\partial \tilde\Xi_{i}^{\text {out }}}&=\epsilon^{\text {out }}(\boldsymbol{x})\frac{\left(1-\tilde\pi_{i}^{\text {out }}\right)\left(r_{i}^T(\boldsymbol{x})\right)^{2}}{2 \Delta^{\text {out }}}.
	\end{aligned}
\end{equation}

Next, we give the detailed calculation for each term of $\Theta(\boldsymbol{x},\boldsymbol{x}^{\prime})$. The recurrent-mean related term is dervied as
\begin{equation}
	\begin{aligned}
		&\sum_{ij} \frac{\partial f(\boldsymbol{x})}{\partial m_{ij}^{\mathrm{rec}}} \frac{\partial f(\boldsymbol{x}^{\prime})}{\partial m_{ij}^{\mathrm{rec}}}
		= \sum_{ij} \frac{\alpha^2 \sigma_{\mathrm{rec}}^2}{N^2} \sum_{t=1}^{T}  \sum_{t^{\prime}=1}^{T} \left(\delta_i^t(\boldsymbol{x})\delta_i^{t^{\prime}}(\boldsymbol{x}^{\prime})\right)\left((1-\tilde\pi_{ij}^{\mathrm{rec}})^2 r_j^{t-1}(\boldsymbol{x})r_j^{t^{\prime}-1}(\boldsymbol{x}^{\prime})\right)\\
		&+\sum_{ij} \frac{\alpha^2 \sigma_{\mathrm{rec}}^2}{N^2} \sum_{t=1}^{T} \sum_{t^{\prime}=1}^{T} \left(\delta_i^t(\boldsymbol{x})\delta_i^{t^{\prime}}(\boldsymbol{x}^{\prime})\right)\left(\epsilon_i^t(\boldsymbol{x})\epsilon_i^{t^{\prime}}(\boldsymbol{x}^{\prime}) \frac{(\tilde\mu_{ij}^{\mathrm{rec}})^2(\tilde\pi_{ij}^{\mathrm{rec}})^2}{\Delta_i^{t-1}(\boldsymbol{x})\Delta_i^{t^{\prime}-1}(\boldsymbol{x}^{\prime})}(r_j^{t-1}(\boldsymbol{x}))^2(r_j^{t^{\prime}-1}(\boldsymbol{x}^{\prime}))^2\right)\\
		&\approx \sum_{t=1}^{T} \sum_{t^{\prime}=1}^{T}   \Pi^{\left(t, t^{\prime}\right)}\left(\boldsymbol{x}, \boldsymbol{x}^{\prime}\right)\alpha^2 \sigma_{\mathrm{rec}}^2\left( \mathrm{F}_{\phi}\left[\boldsymbol{K}^{\left(t, t\right)}\left(\boldsymbol{x}, \boldsymbol{x}^{\prime}\right)\right] + \delta_{\boldsymbol{x}=\boldsymbol{x}^{\prime}}\delta_{t=t^{\prime}} \mathbb{E}\left[(\tilde\pi_{ij}^{\mathrm{rec}})^2\right]\frac{\sigma_{\mathrm{rec}}^2}{N}\frac{2\mathrm{F}_{\phi^2}\left[\boldsymbol{K}^{\left(t, t\right)}\left(\boldsymbol{x}, \boldsymbol{x}\right)\right]}{\mathrm{F}_{\phi}\left[\boldsymbol{K}^{\left(t, t\right)}\left(\boldsymbol{x}, \boldsymbol{x}\right)\right]}\right)\\
		&= \sum_{t=1}^{T}\sum_{t^{\prime}=1}^{T}  \Pi^{\left(t, t^{\prime}\right)}\left(\boldsymbol{x}, \boldsymbol{x}^{\prime}\right)\alpha^2 \sigma_{\mathrm{rec}}^2 \mathrm{F}_{\phi}\left[\boldsymbol{K}^{\left(t, t^{\prime}\right)}\left(\boldsymbol{x}, \boldsymbol{x}^{\prime}\right)\right].
	\end{aligned}
\end{equation}
The recurrent-spike-mass term is given by
\begin{equation}
	\begin{aligned}
		&\sum_{ij} \frac{\partial f(\boldsymbol{x})}{\partial \pi_{ij}^{\mathrm{rec}}} \frac{\partial f(\boldsymbol{x}^{\prime})}{\partial \pi_{ij}^{\mathrm{rec}}}
		= \sum_{ij} \frac{\alpha^2}{N} \sum_{t=1}^{T}  \sum_{t^{\prime}=1}^{T} \left(\delta_i^t(\boldsymbol{x})\delta_i^{t^{\prime}}(\boldsymbol{x}^{\prime})\right)\left((\tilde m_{ij}^{\mathrm{rec}})^2 r_j^{t-1}(\boldsymbol{x})r_j^{t^{\prime}-1}(\boldsymbol{x}^{\prime})\right)\\
		&+\sum_{ij} \frac{\alpha^2}{N} \sum_{t=1}^{T}  \sum_{t^{\prime}=1}^{T} \left(\delta_i^t(\boldsymbol{x})\delta_i^{t^{\prime}}(\boldsymbol{x}^{\prime})\right)\left(\epsilon_i^t(\boldsymbol{x})\epsilon_i^{t^{\prime}}(\boldsymbol{x}^{\prime}) \frac{\left(\left(\tilde m_{i j}^{\mathrm{rec}}\right)^{2}\left(1-2 \tilde \pi_{i j}^{\mathrm{rec}}\right)-\tilde \Xi_{i j}^{\mathrm{rec}}\right)^2}{4\Delta_i^{t-1}(\boldsymbol{x})\Delta_i^{t^{\prime}-1}(\boldsymbol{x}^{\prime})}(r_j^{t-1}(\boldsymbol{x}))^2(r_j^{t^{\prime}-1}(\boldsymbol{x}^{\prime}))^2\right)\\
		&\approx \sum_{t=1}^{T} \sum_{t^{\prime}=1}^{T} \Pi^{\left(t, t^{\prime}\right)}\left(\boldsymbol{x}, \boldsymbol{x}^{\prime}\right)\alpha^2  \left(\sigma_{\mathrm{rec}}^2\mathrm{F}_{\phi}\left[\boldsymbol{K}^{\left(t, t^{\prime}\right)}\left(\boldsymbol{x}, \boldsymbol{x}^{\prime}\right)\right] + \delta_{\boldsymbol{x}=\boldsymbol{x}^{\prime}}\delta_{t=t^{\prime}} \frac{3\sigma_{\mathrm{rec}}^4-\sigma_{\mathrm{rec}}^2+\frac{1}{3}}{N}\frac{\mathrm{F}_{\phi^2}\left[\boldsymbol{K}^{\left(t, t\right)}\left(\boldsymbol{x}, \boldsymbol{x}\right)\right]}{2\mathrm{F}_{\phi}\left[\boldsymbol{K}^{\left(t, t\right)}\left(\boldsymbol{x}, \boldsymbol{x}\right)\right]}\right).\\
	\end{aligned}
\end{equation}
The recurrent-variance term is given by
\begin{equation}
	\begin{aligned}
		\sum_{ij} \frac{\partial f(\boldsymbol{x})}{\partial \Xi_{ij}^{\mathrm{rec}}} \frac{\partial f(\boldsymbol{x}^{\prime})}{\partial \Xi_{ij}^{\mathrm{rec}}}
		&=\sum_{ij} \frac{\alpha^2 }{N^3} \sum_{t=1}^{T}  \sum_{t^{\prime}=1}^{T} \left(\delta_i^t(\boldsymbol{x})\delta_i^{t^{\prime}}(\boldsymbol{x}^{\prime})\right)\\
		&\qquad\left(\epsilon_i^t(\boldsymbol{x})\epsilon_i^{t^{\prime}}(\boldsymbol{x}^{\prime}) \frac{(1-\tilde\pi_{ij}^{\mathrm{rec}})^2}{4\Delta_i^{t-1}(\boldsymbol{x})\Delta_i^{t^{\prime}-1}(\boldsymbol{x}^{\prime})}(r_j^{t-1}(\boldsymbol{x}))^2(r_j^{t^{\prime}-1}(\boldsymbol{x}^{\prime}))^2\right)\\
		&\approx \sum_{t=1}^{T}  \Pi^{\left(t, t\right)}\left(\boldsymbol{x}, \boldsymbol{x}^{\prime}\right)\frac{\alpha^2 }{N}\delta_{\boldsymbol{x}=\boldsymbol{x}^{\prime}}\delta_{t=t'} \frac{\mathrm{F}_{\phi^2}\left[\boldsymbol{K}^{\left(t, t\right)}\left(\boldsymbol{x}, \boldsymbol{x}\right)\right]}{2\mathrm{F}_{\phi}\left[\boldsymbol{K}^{\left(t, t\right)}\left(\boldsymbol{x}, \boldsymbol{x}\right)\right]}.\\
	\end{aligned}
\end{equation}
The input-mean related term is given by
\begin{equation}
	\begin{aligned}
		\sum_{ij} \frac{\partial f(\boldsymbol{x})}{\partial m_{ij}^{\mathrm{in}}} \frac{\partial f(\boldsymbol{x}^{\prime})}{\partial m_{ij}^{\mathrm{in}}}
		&= \sum_{ij} \frac{\alpha^2 \sigma_{\mathrm{in}}^2}{NN_{\mathrm{in}}} \sum_{t=1}^{T}  \sum_{t^{\prime}=1}^{T} \left(\delta_i^t(\boldsymbol{x})\delta_i^{t^{\prime}}(\boldsymbol{x}^{\prime})\right)\left((1-\tilde\pi_{ij}^{\mathrm{in}})^2 x_{t,j} x^{\prime}_{t^{\prime},j}\right)\\
		&\approx \sum_{t=1}^{T}\sum_{t^{\prime}=1}^{T}  \Pi^{\left(t, t^{\prime}\right)}\left(\boldsymbol{x}, \boldsymbol{x}^{\prime}\right)\frac{\alpha^2 \sigma_{\mathrm{in}}^2\langle \boldsymbol{x}_{t}, \boldsymbol{x}^{\prime}_{t^{\prime}}\rangle}{N_{\mathrm{in}}}.
	\end{aligned}
\end{equation}
The output-mean related term is given by
\begin{equation}
	\begin{aligned}\label{fm_out}
		&\sum_{i} \frac{\partial f(\boldsymbol{x})}{\partial m_{i}^{\mathrm{out}}} \frac{\partial f(\boldsymbol{x}^{\prime})}{\partial m_{i}^{\mathrm{out}}}
		= \sum_{i} \frac{ \sigma_{\mathrm{out}}^2}{N}(1-\tilde\pi_{i}^{\mathrm{out}})^2 r_i^{T}(\boldsymbol{x})r_i^{T}(\boldsymbol{x}^{\prime})\\
		&+\sum_{i} \frac{ \sigma_{\mathrm{out}}^2}{N} \left(\epsilon^{\mathrm{out}}(\boldsymbol{x})\epsilon^{\mathrm{out}}(\boldsymbol{x}^{\prime}) \frac{(\tilde\mu_{i}^{\mathrm{out}})^2(\tilde\pi_{i}^{\mathrm{out}})^2}{\Delta_i^{\mathrm{out}}(\boldsymbol{x})\Delta_i^{\mathrm{out}}(\boldsymbol{x}^{\prime})}(r_i^{T}(\boldsymbol{x}))^2(r_i^{T}(\boldsymbol{x}^{\prime}))^2\right)\\
		&\approx \sigma_{\mathrm{out}}^2 \Biggl(\mathrm{F}_{\phi}\left[\boldsymbol{K}^{\left(T+1, T+1\right)}\left(\boldsymbol{x}, \boldsymbol{x}'\right)\right]\\ &+\epsilon^{\mathrm{out}}(\boldsymbol{x})\epsilon^{\mathrm{out}}(\boldsymbol{x}^{\prime}) \mathbb{E}\left[(\tilde\pi_{i}^{\mathrm{out}})^2\right]\frac{\sigma_{\mathrm{out}}^2}{N}\frac{\mathrm{F}_{\phi^2}\left[\boldsymbol{K}^{\left(T+1, T+1\right)}\left(\boldsymbol{x}, \boldsymbol{x}^{\prime}\right)\right]}{\frac{1}{2}\sqrt{\mathrm{F}_{\phi}\left[\boldsymbol{K}^{\left(T+1, T+1\right)}\left(\boldsymbol{x}, \boldsymbol{x}\right)\right]\mathrm{F}_{\phi}\left[\boldsymbol{K}^{\left(T+1, T+1\right)}\left(\boldsymbol{x}^{\prime}, \boldsymbol{x}^{\prime}\right)\right] }}\Biggl)\\
		&=  \sigma_{\mathrm{out}}^2 \mathrm{F}_{\phi}\left[\boldsymbol{K}^{\left(T+1, T+1\right)}\left(\boldsymbol{x}, \boldsymbol{x}'\right)\right]
	\end{aligned}
\end{equation}
The output-spike-mass related term is given by
\begin{equation}
	\begin{aligned}\label{fpi_out}
		&\sum_{i} \frac{\partial f(\boldsymbol{x})}{\partial \pi_{i}^{\mathrm{out}}} \frac{\partial f(\boldsymbol{x}^{\prime})}{\partial \pi_{i}^{\mathrm{out}}}
		= \sum_{i} (\tilde m_{i}^{\mathrm{out}})^2 r_i^{T}(\boldsymbol{x})r_i^{T}(\boldsymbol{x}^{\prime})\\
		&+\sum_{i}  \left(\epsilon^{\mathrm{out}}(\boldsymbol{x})\epsilon^{\mathrm{out}}(\boldsymbol{x}^{\prime}) \frac{\left(\left(\tilde m_{i}^{\text {out }}\right)^{2}\left(1-2 \tilde \pi_{i}^{\text {out }}\right)-\tilde \Xi_{i}^{\text {out }}\right)^2}{4\Delta^{\mathrm{out}}(\boldsymbol{x})\Delta^{\mathrm{out}}(\boldsymbol{x}^{\prime})}(r_i^{T}(\boldsymbol{x}))^2(r_i^{T}(\boldsymbol{x}^{\prime}))^2\right)\\
		&\approx  \sigma_{\mathrm{out}}^2\mathrm{F}_{\phi}\left[\boldsymbol{K}^{\left(T+1, T+1\right)}\left(\boldsymbol{x}, \boldsymbol{x}^{\prime}\right)\right]\\
		&+ \epsilon^{\mathrm{out}}(\boldsymbol{x})\epsilon^{\mathrm{out}}(\boldsymbol{x}^{\prime}) \frac{3\sigma_{\mathrm{out}}^4-\sigma_{\mathrm{out}}^2+\frac{1}{3}}{N}\frac{\mathrm{F}_{\phi^2}\left[\boldsymbol{K}^{\left(T+1, T+1\right)}\left(\boldsymbol{x}, \boldsymbol{x}^{\prime}\right)\right]}{2\sqrt{\mathrm{F}_{\phi}\left[\boldsymbol{K}^{\left(T+1, T+1\right)}\left(\boldsymbol{x}, \boldsymbol{x}\right)\right]\mathrm{F}_{\phi}\left[\boldsymbol{K}^{\left(T+1, T+1\right)}\left(\boldsymbol{x}^{\prime}, \boldsymbol{x}^{\prime}\right)\right] }}.\\
	\end{aligned}
\end{equation}
The output-variance related sum is given by
\begin{equation}
	\begin{aligned}\label{fxi_out}
		\sum_{i} \frac{\partial f(\boldsymbol{x})}{\partial \Xi_{i}^{\mathrm{out}}} \frac{\partial f(\boldsymbol{x}^{\prime})}{\partial \Xi_{i}^{\mathrm{out}}}
		&=\sum_{i} \frac{1}{N^2}\epsilon^{\mathrm{out}}(\boldsymbol{x})\epsilon^{\mathrm{out}}(\boldsymbol{x}^{\prime}) \frac{\left(1- \tilde\pi_{i}^{\text {out }}\right)^2}{4\Delta^{\mathrm{out}}(\boldsymbol{x})\Delta^{\mathrm{out}}(\boldsymbol{x}^{\prime})}(r_i^{T}(\boldsymbol{x}))^2(r_i^{T}(\boldsymbol{x}^{\prime}))^2\\
		&\approx\frac{1}{N}\frac{\mathrm{F}_{\phi^2}\left[\boldsymbol{K}^{\left(T+1, T+1\right)}\left(\boldsymbol{x}, \boldsymbol{x}^{\prime}\right)\right]}{2\sqrt{\mathrm{F}_{\phi}\left[\boldsymbol{K}^{\left(T+1, T+1\right)}\left(\boldsymbol{x}, \boldsymbol{x}\right)\right]\mathrm{F}_{\phi}\left[\boldsymbol{K}^{\left(T+1, T+1\right)}\left(\boldsymbol{x}^{\prime}, \boldsymbol{x}^{\prime}\right)\right] }}.\\
	\end{aligned}
\end{equation}
By collecting all the above terms together, we obtain the formula of RNTK $\Theta\left(\boldsymbol{x}, \boldsymbol{x}^{\prime}\right)$ as follows,
\begin{equation}
	\begin{aligned}
		\Theta\left(\boldsymbol{x}, \boldsymbol{x}^{\prime}\right)
		&= \sum_{t=1}^{T}\sum_{t^{\prime}=1}^{T}  \Pi^{\left(t, t^{\prime}\right)}\left(\boldsymbol{x}, \boldsymbol{x}^{\prime}\right) \Biggl(2\alpha^2 \sigma_{\mathrm{rec}}^2\mathrm{F}_{\phi}\left[\boldsymbol{K}^{\left(t, t^{\prime}\right)}\left(\boldsymbol{x}, \boldsymbol{x}^{\prime}\right)\right] + \frac{\alpha^2 \sigma_{\mathrm{in}}^2\langle \boldsymbol{x}_{t}, \boldsymbol{x}^{\prime}_{t^{\prime}}\rangle}{N_{\mathrm{in}}}\\ 
		&  + \delta_{\boldsymbol{x}=\boldsymbol{x^{\prime}}}\delta_{t=t^{\prime}} \frac{\alpha^2}{2N}\left(3\sigma_{\mathrm{rec}}^4-\sigma_{\mathrm{rec}}^2+\frac{4}{3}\right)\frac{\mathrm{F}_{\phi^2}\left[\boldsymbol{K}^{\left(t, t\right)}\left(\boldsymbol{x}, \boldsymbol{x}\right)\right]}{\mathrm{F}_{\phi}\left[\boldsymbol{K}^{\left(t, t\right)}\left(\boldsymbol{x}, \boldsymbol{x}\right)\right]}\Biggl)+  2\sigma_{\mathrm{out}}^2\mathrm{F}_{\phi}\left[\boldsymbol{K}^{\left(T+1, T+1\right)}\left(\boldsymbol{x}, \boldsymbol{x}^{\prime}\right)\right]\\
		& + \frac{\epsilon^{\mathrm{out}}(\boldsymbol{x})\epsilon^{\mathrm{out}}(\boldsymbol{x}^{\prime})}{2N}\left(3\sigma_{\mathrm{out}}^4-\sigma_{\mathrm{out}}^2+\frac{4}{3}\right)\frac{\mathrm{F}_{\phi^2}\left[\boldsymbol{K}^{\left(T+1, T+1\right)}\left(\boldsymbol{x}, \boldsymbol{x}^{\prime}\right)\right]}{\sqrt{\mathrm{F}_{\phi}\left[\boldsymbol{K}^{\left(T+1, T+1\right)}\left(\boldsymbol{x}, \boldsymbol{x}\right)\right]\mathrm{F}_{\phi}\left[\boldsymbol{K}^{\left(T+1, T+1\right)}\left(\boldsymbol{x}^{\prime}, \boldsymbol{x}^{\prime}\right)\right] }}.
	\end{aligned}
\end{equation}

A careful inspection shows that the fluctuation-related terms has a smaller magnitude as the network size increases. For a finite-sized network,
these terms serve as a correction to the mean-field-limit result.
In the mean-field limit,
we can discard all these terms, and achieve the final compact formula as
\begin{equation}\label{Theta}
	\begin{aligned}
		\Theta\left(\boldsymbol{x}, \boldsymbol{x}^{\prime}\right)
		&= \sum_{t=1}^{T}\sum_{t^{\prime}=1}^{T}  \Pi^{\left(t, t^{\prime}\right)}\left(\boldsymbol{x}, \boldsymbol{x}^{\prime}\right) \Biggl(2\alpha^2 \sigma_{\mathrm{rec}}^2\mathrm{F}_{\phi}\left[\boldsymbol{K}^{\left(t, t^{\prime}\right)}\left(\boldsymbol{x}, \boldsymbol{x}^{\prime}\right)\right] + \frac{\alpha^2 \sigma_{\mathrm{in}}^2\langle \boldsymbol{x}_{t}, \boldsymbol{x}^{\prime}_{t^{\prime}}\rangle}{N_{\mathrm{in}}}\Biggl)\\
		&+  2\sigma_{\mathrm{out}}^2\mathrm{F}_{\phi}\left[\boldsymbol{K}^{\left(T+1, T+1\right)}\left(\boldsymbol{x}, \boldsymbol{x}^{\prime}\right)\right].	
	\end{aligned}
\end{equation}
The recursive formula for kernels during the backward phase [Eq.~(\ref{GammafromPi}) and Eq.~(\ref{PiTT})] can also be simplified as
\begin{equation}
	\begin{aligned}
		\Gamma^{\left(t, t^{\prime}\right)}\left(\boldsymbol{x}, \boldsymbol{x}^{\prime}\right) 
		&= \sigma_{\mathrm{rec}}^{2}\mathrm{F}_{\phi^{\prime}}\left[\boldsymbol{K}^{\left(t+1, t^{\prime}+1\right)}\left(\boldsymbol{x}, \boldsymbol{x}^{\prime}\right)\right]\Pi^{\left(t+1, t^{\prime}+1\right)}\left(\boldsymbol{x}, \boldsymbol{x}^{\prime}\right),\\
		\Pi^{\left(T, T\right)}\left(\boldsymbol{x}, \boldsymbol{x}^{\prime}\right)
		 &= \sigma_{\mathrm{out}}^{2} \mathrm{F}_{\phi^{\prime}}\left[\boldsymbol{K}^{\left(T+1, T+1\right)}\left(\boldsymbol{x}, \boldsymbol{x}'\right)\right].
	\end{aligned}
\end{equation}
Therefore, given two inputs $\boldsymbol{x}$ and $\boldsymbol{x}^{\prime}$, the 
RNTK $\Theta\left(\boldsymbol{x}, \boldsymbol{x}^{\prime}\right)$ converges to a constant independent of the realization of the trainable parameters $\boldsymbol{\theta}$ as well as
standard Gaussian variables $\boldsymbol{\epsilon}^t(\boldsymbol{x})$, $\epsilon^{\mathrm{out}}(\boldsymbol{x})$ during training for RNNs of infinite width.

\subsection{Algorithms for the RNTK computation}
The algorithm for computing the RNTK for a general $\alpha$ ($\alpha\in[0,1]$)
is sketched in the Algorithm~\ref{alg1}.
\begin{algorithm}[H]
	\caption{Compute SaS-RNTK $\Theta(\boldsymbol{x},\boldsymbol{x}^{\prime})$ for $\alpha\in[0,1]$}\label{alg1}
	\begin{algorithmic}[1]
	\Require $N$,~$N_{\mathrm{in}}$,~$T$,~$\alpha$,~$\sigma_{\mathrm{in}}$,~$\sigma_{\mathrm{rec}}$,~$\sigma_{\mathrm{out}}$,~$\sigma_h$,~$\boldsymbol{x}$,~$\boldsymbol{x}^{\prime}$
	\Ensure $\Theta(\boldsymbol{x},\boldsymbol{x}^{\prime})$
	\For{$t=0,1,...,T-1$} \Comment{Forward Pass}
		\For{$t_{\mathrm{in}}=1,2,...,t$}
		\State Compute $\Omega^{\left(t_{\mathrm{in}}, t\right)}\left(\boldsymbol{x}, \boldsymbol{x}^{\prime}\right)$ and $\Omega^{\left(t, t_{\mathrm{in}}\right)}\left(\boldsymbol{x}, \boldsymbol{x}^{\prime}\right)$ according to Eq. (\ref{OmegafromSigma});
		\EndFor
		\For{$t_{\mathrm{in}}=0,1,...,t$}
		\State Compute $\Sigma^{\left(t_{\mathrm{in}}, t\right)}\left(\boldsymbol{x}, \boldsymbol{x}^{\prime}\right)$ and $\Sigma^{\left(t, t_{\mathrm{in}}\right)}\left(\boldsymbol{x}, \boldsymbol{x}^{\prime}\right)$ according to Eq. (\ref{SigmafromOmega}) and Eq.~(\ref{Sigma0});
		\EndFor
	\EndFor
	\For{$t=T,T-1,...,1$} \Comment{Backward Pass}
		\For{$t_{\mathrm{in}}=t,t+1,...,T-1$}
		\State Compute $\Gamma^{\left(t_{\mathrm{in}}, t\right)}\left(\boldsymbol{x}, \boldsymbol{x}^{\prime}\right)$ and $\Gamma^{\left(t, t_{\mathrm{in}}\right)}\left(\boldsymbol{x}, \boldsymbol{x}^{\prime}\right)$ according to Eq. (\ref{GammafromPi});
		\EndFor
		\For{$t_{\mathrm{in}}=t,t+1,...,T$}
		\State Compute $\Pi^{\left(t_{\mathrm{in}}, t\right)}\left(\boldsymbol{x}, \boldsymbol{x}^{\prime}\right)$ and $\Pi^{\left(t, t_{\mathrm{in}}\right)}\left(\boldsymbol{x}, \boldsymbol{x}^{\prime}\right)$ according to Eqs. (\ref{PifromGamma}), (\ref{PiTT}), and (\ref{PiTt});
		\EndFor
	\EndFor	
	\State Compute $\Theta(\boldsymbol{x},\boldsymbol{x}^{\prime})$ according to Eq. (\ref{Theta}).
	\end{algorithmic}
\end{algorithm}

When $\alpha$ equals to $1$, the computation will be greatly simplified. 
More precisely, in the forward pass, the auxiliary kernel $\Omega^{t,t^{\prime}}(\boldsymbol{x},\boldsymbol{x}^{\prime})$ reduces
to the forward pass kernel $\Sigma^{t,t^{\prime}}(\boldsymbol{x},\boldsymbol{x}^{\prime})$,
and the forward pass recursive formulas are simplified as
\begin{equation}\label{Sigma_alpha1}
	\begin{aligned}
		\Sigma^{\left(0, 0\right)}\left(\boldsymbol{x}, \boldsymbol{x}^{\prime}\right) & = \delta_{\boldsymbol{x}=\boldsymbol{x^{\prime}}}\sigma_h^2,\\
		\Sigma^{\left(t, 0\right)}\left(\boldsymbol{x}, \boldsymbol{x}^{\prime}\right) & = 0,\\
		\Sigma^{\left(0, t^{\prime}\right)}\left(\boldsymbol{x}, \boldsymbol{x}^{\prime}\right) & = 0,\\
		\Sigma^{\left(t, t^{\prime}\right)}\left(\boldsymbol{x}, \boldsymbol{x}^{\prime}\right) &= \left(\sigma_{\mathrm{rec}}^{2}+\frac{1}{2}\delta_{\boldsymbol{x}=\boldsymbol{x^{\prime}},t=t^{\prime}}\right) \mathrm{F}_{\phi}\left[\boldsymbol{K}^{\left(t, t^{\prime}\right)}\left(\boldsymbol{x}, \boldsymbol{x}^{\prime}\right)\right]+\frac{\sigma_{\mathrm{in}}^{2}}{M}\left\langle\boldsymbol{x}_{t}, \boldsymbol{x}^{\prime}_{t^{\prime}}\right\rangle.
	\end{aligned}
\end{equation}
Similarly, in the backward pass, the auxiliary kernel $\Gamma^{t,t^{\prime}}(\boldsymbol{x},\boldsymbol{x}^{\prime})$ reduces
to the forward pass kernel $\Pi^{t,t^{\prime}}(\boldsymbol{x},\boldsymbol{x}^{\prime})$, and we obtain the simplified recursive formulas as 
\begin{equation}\label{Pi_alpha1}
	\begin{aligned}
		\Pi^{\left(T, T\right)}\left(\boldsymbol{x}, \boldsymbol{x}^{\prime}\right) 
		&=\sigma_{\mathrm{out}}^{2}\mathrm{F}_{\phi^{\prime}}\left[\boldsymbol{K}^{\left(T+1, T+1\right)}\left(\boldsymbol{x}, \boldsymbol{x}^{\prime}\right)\right] ,\\
		\Pi^{\left(t, T\right)}\left(\boldsymbol{x}, \boldsymbol{x}^{\prime}\right) & =0,\\ 
		\Pi^{\left(t, t^{\prime}\right)}\left(\boldsymbol{x}, \boldsymbol{x}^{\prime}\right)&=\sigma_{\mathrm{rec}}^{2}\mathrm{F}_{\phi^{\prime}}\left[\boldsymbol{K}^{\left(t+1, t^{\prime}+1\right)}\left(\boldsymbol{x}, \boldsymbol{x}^{\prime}\right)\right]\Pi^{\left(t+1, t^{\prime}+1\right)}\left(\boldsymbol{x}, \boldsymbol{x}^{\prime}\right).\\
	\end{aligned}
\end{equation}
From the above recursive formulas, we find that when $t$ does not equal to $t^{\prime}$, 
$\Pi^{t,t^{\prime}}(\boldsymbol{x},\boldsymbol{x}^{\prime})$ will
vanish. Thus, we only need to consider the kernels when $t$ equals to $t^{\prime}$, which saves a large amount of computational cost.
The RNTK $\Theta(\boldsymbol{x},\boldsymbol{x}^{\prime})$ for $\alpha=1$ is thus given by
\begin{equation}\label{Theta_alpha1}
	\begin{aligned}
		\Theta\left(\boldsymbol{x}, \boldsymbol{x}^{\prime}\right)
		&= \sum_{t=1}^{T}  \Pi^{\left(t, t\right)}\left(\boldsymbol{x}, \boldsymbol{x}^{\prime}\right) \left(2 \sigma_{\mathrm{rec}}^2\mathrm{F}_{\phi}\left[\boldsymbol{K}^{\left(t, t\right)}\left(\boldsymbol{x}, \boldsymbol{x}^{\prime}\right)\right] + \frac{\alpha^2 \sigma_{\mathrm{in}}^2\langle \boldsymbol{x}_{t}, \boldsymbol{x}^{\prime}_{t}\rangle}{N_{\mathrm{in}}}\right)\\
		&\quad + 2\sigma_{\mathrm{out}}^2\mathrm{F}_{\phi}\left[\boldsymbol{K}^{\left(T+1, T+1\right)}\left(\boldsymbol{x}, \boldsymbol{x}^{\prime}\right)\right].\\
	\end{aligned}
\end{equation}
In addition, the more efficient algorithm for $\alpha=1$ is sketched by the Algorithm~\ref{alg2}.
\begin{algorithm}[H]
	\caption{Compute SaS-RNTK $\Theta(\boldsymbol{x},\boldsymbol{x}^{\prime})$ for $\alpha=1$}\label{alg2}
	\begin{algorithmic}[1]
		\Require $N$,~$N_{\mathrm{in}}$,~$T$,~$\alpha$,~$\sigma_{\mathrm{in}}$,~$\sigma_{\mathrm{rec}}$,~$\sigma_{\mathrm{out}}$,~$\sigma_h$,~$\boldsymbol{x}$,~$\boldsymbol{x}^{\prime}$
		\Ensure $\Theta(\boldsymbol{x},\boldsymbol{x}^{\prime})$
		\For{$t=0,1,...,T-1$} \Comment{Forward Pass}
		\State Compute $\Sigma^{\left(t, t\right)}\left(\boldsymbol{x}, \boldsymbol{x}^{\prime}\right)$ according to Eq. (\ref{Sigma_alpha1});
		\EndFor
		\For{$t=T,T-1,...,1$} \Comment{Backward Pass}
		\State Compute $\Pi^{\left(t, t\right)}\left(\boldsymbol{x}, \boldsymbol{x}^{\prime}\right)$ according to Eq. (\ref{Pi_alpha1});
		\EndFor
		\State Compute $\Theta(\boldsymbol{x},\boldsymbol{x}^{\prime})$ according to Eq. (\ref{Theta_alpha1}).
	\end{algorithmic}
\end{algorithm}

We finally remark that, for the nonlinear transfer function ReLU, 
the Gaussian integrals $F_{\phi}(\boldsymbol{K})$ and $F_{\phi^{\prime}}(\boldsymbol{K})$ have closed forms as follows
\begin{equation}
	\begin{aligned}
		&\mathrm{F}_{\phi}[\boldsymbol{K}]=\mathbb{E}_{\boldsymbol{z}\sim\mathcal{N}(0,\boldsymbol{K})}[\phi(z_1)\phi(z_2)]=\frac{1}{2 \pi}\left(c(\pi-\arccos (\mathrm{c}))+\sqrt{1-\mathrm{c}^{2}}\right) \sqrt{K_{1} K_{2}}, \\
		&\mathrm{F}_{\phi^{\prime}}[\boldsymbol{K}]=\mathbb{E}_{\boldsymbol{z}\sim\mathcal{N}(0,\boldsymbol{K})}[\phi^{\prime}(z_1)\phi^{\prime}(z_2)]=\frac{1}{2 \pi}(\pi-\arccos (\mathrm{c})),\\
	\end{aligned}
\end{equation}
where $\boldsymbol{K}=\left[\begin{array}{ll} K_{1} & K_{3} \\ K_{3} & K_{2} \end{array}\right]$ and $c=K_3/\sqrt{K_1K_2}$.



\end{document}